\documentclass{article}
\usepackage{amsfonts}
\usepackage{amsmath}
\usepackage{eucal}
\usepackage{amssymb}
\usepackage{graphicx}
\usepackage{graphpap}  % for the diagrams!
\usepackage{epic, xypic} % for the diagrams!
\usepackage{texdraw}
\newcommand{\sq}{\sqrt{3}}
\newcommand{\di}[1]{\frac{#1}{\sqrt{3}}}
\renewcommand{\theequation}{\thesection.\arabic{equation}}
\newcommand{\cvd}{\begin{flushright}$\Box$\end{flushright}}
\newcommand{\ad}{{\rm ad}\;}
\newcommand{\dd}{{\rm d}}
\newcommand{\ee}{{\rm e}}
\newcommand{\tr}{{\rm Tr}\;}
\newcommand{\sgn}{{\rm sgn}\,}
\newcommand{\rif}[1]{(\ref{#1})}
\newcommand{\der}[2]{{\partial #1\over\partial #2}}
\newcommand{\bra}[1]{\left< #1\right|}
\newcommand{\ket}[1]{\left| #1\right>}
\newcommand{\sign}{{\mathrm{sign}}}
\newcommand{\eq}{\begin{equation}}
\newcommand{\feq}{\end{equation}}
\newcommand{\eqn}{\begin{eqnarray}}
\newcommand{\feqn}{\end{eqnarray}}
\newcommand{\arr}{\begin{eqnarray*}}
\newcommand{\farr}{\end{eqnarray*}}
\newcommand{\bea}{\begin{array}}
\newcommand{\ea}{\end{array}}
\newcommand{\dem}{{\frac 12}}
\newcommand{\inv}[1]{{1\over #1}}
\newcommand{\intS}{\int_\Sigma}
\newcommand{\intM}{\int_{\cal M}}
\newcommand{\intdM}{\int_{\partial\cal M}}
\newcommand{\dM}{{\partial\cal M}}
\newcommand{\M}{{\cal M}}
\newcommand{\RR}{{\cal R}}
\newcommand{\DD}{{\cal D}}
\newcommand{\HH}{{\cal H}}
\newcommand{\VV}{{\cal V}}
\newcommand{\lie}{{\cal L}}
\newcommand{\G}{{\cal G}}
\newcommand{\p}{\partial}
\newcommand{\w}{\wedge}
\newcommand{\rerg}{r_{\rm erg}}
\newcommand{\ig}{{g^{-1}}}
\newcommand{\ih}{{h^{-1}}}
\newcommand{\At}{{\tilde A}}
\newcommand{\lp}{\left(}
\newcommand{\rp}{\right)}
\newcommand{\de}{\partial}
\newcommand{\ld}{\ldots}
\newcommand{\al}{\alpha}
\newcommand{\tc}{\tilde c}
\newcommand{\cosech}{{\mathrm{cosech}}}
\font\mybb=msbm10 at 12pt
\def\bb#1{\hbox{\mybb#1}}
\def\bZ {\bb{Z}}
\def\bR {\bb{R}}
\def\bE {\bb{E}}
\def\bT {\bb{T}}
\def\bM {\bb{M}}
\def\bC {\bb{C}}
\def\bO {\bb{O}}
\def\bP {\bb{P}}
\newcommand{\cG}{{\cal G}}
\newcommand{\cHH}{{\cal H}}
\newcommand{\cP}{{\cal P}}

\newcommand{\SSu}{\rm Sum}
\newcommand{\Coe}{\rm Coefficient}
\newcommand{\Part}{\rm Part}
\newcommand{\QQ}{\rm QQ}
\newcommand{\Qm}{\rm Qm}
\newcommand{\QP}{\rm QP}
\newcommand{\Conj}{\rm Conj}
\newcommand{\Octp}{\rm OctP}
\newcommand{\Octps}{\rm OctPS}
\newcommand{\Append}{\rm Append}
\newcommand{\Do}{\rm Do}
\newcommand{\Array}{\rm Array}

\begin{document}
%%%%%%%%%%%%%%%%%%%%%%%%%%%%%%%%%%%%%%%%%%%%%%%%%%%%%%%%%%%%%%%%%%%%%%%%%%%%%%%%%%%%%%%%%%%%%%%%%
\begin{titlepage}
%\hfill{}
\begin{flushright}
UCB-PTH-07/07\\ IFIC/07$-$21\\ FTUV/07$-$0527\\
\end{flushright}
\vskip 10mm
\begin{center}
\renewcommand{\thefootnote}{\fnsymbol{footnote}}
{\Large \bf Mapping the geometry of the \boldmath{$F_4$} group}

\vskip 10mm {\large \bf { Fabio
Bernardoni$^{1}$\footnote{Fabio.Bernardoni@ific.uv.es}, Sergio
L.~Cacciatori$^{2}$\footnote{sergio.cacciatori@uninsubria.it},
Bianca L.~Cerchiai$^{3}$\footnote{BLCerchiai@lbl.gov} and
Antonio Scotti$^{4}$\footnote{antonio.scotti@gmail.com}}}\\
\renewcommand{\thefootnote}{\arabic{footnote}}
\setcounter{footnote}{0} \vskip 10mm {\small $^1$Departament de
F\'\i sica Te\`orica, IFIC, Universitat de Val\`encia - CSIC
\\ Apt. Correus 22085, E-46071 Val\`encia, Spain. \\

\vspace*{0.5cm}

$^2$ Dipartimento di Scienze Fisiche e Matematiche, \\
\hspace*{0.15cm} Universit\`a dell'Insubria, \\
\hspace*{0.15cm} Via Valleggio 11, I-22100 Como. \\
\vspace*{0.5cm}

$^3$ Lawrence Berkeley National Laboratory\\
Theory Group, Bldg 50A5104\\
1 Cyclotron Rd, Berkeley CA 94720 USA\\
\vspace*{0.5cm}

$^4$
Dipartimento di Matematica dell'Universit\`a di Milano,\\
Via Saldini 50, I-20133 Milano, Italy. \\

\vspace*{0.5cm}

$^5$ INFN, Sezione di Milano, Via Celoria 16, I-20133 Milano.

}
\end{center}
\vspace{0.8cm}
\begin{center}
{\bf Abstract}
\end{center}
{\small { In this paper we present a construction of the compact
form of the exceptional Lie group $F_4$ by exponentiating the
corresponding Lie algebra $f_4$. We realize $F_4$ as the
automorphisms group of the exceptional Jordan algebra, whose
elements are $3\times 3$ hermitian matrices with octonionic entries.
We use a parametrization which generalizes the Euler angles for
$SU(2)$ and is based on the fibration of $F_4$ via a $Spin(9)$
subgroup as a fiber. This technique allows us to determine an
explicit expression for the Haar invariant measure on the $F_4$
group manifold. Apart from shedding light on the structure of $F_4$
and its coset manifold $\bO\bP^2=F_4/Spin(9)$, the octonionic
projective plane, these results are a prerequisite for the study of
$E_6$, of which $F_4$ is a (maximal) subgroup.

}}

\end{titlepage}
%%%%%%%%%%%%%%%%%%%%%%%%%%%%%%%%%%%%%%%%%%%%%%%%%%%%%%%%%%%%%%%%%%%%%%%%%%%%%%
\section{Introduction.}
Simple Lie groups are well understood, starting from their complete
classification. However, often one encounters some points which require
a more detailed discussion or a new perspective. Our main interest,
as an application to physics, is the construction of the $E_6$ group
in a suitable parametrization adapted to perform non perturbative computations
in GUT theories. While searching for such a construction, we have found it
convenient to first determine an analog construction for its maximal subgroup
$F_4$, which deserves a complete analysis by itself.
Even though $F_4$ does not have a direct application to GUT theories, there
are other motivations to consider $F_4$ separately. For example, the
construction of integrable models on exceptional Lie groups and the
corresponding coset manifolds could give rise to new families of integrable
hierarchies. The interest for such problems is related to the fact that
these groups are exceptional, which contrasts with the infinity of the classical
series $A_n, B_n, C_n, D_n$. Of particular interest, from the mathematical
point of view, is the coset manifold $\bO\bP^2=F_4/Spin(9)$, the octonionic
projective plane.\\
However, our paper must be mainly thought of as a preparation for the
construction of the $E_6$ group, which will be presented in a separated
article. As the form of this group relevant for physics is the compact one,
we need in particular the compact form of the $F_4$ group.\\
Here we realize $F_4$ as the group of automorphisms of the exceptional
Jordan algebra. This is a $27$ dimensional abelian algebra whose elements
are $3\times 3$ hermitian matrices with octonionic entries.
The abelian product is obtained by symmetrizing the usual matrix product
(which takes into account the octonionic product). In section
\ref{algebra} we describe shortly the exceptional Jordan algebra and the
corresponding algebra $f_4$ of the infinitesimal automorphisms.\\
In section \ref{group} we construct the group $F_4$ by exponentiating the
algebra in an suitable way. The main idea is to obtain a generalized Euler
parametrization of the group, in the same spirit of our previous
papers \cite{noi}. However, here we also clarify the general strategy of the
construction and some technical points. In particular we show the
surjectivity of our map, a fact which we assumed to be true without
proof in our previous papers. Some technical details are put in the appendices,
including the fundamental Mathematica programs we used to compute the algebra.
All other calculations can be done by hand, as we have indeed done, so we do
not include the Mathematica programs we used to check them.\\
Some possible applications are reported in the conclusions.

%%%%%%%%%%%%%%%%%%%%%%%%%%%%%%%%%%%%%%%%%%%%%%%%%%%%%%%%%%%%%%%%%%%%%%%%%%%%%%%%%%%%%%%%%%%%%%%%%
\section{Construction of the \boldmath{$f_4$} algebra.} \label{algebra}
\setcounter{equation}{0}
The compact form of the $F_4$ exceptional Lie group can be realized as the automorphism group of the Jordan
algebra $J_3$ \cite{freudenthal,adams}, that is the algebra of $3\times 3$ octonionic hermitian matrices with product $\circ$ defined by
\eqn
A\circ B :=\frac 12 (A\cdot B+B\cdot A) \ ,
\feqn
where $A,B \in J_3$ and the dot is the usual product between matrices. Note that the generic $J_3$ matrix has the form
\eqn \label{A}
A=\lp
\begin{array}{ccc}
a_1 & o_1 & o_2 \\
o_1^* & a_2 & o_3 \\
o_2^* & o_3^* & a_3
\end{array}
\rp \ ,
\feqn
where $a_i$ are real numbers and $o_i$ are octonions.
Thus, in this way we obtain a $27$ dimensional representation for $F_4$. The irreducible $26$ dimensional representation
can be easily obtained by restricting the $27$ dimensional one to $ker (\ell)$ \cite{adams}, where $\ell$ is the linear operator
\eqn
\ell : J_3 \longrightarrow \bR\ , \quad A \mapsto \sum_{i=1}^3 A_{ii} \ .
\feqn
However, the $27$ dimensional representation is interesting because it can be  extended in a natural way to the $27$ dimensional
irreducible representation of the exceptional Lie group $E_6$. We will consider this extension in a future work.\\
If a Lie group is realized as the automorphism group of an algebra {$\cal A$}, its Lie algebra is then realized as the
algebra of derivations on {$\cal A$}. To obtain the matrix representation of the $f_4$ algebra we first define the linear
isomorphism
\eqn
&& \Phi : J_3 \longrightarrow \bR^{27} \ , \quad A\mapsto \Phi(A) \ , \cr
&& \Phi(A):=\lp
\begin{array}{c}
a_1 \\
\rho (o_1) \\
\rho (o_2) \\
a_2 \\
\rho (o_3) \\
a_3
\end{array}
\rp \ ,
\feqn
where $A$ is as in (\ref{A}) and $\rho$ is the linear isomorphism between the octonions $\bO$ and $\bR^8$
given by\footnote{our conventions about octonions are explained in App. \ref{app:ottonioni}}
\eqn
&& \rho: \bO \longrightarrow \bR^8 \ , \quad o=o^0 +\sum_{i=1}^7 o^i i_i \mapsto \rho (o) \ ,\cr
&& \rho(o):=\lp
\begin{array}{c}
o^0 \\ o^1 \\ o^2 \\ o^3 \\ o^4 \\ o^5 \\ o^6 \\ o^7
\end{array}
\rp \ .
\feqn
Next we define a $\circ$ product in $\bR^{27}$ by means of $\Phi$:
\eqn
x\circ y := \Phi (\Phi^{-1}(x) \circ \Phi^{-1}(y))\ , \quad \forall x,y \in \bR^{27} \ .
\feqn
The derivations on $J_3$ are then represented by matrices $M \in M(\bR,27)$ which must satisfy the condition
\eqn
M (x\circ y) =(Mx)\circ y +x\circ (My) \ , \quad \forall x,y \in \bR^{27} \ .
\feqn
These equations can be solved by means of Mathematica which gives in fact $52$ independent solutions $M_i$,
$i=1,\ldots 52$, which we choose to normalize with respect to the condition
\linebreak
$-\frac 16 Trace (M_i M_j)=\delta_{ij}$ and $[M_i, M_j]=-\sum_{k=1}^3\epsilon_{ijk} M_k$ for $i,j\in\{1,2,3\}$.
Let $\{e_a \}_{a=1}^{27}$ be the canonical base of $\bR^{27}$.
Since the irreducible representation is realized on $ker (\ell)$, we expect the linear combination
$(e_1+ e_{18}+e_{27})/\sqrt {3}$, which we will call $f_{27}$, to be in the kernel of all
the $M_i$, $i=1,\ldots,52$, as, in fact, can be easily
checked. It is then convenient to express the matrices with respect to the new base $\{f_a \}_{a=1}^{27}$ of $\bR^{27}$
defined by
\eqn
&& f_1:= (e_1-e_{18})/\sqrt 2 \ , \\
&& f_{18}:= (e_1+e_{18}-2e_{27})/\sqrt 6 \ , \\
&& f_{27}:= (e_1+ e_{18}+e_{27})/\sqrt {3} \ , \\
&& f_a:= e_a\ , \mbox{  in the other cases, }
\feqn
in order to explicitly exhibit the $26$ dimensional representation. We will call $c_i$,
$i=1,\ldots,52$, the resulting $27 \times 27$ matrices.

In App. \ref{app:matrici} we present the program used to construct these matrices.
The $26 \times 26$ representation is then obtained
by deleting from each matrix the last row and the last column, which, in fact, vanish.
The corresponding structure constants, which characterize the algebra and also realize
the adjoint representation, are shown in App.\ref{app:struttura}.\\
To check that indeed we obtained the generators of an $f_4$ algebra, we computed the corresponding roots. If $C_i$ denotes the $i$th matrix in the adjoint representation, we
use $C_1,C_6,C_{15},C_{30}$ as generators of a Cartan subalgebra to calculate
the roots.
These turn out to be the generators of an $f_4$ algebra, as expected.
Moreover the corresponding Killing form is
negative definite and proportional to the trace product defined by
\eqn
\langle a, b \rangle :=-\frac 16 Trace (ab) \ ,
\feqn
where $a$ and $b$ are arbitrary $\bR-$linear combinations of the matrices $c_i$.
Thus we have obtained a compact form of
$f_4$.\\
By direct inspection of the structure constants one can easily recognize a chain of subalgebras. The first $21$ matrices generate an
$so(7)$ subalgebra, whose $so(i)$ subalgebras, with $i=6,5,4,3$, are generated by the first $i(i-1)/2$
matrices, respectively. Again this can be checked computing the roots of the subalgebras. A possible choice for the Cartan subalgebra is
$C_1$ for $so(3)$, $C_1,C_6$ for $so(4)$
and $so(5)$ and $C_1,C_6,C_{15}$ for $so(6)$ and $so(7)$. Adding to $so(7)$ the matrices $c_i$ with $i=30,\ldots,36$ we
obtain an $so(8)$ subalgebra. This corresponds to the Lie algebra of the $Spin(8)$ subgroup of $F_4$ which leaves invariant
the three matrices $J_i$, $i=1,2,3$, where $J_i$ has $J_{i,ii}=1$ as the unique non-vanishing entry. To check this, one can notice that the $J_i$ ($i=1,2,3$) correspond to the
vectors $e_i$ of $\bR^{27}$, which are in the kernel of the given
subset of matrices. \\
Finally there are three evident $so(9)$ subalgebras:
\begin{enumerate}
\item $so(9)_{_1}$ obtained adding $c_{45},\ldots,c_{52}$ to $so(8)$. This corresponds to the subgroup
$Spin(9)_{_1}$ of $F_4$
which leaves $J_1$ invariant\footnote{$Spin(9)$ appears as the subgroup of $F_4$ which fixes a matrix $J$ of the Jordan
algebra};
\item $so(9)_{_2}$ obtained adding $c_{37},\ldots,c_{44}$ to $so(8)$. This corresponds to the subgroup
$Spin(9)_{_2}$ of $F_4$
which leaves $J_2$ invariant;
\item $so(9)_{_3}$ obtained adding $c_{22},\ldots,c_{29}$ to $so(8)$. This corresponds to the subgroup
$Spin(9)_{_3}$ of $F_4$
which leaves $J_3$ invariant.
\end{enumerate}
Again this can be checked applying the given matrices to $e_1$, $e_2$ and $e_3$ respectively.
We will use $Spin(9)_{_1}$ and will refer to it simply as $Spin(9)$.\\
To end this section let us call $p$ the linear complement of $so(9)$ in $f_4$. Looking at the structure constants
we find
\eqn
&& [so(9),{p}] \subset { p} \ , \\
&& [{p},{p}]\subset so(9) \ ,
\feqn
which show a structure of direct product. We don't need to look at the structure constants to discover such a structure. It follows from
the fact that the trace product is ad-invariant (therefore proportional to the Killing form, $F_4$ being simple) and the base of matrices is
orthogonal.

%%%%%%%%%%%%%%%%%%%%%%%%%%%%%%%%%%%%%%%%%%%%%%%%%%%%%%%%%%%%%%%%%%%%%%%%%%%%%%%%%%%%%%%%%%%%%%%%%
\section{Construction of the group \boldmath{$F_4$}} \label{group}
\setcounter{equation}{0}
For connected compact Lie groups the exponential map is surjective \cite{duisterkolk}. This means that we could introduce $52$ parameters $x^i$ and simply write
\eqn
g=g(x^1,\ldots,x^{52}) =\exp(x^i c_i) \ ,
\feqn
for any given element $g\in F_4$. However we are searching for a different kind of parametrization, in the spirit of
\cite{noi,sudar}. The point is that, whereas there is no difficulty in computing the volume using the exponential map parametrization,
the hard problem is the determination of the range of parameters. Moreover, the difficulties increase rapidly if one needs to compute
the left invariant $1$-forms $g^{-1}dg$.
These problems are both resolved by means of an Euler type parametrization, which gives all the quantities
in terms of trigonometric functions, instead of the $\sin x/x$ functions appearing when the exponential parametrization is used.

%%%%%%%%%%%%%%%%%%%%%%%%%%%%%%%%%%%%%%
\subsection{The generalized Euler construction.}\label{general}
We would like to explain our general strategy for constructing a Euler type parametrization.
Let $G$ be an $n-$dimensional simple Lie group and $H$ be one of its closed subgroups.
Let $\lambda_i$ be
a base for $\cG:=Lie(G)$, orthonormal with respect to the Killing form. Let us assume that the first $m:= dim H$ generators
are a base for
$\cHH:=Lie H$ and let us call $\cP$ the subspace generated by the remaining generators so that $[Lie(H),\cP]\subset \cP$. This means that $G/H$ is reductive.
Then it follows that any $g\in G$ can be written in the form
\eqn
g=\exp a \exp b \ , \quad \ a\in \cP\ ,\ b\in \cHH \ .
\feqn
It is an established fact that for compact simple Lie groups such a parametrization is surjective.
This fact is not, generally, well known. We give another proof of it \cite{spigola} in appendix \ref{proof},
because it constitutes an important step in our derivation.\\
The next step consists in finding a subset of linearly free elements $\tau_1,\ldots,\tau_k\in \cP$ with
the following properties
\begin{itemize}
\item if $V$ is the linear subspace generated by $\tau_i$, $i=1,\ldots,k$, then $\cP=Ad_H (V)$, that is, the whole $\cP$ is generated from $V$ through the
adjoint action of $H$;
\item $V$ is minimal, in the sense that it does not contain any proper subspaces with the previous property.
\end{itemize}
This means that the general element $g$ of $G$ can be written in the form
\eqn
g= \exp (\tilde h) \exp (v) \exp (h)\ , \quad h,\tilde h \in \cHH\ ,\ v\in V \ .
\feqn
This way of writing $g$ is surjective but redundant. The redundancy will be $r=2m+k-n$
dimensional, where $n=dim(G)$, $m=dim(H)$ and $k=dim (V)$. The point is
that in general we need less then the whole $H$ to generate the whole $V$ by adjunction. In fact $H$ will contain some subgroup $K$ generating
automorphisms of $V$
\eqn
Ad_K : V\longrightarrow V \ .
\feqn
Then $K$ must be $r$-dimensional and  the generalized Euler decomposition with respect to $H$
\eqn
G=B \exp (V) H \ , \label{eulero}
\feqn
where $B:=H/K$.\\
In general the technical difficulties arise in the construction of $B$. In order to minimize such difficulties it is convenient to choose for $H$ the biggest subgroup of $G$.

%%%%%%%%%%%%%%%%%%%%%%%%%%%%%%%%%%%%%%%%%%%%%%%%%%%%%%
\subsection{The set up for \boldmath{$F_4$}.}
The maximal subgroup of $F_4$ is $H=Spin(9)$. In section \ref{algebra} we have found three $Spin(9)$ subgroups. As we said there, we choose $H=Spin(9)_{_1}$
which we will call simply $Spin(9)$. Then $\cP$ is the $16$ dimensional real vector space generated by the matrices $c_i$,
with $i=22,\ldots,29,37,\ldots,44$. Looking at the structure constants, we see that we can take as $V$ any one dimensional subspace of $\cP$.
We choose $c_{22}$ as a base for $V$. Thus $r=21$. Since $V$ is one dimensional, we expect the subgroup $K$ to commute with $c_{22}$ and
its dimension suggests that it could be a $Spin(7)$ subgroup of $Spin(9)$. Indeed, we can check that this is true. We know that the first $21$ matrices
generate an $so(7)$ algebra. We will now construct a new set of $21$ generators $\tilde c_i$, $i=1,\ldots,21$ commuting with $c_{22}$ and having
the same structure constants as the previous ones. To this end let us look at the $so(8)$ subalgebra: $c_I$, $I=1,\ldots,21,30,\ldots,36$. In particular
let us start with $c_\alpha$, $\alpha=30,\ldots,36$. Then from App.\ref{app:struttura} we see that the remaining first $21$ matrices can be generated
as follows
\eqn
c_{\frac {k(k-1)}2+i+1}=[c_{30+i},c_{30+k}] \ , k=1,\ldots,6, \ i=0,\ldots,k-1 \ .
\feqn
Next, we notice that for $a,b\in\{22,\ldots,29\}$ the commutator $[c_a,c_b]$ is a combination of four elements of $so(8)$, each of which has the
same commutator with $c_{22}$. Using this, we define
\eqn
\tilde c_{30+i}:=-[c_{22},c_{23+i}] \ , \quad i=0,\ldots,7 \ ,
\feqn
and then
\eqn
\tilde c_{\frac {k(k-1)}2+i+1}=[\tilde c_{30+i},\tilde c_{30+k}] \ , k=1,\ldots,6, \ i=0,\ldots,k-1 \ .
\feqn
The surprising fact, for which we have no  explanation, is that the matrices $\tilde c_I$, with
\\ $I=1,\ldots,21$, $30,\ldots,36$ have exactly the same
structure constants of $c_I$ and $[\tilde c_i, c_{22}]=0$ for $i=1,\ldots,21$. This is exactly the $so(7)$ we were searching for. We will call it
$\widetilde {so}(7)$, so that $K=\exp (\widetilde {so}(7))$.\\
At this point let us note that in order to construct the Euler parametrization we proceed by induction: Together with $B$ one needs to give the parametrization
of the maximal subgroup $H=Spin(9)$. Again, this can be done applying the generalized Euler parametrization with respect to the maximal subgroup
$Spin(8)$. Next, $Spin(8)$ could be decomposed with respect to $Spin(7)$ and so on. In conclusion it is convenient to start from $SU(2)$,
to construct $Spin(n)$ up to $n=9$ and finally $F_4$. At any step (ensured the surjectivity) the range of parameters can be determined by means
of the topological method explained in \cite{noi}. To simplify our exposition we will give details only for the most interesting case $F_4$,
and limit the $Spin(j)$ subgroups to a list. The details can be easily reproduced in the same way as for $F_4$.

%%%%%%%%%%%%%%%%%%%%%%%%%%%%%%%%%%%%%%%%%%%%%%%%%%%%
\subsection{The list of subgroups.}
Here we give the results for the subgroups. The details could be considered as an exercise. Rational homology groups and roots, necessary ingredients
for the Macdonald formula, are given in App.\ref{app:macdonald}.

%%%%%%%%%%%%%%%%%%%%%%%%%%%%%%
\subsubsection{\boldmath{$SU(2)$}.}
The generators are $c_{i},\ i=1,2,3$. We have
\eqn
SU(2)[x_1,x_2,x_3]= e^{x_1 c_3} e^{x_2 c_2} e^{x_3 c_3} \ ,
\feqn
with range
\eqn
x_1 \in [0,2\pi] \ , \quad x_2 \in [0,\pi] \ , \quad x_3 \in [0,4\pi] \ .
\feqn
Note that $4\pi$ is the period of $\exp (xc_i)$ for every $i=1,\ldots,3$.
The invariant measure is
\eqn
d\mu_{SU(2)}[x_1,x_2,x_3]=\sin x_2 dx_1 dx_2 dx_3 \ .
\feqn

%%%%%%%%%%%%%%%%%%%%%%%%%%%%%
\subsubsection{\boldmath{$Spin(4)$}.}
The generators are $c_i,\ i=1,\ldots,6$.
We take $H=SU(2)$ generated by $c_5, c_6, c_3$ which can be obtained from the previous one by simple substitutions.
$V$ is one dimensional and we can take $c_4$ as generator. $r=1$ and $K=U(1)=e^{xc_3}$ so that
$B[x,y]=H/K=e^{x c_3} e^{y c_5}$, with $x\in [0,2\pi]$ and $y\in[0,\pi]$. Then
\eqn
Spin(4)[x_1,\ldots,x_6]=e^{x_1 c_3} e^{x_2 c_5} e^{x_3 c_4}SU(2)[x_4,x_5,x_6] \ .
\feqn
The invariant measure is
\eqn
d\mu_{_{Spin(4)}} = \sin x_2 \sin^2 x_3 dx_1 dx_2 dx_3 d\mu_{_{SU(2)}}[x_4,\ldots,x_6] \ ,
\feqn
and the range of parameters
\eqn
x_1 \in [0,2\pi] \ , \quad x_2 \in [0,\pi] \ , \quad x_3 \in [0,\pi] \ ,
\feqn
the others being the ones of $SU(2)$.

%%%%%%%%%%%%%%%%%%%%%%%%%%%%%
\subsubsection{\boldmath{$Spin(5)$}.}
The generators are $c_i,\ i=1,\ldots,10$.
The subgroup is $H=Spin(4)$ as before.
$V$ is one dimensional and we can take $c_{7}$ as generator. $r=3$ and $K=SU(2)$ generated by $c_\alpha$, $\alpha=3,5,6$ commute with $c_7$ so that
$B_5[x_1,x_2,x_3]=H/K=e^{x_1 c_3} e^{x_2 c_5} e^{x_3 c_4}$. Then
\eqn
Spin(5)[x_1,\ldots,x_{10}]=B_5[x_1,\ldots,x_3] e^{x_4 c_7} Spin(4)[x_5,\ldots,x_{10}] \ .
\feqn
The invariant measure is
\eqn
d\mu_{_{Spin(5)}}[x_1,\ldots,x_{10}] = \sin x_2 \cos^2 x_3 \sin^3 x_4 dx_1 dx_2 dx_3 dx_4 d\mu_{_Spin(4)}[x_5,\ldots,x_{10}] \ .
\feqn
and the range of parameters
\eqn
x_1 \in [0,2\pi] \ , \quad x_2 \in [0,\pi] \ , \quad x_3 \in [-\frac \pi2,\frac \pi2] \ , \quad x_4\in [0,\pi]
\feqn
the others being the ones of $Spin(4)$.

%%%%%%%%%%%%%%%%%%%%%%%%%%%%%
\subsubsection{\boldmath{$Spin(6)$}.}
The generators are $c_i,\ i=1,\ldots,15$.
The subgroup is $H=Spin(5)$ as before.
$V$ is one dimensional and we can take $c_{11}$ as generator. $r=6$ and $K=Spin(4)$ generated by $c_\alpha$, $\alpha=3,5,6,8,9,10$ commute with
$c_{11}$ so that
$$
B_6[x_1,\ldots,x_4]=H/K=e^{x_1 c_3} e^{x_2 c_5} e^{x_3 c_4} e^{x_4 c_7} \ .
$$
Then
\eqn
Spin(6)[x_1,\ldots,x_{15}]=B_6[x_1,\ldots,x_4] e^{x_5 c_{11}} Spin(5)[x_{6},\ldots,x_{15}] \ .
\feqn
The invariant measure is
\eqn
d\mu_{_{Spin(6)}}[x_1,\ldots,x_{15}] =
\sin x_2 \cos^2 x_3 \cos^3 x_4 \sin^4 x_5 dx_1 dx_2 dx_3 dx_4 dx_5 d\mu_{_{Spin(5)}}[x_6,\ldots,x_{15}]  \ ,
\feqn
and the range of parameters
\eqn
&& x_1 \in [0,2\pi] \ , \quad x_2 \in [0,\pi] \ , \quad x_3 \in [-\frac \pi2,\frac \pi2] \ , \quad x_4 \in [-\frac \pi2,\frac \pi2] \ , \cr
&& x_5\in [0,\pi] \ ,
\feqn
the others being the ones of $Spin(5)$.

%%%%%%%%%%%%%%%%%%%%%%%%%%%%%
\subsubsection{\boldmath{$Spin(7)$}.}
The generators are $c_i,\ i=1,\ldots,21$.
The subgroup is $H=Spin(6)$ as before.
$V$ is one dimensional and we can take $c_{16}$ as generator. $r=10$ and $K=Spin(5)$ generated by $c_\alpha$, $\alpha=3,5,6,8,9,10,12,13,14,15$
commute with $c_{16}$ so that
$$
B_7[x_1,\ldots,x_5]=H/K=e^{x_1 c_3} e^{x_2 c_5} e^{x_3 c_4} e^{x_4 c_7} e^{x_5 c_{11}} \ .
$$
Then
\eqn
Spin(7)[x_1,\ldots,x_{21}]=B_7[x_1,\ldots,x_5] e^{x_6 c_{16}} Spin(6)[x_{7},\ldots,x_{21}] \ .
\feqn
The invariant measure is
\eqn
&& d\mu_{_{Spin(7)}}[x_1,\ldots,x_{21}] =\sin x_2 \cos^2 x_3 \cos^3 x_4 \cos^4 x_5
\sin^5 x_6 dx_1 dx_2 dx_3 dx_4 dx_5 dx_6 \cdot \cr
&& \qquad\qquad\qquad \qquad \qquad \cdot d\mu_{_{Spin(6)}}[x_7,\ldots,x_{21}]  \ ,
\feqn
and the range of parameters
\eqn
&& x_1 \in [0,2\pi] \ , \quad x_2 \in [0,\pi] \ , \quad x_3 \in [-\frac \pi2,\frac \pi2] \ , \quad x_4 \in [-\frac \pi2,\frac \pi2] \ , \cr
&& x_5 \in [-\frac \pi2,\frac \pi2] \ , \quad x_6\in [0,\pi] \ ,
\feqn
the others being the ones of $Spin(6)$.

%%%%%%%%%%%%%%%%%%%%%%%%%%%%%
\subsubsection{\boldmath{$Spin(8)$}.}
The generators are $c_i,\ i=1,\ldots,21,30,\ldots,36$.
The subgroup is $H=Spin(7)$ as before.
$V$ is one dimensional and we can take $c_{30}$ as generator. $r=15$ and $K=Spin(6)$ generated by $c_\alpha$,
$\alpha=3,5,6,8,9,10,12,13,14,15,17,18,19,20,21$.
Up to now we have proceeded in a systematical way. One could proceed in this way but technical difficulties increase with the dimension of the group.
In particular the computation of the invariant measure becomes too hard for $F_4$. To solve this problem we have found it convenient to change the parameterization of the quotient $B$. The simplification consists in using as many commuting matrices as possible to realize $B$. This must be compatible
with the fact that $B\cdot K$ must cover the whole $Spin(7)$ group. There are many possibilities. We have chosen
\eqn
B_8[x_1,\ldots,x_6] =e^{x_1 c_3} e^{x_2 c_{16}} e^{x_3 c_{15}} e^{x_4 c_{35}} e^{x_5 c_5} e^{x_6 c_1} \ .
\feqn
The fact that it works can be checked by doing the previous analysis backward.
Then
\eqn
Spin(8)[x_1,\ldots,x_{28}]=B_8[x_1,\ldots,x_6] e^{x_7 c_{30}} Spin(7)[x_{8},\ldots,x_{28}] \ .
\feqn
The invariant measure is
\eqn
&& d\mu_{_{Spin(8)}}[x_1,\ldots,x_{28}] =\sin x_4 \cos x_5 \cos x_6 \sin^2 x_6 \cos^2 x_7 \sin^4 x_7
\prod_{i=1}^7 dx_i\cdot \cr
&& \qquad\qquad\qquad \qquad \qquad \cdot d\mu_{_{Spin(7)}}[x_8,\ldots,x_{28}]  \ ,
\feqn
and the range of parameters
\eqn
&& x_1 \in [0,2\pi] \ , \quad x_2 \in [0,2\pi] \ , \quad x_3 \in [0,2\pi] \ , \quad x_4 \in [0,\pi] \ , \cr
&& x_5 \in [-\frac \pi2,\frac \pi2] \ , \quad x_6\in [0,\frac \pi2] \ , \quad x_7\in [0,\frac \pi2] \ ,
\feqn
the others being the ones of $Spin(7)$.

%%%%%%%%%%%%%%%%%%%%%%%%%%%%%

\subsubsection{\boldmath{$Spin(9)$}.}
Here we used the same procedure as for $Spin(8)$.
The generators are $c_i$, with\\ $i=1,\ldots,21,30,\ldots,36,45,\ldots,52$.
The subgroup is $H=Spin(8)$ as before.
$V$ is one dimensional and we can take $c_{45}$ as generator. $r=21$ and $K=Spin(7)$ generated by $c_\alpha$,
$\alpha=3,5,6,8,9,10,12,13,14,15,17,18,19,20,21,31,32,33,34,35,36$.
The choice of $B$ can be deduced from the one for $Spin(8)$. Then
\eqn
B_9[x_1,\ldots,x_7] =e^{x_1 c_3} e^{x_2 c_{16}} e^{x_3 c_{15}} e^{x_4 c_{35}} e^{x_5 c_5} e^{x_6 c_1} e^{x_7 c_{30}}\ .
\feqn
and
\eqn
Spin(9)[x_1,\ldots,x_{36}]=B_9[x_1,\ldots,x_7] e^{x_8 c_{45}} Spin(8)[x_{9},\ldots,x_{36}] \ .
\feqn
The invariant measure is
\eqn
&& d\mu_{_{Spin(9)}}[x_1,\ldots,x_{36}] =\sin x_4 \cos x_5 \cos x_6 \sin^2 x_6 \cos^4 x_7 \sin^2 x_7 \sin^7 x_8
\prod_{i=1}^8 dx_i \cdot \cr
&& \qquad\qquad\qquad \qquad \qquad \cdot d\mu_{_{Spin(8)}}[x_9,\ldots,x_{36}]  \ ,
\feqn
and the range of parameters
\eqn
&& x_1 \in [0,2\pi] \ , \quad x_2 \in [0,2\pi] \ , \quad x_3 \in [0,2\pi] \ , \quad x_4 \in [0,\pi] \ , \cr
&& x_5 \in [-\frac \pi2,\frac \pi2] \ , \quad x_6\in [0,\frac \pi2] \ , \quad x_7\in [0,\frac \pi2] \ , \quad x_8\in [0,\pi] \ ,
\feqn
the others being the ones of $Spin(8)$.

%%%%%%%%%%%%%%%%%%%%%%%%%%%%%%%%%%%%%%%%%%%%%%%%%%%%%%%%%%%%%%%%%
\subsection{Construction of \boldmath{$F_4$}.}
We are now ready to realize the construction of the group $F_4$. The maximal subgroup is a $Spin(9)$ subgroup which
we choose to be $Spin(9)_{_1}$. Then we know that the generic element of $F_4$ can be written formally as
\eqn
F_4=e^{\cP} e^{so(9)} \ ,
\feqn
where $\cP$ is the linear space generated by the matrices $c_i$, with $i=22,\ldots,29,37,\ldots,44$. Looking at the
structure constants we can see that
\eqn
\exp (-2x c_i) c_{22} \exp (2x c_i) =\cos x c_{22} \pm \sin x c_j \ ,
\feqn
where $j=29,25,28,23,27,26,24$ respectively, if $i=30,\ldots,36$ and $j=44,40,43,38,42,41,39,37$ for $i=45,\ldots,52$.
This ensures that $\cP$ can be generated acting on $c_{22}$ by adjunction with $H$.\footnote{More precisely this means
that we could write the general element of $F_4$ in the form\\ $F_4=B[x_1,\ldots,x_{15}]e^{x_{16}c_{22}} Spin(9)$,
with $B=\prod_{i=1}^{15} e^{x_i c_{j_i}}$, $j_i=30,\ldots,36,45,\ldots,52$. However such a realization is not sufficiently
simple to allow technical computations.} Thus the generic element of $F_4$ is
\eqn
g=e^a e^{xc_{22}} e^b \ , \label{larderello}
\feqn
with
$a,b\in so(9)$. We know that the $21$ dimensional redundancy of such a parametrization is due to a $Spin(7)$
subgroup of $Spin(9)$ which commutes with $c_{22}$. This subgroup is generated by the matrices $\tilde c_i$ with
$i=1,\ldots 21$, which satisfy the same commutation relations of the corresponding $c_i$. In fact this is true also adding the
$\tilde c_i$, $i=30,\ldots,36$ and moreover, for the whole $Spin(9)$ subgroup, if we define $\tilde c_i=c_i$ for
$i=45,\ldots,52$. Thus we can use $\tilde c_i$ in place of $c_i$ to construct $Spin(9)$, at least for the left factor
in (\ref{larderello}).
Now in our construction
\eqn
&& Spin(9)[x_1,\ldots,x_{36}]=B_9 [x_1,\ldots,x_7] e^{x_8 \tilde c_{45}} Spin(8)[x_9,\ldots,x_{36}] \cr
&& \qquad\qquad\qquad\ =B_9 [x_1,\ldots,x_7] e^{x_8 \tilde c_{45}} B_8 [x_9,\ldots,x_{14}] e^{x_{15} \tilde c_{30}}
Spin(7)[x_{16},\ldots,x_{36}] \ ,
\feqn
so that
\eqn
B_{F_4}[x_1,\ldots,x_{15}]=B_9 [x_1,\ldots,x_7] e^{x_8 \tilde c_{45}} B_8 [x_9,\ldots,x_{14}] e^{x_{15} \tilde c_{30}} \ ,
\feqn
and
\eqn
F_4[x_1,\ldots,x_{52}]=B_{F_4}[x_1,\ldots,x_{15}]e^{x_{16} c_{22}} Spin(9)[x_{17},\ldots,x_{52}] \ .
\feqn
We also know that for the ranges determined for $Spin(9)$, $Spin(8)$ and $Spin(7)$, $B_{F_4}$ covers the whole
$H/K$, so that only the range for $x_{16}$ remains to be determined. However we will now determine the range of
all parameters.

%%%%%%%%%%%%%%%%%%%%%%%%%%%%%%%%%%%%%%%%%%%%%%%%%%%%%%%%%%%%%%%%
\subsection{Determination of the range of parameters.}
To determine the range we will use the topological method introduced in \cite{noi}.
For convenience let us recall here how it works. The first step consists in the determination of the invariant
measure. It will depend explicitly on some of the parameters.
One can then construct a closed variety, having the same
dimension of the whole group, simply by choosing for these variables the
maximal range which still allows the measure to be well defined, while for
the remaining variables the range should coincide with their period.\footnote{Each of the $x_i$ appears
in the parametrization in the form $e^{x_i c_i}$ or $e^{x_i \tilde c_i}$,
and is therefore periodic as a consequence of compactness of the group. With our
normalization, we find that all periods are equal to $4\pi$.}
If the parametrization adopted is surjective and the group is connected, then surely the variety obtained in this way covers the
whole group. Here is where surjectivity is crucial!\\
At this point it is possible that, with this choice of parameters, we cover some points of the the group more then once.
Fortunately one can check this by means of the Macdonald formula \cite{burgy,mac} which gives the volume of a compact
Lie group with respect to an invariant measure induced on the group by a Lebesgue measure on the Lie algebra.
If the resulting number of covering is higher than $1$, the range of parameters must be further reduced using
some automorphism of the space of parameters which leaves the group invariant under reparametrization. See \cite{noi}
for more details.

%%%%%%%%%%%%%%%%%%%%%%%%%%%%%%%%%%%%%%
\subsubsection{The volume of \boldmath{$F_4$}.}
Let us compute the volume of $F_4$ by means of the Macdonald formula. The Betty numbers of the exceptional Lie groups
were computed in \cite{chevalley}. For $F_4$ there are four free generators for the rational homology. Their dimensions are
\eqn
&& d_1=3, \ d_2=11, \ d_3=15, \ d_4=23 \ .
\feqn
The simple roots are \cite{Fulton-harris}
\eqn
&& r_1 =L_2-L_3 \\
&& r_2 =L_3-L_4 \\
&& r_3 =L_4 \\
&& r_4 =\frac {L_1-L_2-L_3-L_4}2
\feqn
where $L_i$, $i=1,\ldots,4$ is an orthonormal base for the Cartan algebra.
The volume of the fundamental region is then
\eqn
Vol(f_{F_4})= \frac 12 \ .
\feqn
Furthermore there are $24$ positive roots, $12$ of which have length $1$, and $12$ have length
$\sqrt 2 $ \cite{Fulton-harris}. We found explicitly these roots, as explained
in section \ref{algebra}, with $L_i=e_i$, the canonical base of $\mathbb{R}^4$.
The volume of $F_4$ is then
\eqn
Vol(F_4) =\frac {2^{26} \cdot \pi^{28}}{3^7 \cdot 5^4 \cdot 7^2 \cdot 11} \ . \label{volumeF4}
\feqn

%%%%%%%%%%%%%%%%%%%%%%%%%%%%%%%%%%%%%%%%%
\subsubsection{The invariant measure on \boldmath{$F_4$}.}
The invariant measure on $F_4$ decomposes in the product of the measure on $Spin(9)$ and the
one on $M=F_4/Spin(9)$. This was shown in general in \cite{noi} but let us rewrite it in terms of (\ref{eulero}). If we define
\eqn
J_H:=H^{-1}dH \ , \qquad \ J_M:=\pi_{\cP} (e^{-x_{16}c_{22}} B_{F_4}^{-1} d (B_{F_4}e^{x_{16}c_{22}})) \ ,
\feqn
with $H=Spin(9)$, then
\eqn
ds^2_M=-\frac 16 Trace(J_M \otimes J_M) \ ,
\feqn
is the induced invariant measure on $M$ and
\eqn
d\mu_{F_4}=|det(J_{Mi}^j)| d\mu_{Spin(9)} \prod_{l=1}^{16} dx_l \ ,
\feqn
where $J_{Mi}^j$ is the $16\times 16$ matrix defined by
\eqn
J_M =\sum_{i,j=1}^{16} J_{Mi}^j c_{i_j} dx^i \ ,
\feqn
where $c_{i_j}$ is the base $\{c_{22},\ldots,c_{29},c_{37},\ldots,c_{44}\}$ of $\cP$.\\
Let us now introduce the notation
\eqn
&& M_8[x_1,\ldots,x_8]:=B_9[x_1,\ldots,x_7]e^{x_8 \tilde c_{45}} \ , \\
&& M_7[x_9,\ldots,x_{15}]:=B_8[x_9,\ldots,x_{14}]e^{x_{15} \tilde c_{30}} \ .
\feqn
Then
\eqn
&& J_M = dx_{16} c_{22} +e^{-x_{16} c_{22}} M_7^{-1} dM_7 e^{x_{16} c_{22}}
          + e^{-x_{16} c_{22}} M_7^{-1} M_8^{-1}dM_8 M_7 e^{x_{16} c_{22}} \cr
          && \ \ \quad =: dx_{16} c_{22} +e^{-x_{16} c_{22}} J_7 e^{x_{16} c_{22}}
          + e^{-x_{16} c_{22}} M_7^{-1} J_8 M_7 e^{x_{16} c_{22}} \ .
\feqn
Some remarks are in order now
\begin{enumerate}
\item $M_7 \in Spin(8)$ corresponding to the algebra of matrices $c_i$ with $i=1,\ldots,21,30,\ldots,36$;
\item $M_8 \in Spin(9)$ corresponding to the algebra of matrices $c_i$ with \\ $i=1,\ldots,21,30,\ldots,36,45,\ldots,52$;
\item looking at commutators we see that the adjoint action of $e^{\al_{16} c_{22}}$ on $so(8)$ generate linear
      combination of the matrices of $so(8)$ itself, adding also combination of the matrices
      $c_j$ with $j=23,\ldots,29$;
\item the adjoint action of $Spin(8)$ on $so(9)$ restricted to the linear subspace generated by \\ $c_{45},\ldots, c_{52}$
      is a rotation, that is $Spin(8)^{-1} c_i Spin(8) = \sum_{j=45}^{52} {R_i}^j c_j$ with $i=45,\ldots,52$, where
      ${R_i}^j$ is a rotation matrix. In particular $|\det {R_I}^J| =1$;
\item the adjoint action of $e^{x c_{22}}$ on $c_i$ with $i=45,\ldots, 52$ is a rotation of the form
      $c_i \longmapsto \cos \frac x2  c_i \pm \sin \frac x2 c_{\tilde i}$, where
      $\tilde i \in \{37,\ldots 44 \}  $.
\end{enumerate}
{F}rom these remarks one can deduce
\begin{enumerate}
\item $dx_{16}$ is the only coefficient of $c_{22}$;
\item the projection of $e^{-x_{16} c_{22}} J_7 e^{x_{16} c_{22}}$ on $c_i$, $i=22,37,\ldots,44$, vanishes,
      so that it gives rise to a $7 \times 7$ diagonal block. In other words, if the columns give the projections on
      $c_i$ (with ordering $i=22,\ldots,29,37,\ldots,44$) and the rows are the differentials $d\al_j$ (with
      $J=1,\ldots,16$ starting from below) then we must compute the determinant of the matrix
\eqn
\left(
\begin{array}{ccc}
1 & 0 & 0 \\
* & A & 0 \\
* & * & B
\end{array}
\right)
\feqn
where $A$ is the $7\times7$ diagonal block given above and $B$ is a $8\times8$ block obtained by projecting \par
$e^{-x_{16} c_{22}} M_7^{-1} J_8 M_7 e^{x_{16} c_{22}}$ on $c_{37},\ldots,c_{44}$. The $*$ blocks are irrelevant
for the computation of the determinant which in fact will be $\det A \cdot \det B$;
\item from point $5$ of the remarks it follows $\det B =\sin^8 \frac {x_{16}}2 \det \tilde B$, where $\tilde B$
      is the projection of $M_7^{-1} J_8 M_7$ on $c_{45},\ldots,c_{52}$. On the other hand, from the remaining remarks
      it follows $\det \tilde B = \det R \cdot \det \hat B$, where $R$ is the orthogonal matrix introduced in
      remark $4$ and does not contribute to the determinant, whereas $\hat B$ is the projection of $J_8$ on
      $c_{45}, \ldots,c_{52}$. In particular
      $\det \hat B=\sin x_4 \cos x_5 \cos x_6 \sin^2 x_6 \cos^4 x_7 \sin^2 x_7 \sin^7 x_8$.
\end{enumerate}
Thus we can quite easily compute the invariant measure for $M$, which turns out to be
\eqn
&& d\mu_{M}=2^7 \cos^7 \frac {x_{16}}2  \sin^{15} \frac {x_{16}}2 \sin x_4 \cos x_5 \cos x_6 \sin^2 x_6 \cos^4 x_7 \sin^2 x_7 \sin^7 x_8 \cdot \cr
&& \qquad\ \cdot\sin x_{12} \cos x_{13} \cos x_{14} \sin^2 x_{14} \cos^2 x_{15} \sin^4 x_{15} \prod_{i=1}^{16} dx_i \ .
\feqn
Note that the periods of the variables are $4\pi$ so that one should take the range $x_i=[0,4\pi]$ for $i=1,2,3$
and $i=9,10,11$. However it is easy to show directly from the parametrization that they can all be restricted
to $[0,2\pi]$. In fact for all $\tilde c_i \in so(7)$ we have that $e^{2\pi \tilde c_i}$ commute with $\tilde c_j$ and
with $c_{22}$, so that it can be reabsorbed in the $Spin(9)$ factor of $F_4$. The range of
$x_i$ is then
\eqn
&& x_1 \in [0,2\pi] \ , \quad x_2 \in [0,2\pi] \ , \quad x_3 \in [0,2\pi] \ , \quad x_4 \in [0,\pi] \ , \cr
&& x_5 \in [-\frac \pi2,\frac \pi2] \ , \quad x_6\in [0,\frac \pi2] \ , \quad x_7\in [0,\frac \pi2] \ , \quad x_8\in [0,\pi] \ , \cr
&& x_9 \in [0,2\pi] \ , \quad x_{10} \in [0,2\pi] \ , \quad x_{11} \in [0,2\pi] \ , \quad x_{12} \in [0,\pi] \ , \cr
&& x_{13} \in [-\frac \pi2,\frac \pi2] \ , \quad x_{14}\in [0,\frac \pi2] \ , \quad x_{15}\in [0,\frac \pi2] \ ,
\quad x_{16}\in [0,\pi] \ .
\feqn
Note that the firsts $15$ are exactly the ones predicted by $Spin(9)/Spin(7)$.
The remaining parameters $x_j$, $j=17,\ldots,52$, will run over the range for $Spin(9)$.
The volume of the whole closed cycle $V$
so obtained is then
\eqn
Vol (V)=Vol(Spin(9))\int_R d\mu_M =\frac {2^{26} \cdot \pi^{28}}{3^7 \cdot 5^4 \cdot 7^2 \cdot 11} \ ,
\feqn
where $R$ is the range of parameters $x_i$, $i=1,\ldots,16$.
This is the volume of $F_4$, so that we cover the group exactly one time.\footnote{Obviously there is a
subset of vanishing measure multiply covered.}

%%%%%%%%%%%%%%%%%%%%%%%%%%%%%%%%%%%%%%%%%%%%%%%%%%%%%%%%%%%%%%%%%%%%%%%%%%%%%%%%%%%%%%%%%%%%%%%%%
\section{Conclusions.}
\setcounter{equation}{0}
In this paper we have considered the problem of giving an explicit construction
of the $F_4$ simple Lie group and in particular of its compact form.
The main motivation is that we are interested in studying the Lie group $E_6$ in a future paper,
because in its compact realization it is the most promising exceptional Lie group for unification in GUT theories \cite{caramor}.
In particular to perform non perturbative calculations a parameterization is needed which,
on the one hand should yield the most simple expression for the invariant measure on
the group, while at the same time still being able of providing an explicit expression
for the range of the parameters. Both these requirements are necessary in order to
minimize the computation power needed for computer simulations of lattice models. \\
It seems that the best solution to both of these problems is the determination of Euler
like angles. In section \ref{general} we have explained in detail the general strategy for
defining such a parametrization and shown that it turns out to be surjective for each
compact Lie group. It is clear from the construction that the Euler angles for a given
group are not uniquely defined, but that they can depend for example on the choice of a subgroup,
which can be fixed according to the requirements. The surjectivity of the map allows the
use of a topological method to determine the range of parameters.\\
The first application of our results will be the construction of the generalized Euler
parametrization of the $E_6$ group associated with the $F_4$ subgroup, along the lines
of section \ref{general}. However another immediate interesting application could be the
explicit construction of the $F_4$ invariant metric for $\bO\bP^2=F_4/Spin(9)$.

%%%%%%%%%%%%%%%%%%%%%%%%%%%%%%%%%%%%%%%%%%%%%%%%%%%%%%%%%%
\subsection*{Acknowledgments}
We are grateful to Stefano Pigola for explaining the proof given in App. \ref{proof}.
BLC would like to thank O. Ganor for useful discussions. We would like to thank
A. Garrett Lisi for pointing out some typos in App. \ref{app:matrici}.
This work has been supported in part by the Spanish Ministry of Education and Science (grant AP2005-5201) and in part
by the Director, Office of Science, Office of High Energy and Nuclear Physics of the U.S. Department of Energy
under Contract DE-AC02-05CH11231.

%%%%%%%%%%%%%%%%%%%%%%%%%%%%%%%%%%%%%%%%%%%%%%%%%%%%%%%%%%%%%%%%%%%%%%%%%%%%%%%%%%%%%%%%%%%%%%%%%
\newpage
\begin{appendix}
%%%%%%%%%%%%%%%%%%%%%%%%%%%%%%%%%%%%%%%%%%%%%%%%%%%%%%%%%%%%%%%%%%%%%%%%
\section{The octonionic algebra.}\label{app:ottonioni}
\setcounter{equation}{0}
The octonionic algebra is obtained from the eight dimensional real vector space $\bO$ generated by a
{\it real unit} $1$ and seven {\it imaginary units} $i_i$, $i=1,\ldots,7$. The structure of a non-abelian,
non-associative division algebra is obtained introducing a distributive product
\eqn
\cdot : \bO \times \bO \longrightarrow \bO \ , \quad (a,b) \mapsto ab \ , \nonumber
\feqn
by means of the following rules:
\begin{itemize}
\item $1$ is the identity for the product;
\item $i_i^2=-1$ for $i=1,\ldots,7$ and $i_i i_j=-i_j i_i$ for $1\leq i_i<i_j\leq 7$;
\item the units $i_1, i_2, i_3$ generate a quaternionic subalgebra;
\item the remaining independent products between the imaginary units are given in the next program, where
$e[1]=1$ and $e[i+1]=i_i$, $i=1,\ldots,7$.
%\begin{center}
%$$
%\begin{picture}(150, 170)(-70,-85)
%\put(-75,-42){\line(1,0){150}}
%\put(-75,-42){\line(3,5){75}}
%\put(75,-42){\line(-3,5){75}}
%\drawline(-75,-42)(37.5, 21)
%\drawline(75,-42)(-37.5, 21)
%\drawline(0,-42)(0, 84)
%\put(0,0){\circle*{4}}
%\put(0,-42){\circle*{4}}
%\put(-75,-42){\circle*{4}}
%\put(75,-42){\circle*{4}}
%\put(0,84){\circle*{4}}
%\put(-37.5,21){\circle*{4}}
%\put(37.5,21){\circle*{4}}
%\end{picture}
%$$
%\end{center}
\end{itemize}

%\begin{texdraw}
%\move(0 0)
%\lcir r:1
%\end{texdraw}

%\newpage
%%%%%%%%%%%%%%%%%%%%%%%%%%%%%%%%%%%%%%%%%%%%%%%%%%%%%%%%%%%%%%%%%%%%%%%%
\section{The \boldmath{$f_4$} matrices.} \label{app:matrici}
\setcounter{equation}{0}
The matrices we found using Mathematica, and orthonormalized with respect to the scalar product
$\langle a , b\rangle :=-\frac 16 Trace (ab)$, were computed by means of the followings programs.

%%%%%%%%%%%%%%%%%%%%%%%%%%%%%%%%%%%%%%%
\subsection{Construction of the matrices.}
The following program gives the $27\times 27$ matrices of $F_4$ before and after the $26$ dimensional reduction.
\begin{eqnarray*}
\rm
&& \%\%\%\% \mbox{    the octonionic products}\\
&& \QQ[1,1]=e[1]; \qquad \QQ[1,2]=e[2]; \qquad \QQ[1,3]=e[3];\\
&& \QQ[1,4]=e[4]; \qquad \QQ[1,5]=e[5]; \qquad \QQ[1,6]=e[6];\\
&& \QQ[1,7]=e[7]; \qquad \QQ[1,8]=e[8]; \qquad \QQ[2,1]=e[2];\\
&& \QQ[2,2]=-e[1]; \qquad \QQ[2,3]=e[5]; \qquad \QQ[2,4]=e[8];\\
&& \QQ[2,5]=-e[3]; \qquad \QQ[2,6]=e[7]; \qquad \QQ[2,7]=-e[6];\\
&& \QQ[2,8]=-e[4]; \qquad \QQ[3,1]=e[3]; \qquad \QQ[3,2]=-e[5];\\
&& \QQ[3,3]=-e[1]; \qquad \QQ[3,4]=e[6]; \qquad \QQ[3,5]=e[2];\\
&& \QQ[3,6]=-e[4]; \qquad \QQ[3,7]=e[8]; \qquad \QQ[3,8]=-e[7];\\
&& \QQ[4,1]=e[4]; \qquad \QQ[4,2]=-e[8]; \qquad \QQ[4,3]=-e[6];\\
&& \QQ[4,4]=-e[1]; \qquad \QQ[4,5]=e[7]; \qquad \QQ[4,6]=e[3];\\
&& \QQ[4,7]=-e[5]; \qquad \QQ[4,8]=e[2]; \qquad \QQ[5,1]=e[5];\\
&& \QQ[5,2]=e[3]; \qquad \QQ[5,3]=-e[2]; \qquad \QQ[5,4]=-e[7];\\
&& \QQ[5,5]=-e[1]; \qquad \QQ[5,6]=e[8]; \qquad \QQ[5,7]=e[4];\\
&& \QQ[5,8]=-e[6]; \qquad \QQ[6,1]=e[6]; \qquad \QQ[6,2]=-e[7];
\end{eqnarray*}
\begin{eqnarray*}\rm
&& \QQ[6,3]=e[4]; \qquad \QQ[6,4]=-e[3]; \qquad \QQ[6,5]=-e[8];\\
&& \QQ[6,6]=-e[1]; \qquad \QQ[6,7]=e[2]; \qquad \QQ[6,8]=e[5];\\
&& \QQ[7,1]=e[7]; \qquad \QQ[7,2]=e[6]; \qquad \QQ[7,3]=-e[8];\\
&& \QQ[7,4]=e[5]; \qquad \QQ[7,5]=-e[4]; \qquad \QQ[7,6]=-e[2];\\
&& \QQ[7,7]=-e[1]; \qquad \QQ[7,8]=e[3]; \qquad \QQ[8,1]=e[8];\\
&& \QQ[8,2]=e[4]; \qquad \QQ[8,3]=e[7]; \qquad \QQ[8,4]=-e[2];\\
&& \QQ[8,5]=e[6]; \qquad \QQ[8,6]=-e[5]; \qquad \QQ[8,7]=-e[3];\\
&& \QQ[8,8]=-e[1];\\
&& \%\%\%\% \mbox{    the Jordan algebra product}\\
%%%%%%%%%%%%%%%%%%%%%%%%%%%%%%%%%%%%%%%%%%%%%%%%%%%%
&& \Qm[x_\_,y_\_] :=\SSu[\SSu[x[[i]]y[[j]]\QQ[i,j],\{i,8\}],\{j,8\}]\\
&& \QP[x_\_,y_\_] :=\left\{\Coe[\Qm[x,y],e[1]],\ \Coe[\Qm[x,y],e[2]], \right.\\
&& \qquad \Coe [\Qm[x,y],e[3]],\ \Coe [\Qm[x,y],e[4]],\ \Coe [\Qm[x,y],e[5]], \\
&& \left. \qquad \Coe [\Qm[x,y],e[6]],\ \Coe [\Qm[x,y],e[7]],\ \Coe [\Qm[x,y],e[8]]\right\}\\
&& {\rm o1=\{a1,\ a2,\ a3,\ a4,\ a5,\ a6,\ a7,\ a8 \}};\\
&& {\rm o2=\{b1,\ b2,\ b3,\ b4,\ b5,\ b6,\ b7,\ b8 \}};\\
%%%%%%%%%%%%%%%%%%%%%%%%%%%%%%%%%%%%%%%%%%%%%%%%%%%%%%%%
&& \Conj[x_\_]:= \{ x[[1]],\ -x[[2]],\ -x[[3]],\ -x[[4]],\ -x[[5]],\ -x[[6]],\ -x[[7]],\ -x[[8]]\}\\
%%%%%%%%%%%%%%%%%%%%%%%%%%%%%%%%%%%%%%%%%%%%%%%%%%%%%%%%
&& \Octp[a_\_,b_\_]:= \left\{\left\{\SSu [\QP[\Part[\Part[a,1],i],\Part[\Part[b,i],1]],\{i,3\}]\right.\right. ,\\
&& \qquad \SSu [\QP[\Part[\Part[a,1],i],\Part[\Part[b,i],2]],\{i,3\}],\\
&& \left. \qquad \SSu [\QP[\Part[\Part[a,1],i],\Part[\Part[b,i],3]],\{i,3\}]\right\} ,\\
&& \quad\ \ \left\{ \SSu [\QP[\Part[\Part[a,2],i],\Part[\Part[b,i],1]],\{i,3\}], \right. \\
&& \qquad \SSu [\QP[\Part[\Part[a,2],i],\Part[\Part[b,i],2]],\{i,3\}],\\
&& \left. \qquad \SSu [\QP[\Part[\Part[a,2],i],\Part[\Part[b,i],3]],\{i,3\}]\right\} ,\\
&& \quad\ \ \left\{ \SSu [\QP[\Part[\Part[a,3],i],\Part[\Part[b,i],1]],\{i,3\}], \right. \\
&& \qquad \SSu [\QP[\Part[\Part[a,3],i],\Part[\Part[b,i],2]],\{i,3\}],\\
&& \left. \left. \qquad \SSu [\QP[\Part[\Part[a,3],i],\Part[\Part[b,i],3]],\{i,3\}]\right\}\right\}\\
%%%%%%%%%%%%%%%%%%%%%%%%%%%%%%%%%%%%%%%%%%%%%%%%%%%%%%
&& \Octps[a_\_,b_\_]:=1/2 (\Octp[a,b]+\Octp[b,a])\\
%%%%%%%%%%%%%%%%%%%%%%%%%%%%%%%%%%%%%%%%%%%%%%%%%%%%%%
&& \%\%\%\% \mbox{    correspondence between the Jordan algebra and $\bR^{27}$}\\
&& {\rm A} =\left\{ \left\{ \{\rm a[1],\ 0,\ 0,\ 0,\ 0,\ 0,\ 0,\ 0\},\ \{a[2],\ a[3],\ a[4],\ a[5],\ a[6],\ a[7],\ a[8],\ a[9] \}, \right.\right.\\
&& \rm \left. \qquad\ \ \{ a[10],\ a[11],\ a[12],\ a[13],\ a[14],\ a[15],\ a[16],\ a[17] \} \right\},\\
&& \rm \qquad \left. \left\{ \{ a[2],\ -a[3],\ -a[4],\ -a[5],\ -a[6],\ -a[7],\ -a[8],\ -a[9] \},\ \{a[18],\ 0,\ 0,\ 0,\ 0,\ 0,\ 0,\ 0\}, \right. \right. \\
&& \rm \left. \qquad\ \ \{ a[19],\ a[20],\ a[21],\ a[22],\ a[23],\ a[24],\ a[25],\ a[26] \} \right\},\\
&& \rm \qquad \left\{ \{ a[10],\ -a[11],\ -a[12],\ -a[13],\ -a[14],\ -a[15],\ -a[16],\ -a[17] \},\ \left\{a[19],\ -a[20], \right. \right.\\
&& \rm \left.\left.\left. \qquad\ \ -a[21],\ -a[22],\ -a[23],\ -a[24],\ -a[25],\ -a[26] \right\},\ \{a[27],\ 0,\ 0,\ 0,\ 0,\ 0,\ 0,\ 0\} \right\}
   \right\};\\
%%%%%%%%%%%%%%%%%%%%%%%%%%%%%%%%%%%%%%%%%%%%%%%%%%%%%%%
&& \rm {\rm B} =\left\{ \left\{ \{ b[1],\ 0,\ 0,\ 0,\ 0,\ 0,\ 0,\ 0\},\ \{b[2],\ b[3],\ b[4],\ b[5],\ b[6],\ b[7],\ b[8],\ b[9] \}, \right.\right.\\
&& \rm \left. \qquad\ \ \{ b[10],\ b[11],\ b[12],\ b[13],\ b[14],\ b[15],\ b[16],\ b[17] \} \right\},\\
&& \rm \qquad \left.\left\{ \{ b[2],\ -b[3],\ -b[4],\ -b[5],\ -b[6],\ -b[7],\ -b[8],\ -b[9] \},\ \{b[18],\ 0,\ 0,\ 0,\ 0,\ 0,\ 0,\ 0\}, \right. \right.
\end{eqnarray*}
\begin{eqnarray*}
&& \rm \left. \qquad\ \ \{ b[19],\ b[20],\ b[21],\ b[22],\ b[23],\ b[24],\ b[25],\ b[26] \} \right\},\\
&& \rm \qquad \left\{ \{ b[10],\ -b[11],\ -b[12],\ -b[13],\ -b[14],\ -b[15],\ -b[16],\ -b[17] \},\ \left\{b[19],\ -b[20], \right. \right.\\
&& \rm \left.\left.\left. \qquad\ \ -b[21],\ -b[22],\ -b[23],\ -b[24],\ -b[25],\ -b[26] \right\},\ \{b[27],\ 0,\ 0,\ 0,\ 0,\ 0,\ 0,\ 0\} \right\}
   \right\};\\
%%%%%%%%%%%%%%%%%%%%%%%%%%%%%%%%%%%%%%%%%%%%%%%%%%%%%%%%
&& {\rm FF}[{\rm AA}_\_]:=\left\{ \Part[\{\Part[\Part[\Part[{\rm AA},1],\ 1],\ 1]\},\ 1],\right.\\
&& \qquad \Part[\{\Part[\Part[\Part[{\rm AA},1],\ 2],\ 1]\},\ 1],\ \Part[\{\Part[\Part[\Part[{\rm AA},1],\ 2],\ 2]\},\ 1],\\
&& \qquad \Part[\{\Part[\Part[\Part[{\rm AA},1],\ 2],\ 3]\},\ 1],\ \Part[\{\Part[\Part[\Part[{\rm AA},1],\ 2],\ 4]\},\ 1],\\
&& \qquad \Part[\{\Part[\Part[\Part[{\rm AA},1],\ 2],\ 5]\},\ 1],\ \Part[\{\Part[\Part[\Part[{\rm AA},1],\ 2],\ 6]\},\ 1],\\
&& \qquad \Part[\{\Part[\Part[\Part[{\rm AA},1],\ 2],\ 7]\},\ 1],\ \Part[\{\Part[\Part[\Part[{\rm AA},1],\ 2],\ 8]\},\ 1],\\
&& \qquad \Part[\{\Part[\Part[\Part[{\rm AA},1],\ 3],\ 1]\},\ 1],\ \Part[\{\Part[\Part[\Part[{\rm AA},1],\ 3],\ 2]\},\ 1],\\
&& \qquad \Part[\{\Part[\Part[\Part[{\rm AA},1],\ 3],\ 3]\},\ 1],\ \Part[\{\Part[\Part[\Part[{\rm AA},1],\ 3],\ 4]\},\ 1],\\
&& \qquad \Part[\{\Part[\Part[\Part[{\rm AA},1],\ 3],\ 5]\},\ 1],\ \Part[\{\Part[\Part[\Part[{\rm AA},1],\ 3],\ 6]\},\ 1],\\
&& \qquad \Part[\{\Part[\Part[\Part[{\rm AA},1],\ 3],\ 7]\},\ 1],\ \Part[\{\Part[\Part[\Part[{\rm AA},1],\ 3],\ 8]\},\ 1],\\
&& \qquad \Part[\{\Part[\Part[\Part[{\rm AA},2],\ 2],\ 1]\},\ 1],\ \Part[\{\Part[\Part[\Part[{\rm AA},2],\ 3],\ 1]\},\ 1],\\
&& \qquad \Part[\{\Part[\Part[\Part[{\rm AA},2],\ 3],\ 2]\},\ 1],\ \Part[\{\Part[\Part[\Part[{\rm AA},2],\ 3],\ 3]\},\ 1],\\
&& \qquad \Part[\{\Part[\Part[\Part[{\rm AA},2],\ 3],\ 4]\},\ 1],\ \Part[\{\Part[\Part[\Part[{\rm AA},2],\ 3],\ 5]\},\ 1],\\
&& \qquad \Part[\{\Part[\Part[\Part[{\rm AA},2],\ 3],\ 6]\},\ 1],\ \Part[\{\Part[\Part[\Part[{\rm AA},2],\ 3],\ 7]\},\ 1],\\
&& \left.\qquad \Part[\{\Part[\Part[\Part[{\rm AA},2],\ 3],\ 8]\},\ 1],\ \Part[\{\Part[\Part[\Part[{\rm AA},3],\ 3],\ 1]\},\ 1]\right\}\\
%%%%%%%%%%%%%%%%%%%%%%%%%%%%%%%%%%%%%%%%%%%%%%%%%%%%%%%%%
&& {\rm FFi}[{\rm vv}_\_]:=\\
&& \ \ \left\{ \left\{ \{ \Part[{\rm vv},\ 1],\ 0,\ 0,\ 0,\ 0,\ 0,\ 0,\ 0\},\ \left\{ \Part[{\rm vv},\ 2],\ \Part[{\rm vv},\ 3],\
   \Part[{\rm vv},\ 4],\ \Part[{\rm vv},\ 5], \right. \right. \right. \\
&& \left. \qquad  \Part[{\rm vv},\ 6],\ \Part[{\rm vv},\ 7],\ \Part[{\rm vv},\ 8],\ \Part[{\rm vv},\ 9]\right\},\ \left\{
   \Part[{\rm vv},\ 10],\ \Part[{\rm vv},\ 11], \right. \\
&& \left.\left. \qquad  \!\Part[{\rm vv},\ 12],\ \Part[{\rm vv},\ 13],\ \Part[{\rm vv},\ 14],\ \Part[{\rm vv},\ 15],\
   \Part[{\rm vv},\ 16],\ \Part[{\rm vv},\ 17]\right\}\right\},\\
&& \quad \left\{\left\{ \Part[{\rm vv},\ 2],\ -\Part[{\rm vv},\ 3],\ -\Part[{\rm vv},\ 4],\ -\Part[{\rm vv},\ 5],\ -\Part[{\rm vv},\ 6],
   \right.\right. \\
&& \left. \qquad  -\Part[{\rm vv},\ 7],\ -\Part[{\rm vv},\ 8],\ -\Part[{\rm vv},\ 9]\right\},\ \{
   \Part[{\rm vv},\ 18],\  0,\ 0,\ 0,\ 0,\ 0,\ 0,\ 0\}, \\
&& \quad \ \ \left\{ \Part[{\rm vv},\ 19],\ \Part[{\rm vv},\ 20],\ \Part[{\rm vv},\ 21],\ \Part[{\rm vv},\ 22],\right.\\
&& \left.\left. \qquad \! \Part[{\rm vv},\ 23],\ \Part[{\rm vv},\ 24],\ \Part[{\rm vv},\ 25],\ \Part[{\rm vv},\ 26]\right\}\right\},\\
&& \quad \left\{\left\{ \Part[{\rm vv},\ 10],\ -\Part[{\rm vv},\ 11],\ -\Part[{\rm vv},\ 12],\ -\Part[{\rm vv},\ 13], \right.\right. \\
&& \left. \qquad  -\Part[{\rm vv},\ 14],\ -\Part[{\rm vv},\ 15],\ -\Part[{\rm vv},\ 16],\ -\Part[{\rm vv},\ 17] \right\},\\
&& \quad \ \ \left\{ \Part[{\rm vv},\ 19],\ -\Part[{\rm vv},\ 20],\ -\Part[{\rm vv},\ 21],\ -\Part[{\rm vv},\ 22],\ -\Part[{\rm vv},\ 23],
   \right.\\
&& \left.\left.\left. \qquad \! -\Part[{\rm vv},\ 24],\ -\Part[{\rm vv},\ 25],\ -\Part[{\rm vv},\ 26]\right\},\ \{
   \Part[{\rm vv},\ 27],\  0,\ 0,\ 0,\ 0,\ 0,\ 0,\ 0\}\right\}\right\}\\
&& \%\%\%\% \mbox{     construction of the matrices}\\
&& {\rm MM}=\Array[{\rm mm}, \{27,27\}];\\
&& {\rm vaa}={\rm FF[A]};\\
&& {\rm vbb}={\rm FF[B]};\\
&& {\rm v1aa}={\rm MM.vaa};\\
&& {\rm v1bb}={\rm MM.vbb};\\
&& {\rm AA1}={\rm FFi[v1aa]};\\
&& {\rm BB1}={\rm FFi[v1bb]};
\end{eqnarray*}
\begin{eqnarray*}
&& {\rm V1}={\rm FF[OctPS[AA1,\ B]]};\\
&& {\rm V2}={\rm FF[OctPS[A,\ BB1]]};\\
&& {\rm AB}={\rm OctPS[AB]};\\
&& {\rm V}={\rm FF[AB]};\\
&& {\rm VV=MM.V};\\
&& {\rm diff =VV-V1-V2};\\
&& \Do\left[\phantom{}\right.\\
&& \qquad \left. \Do[\Do[{\rm ff}[i,j,k]=\Coe [\Part[{\rm diff},\ k], a[i]b[j]],\ \{i,\ 27\}],\ \{j,\ 27\}], \{k,\ 27\}\right];\\
&& {\rm n=0};\\
&& \Do\left[ \Do\left[ \Do\left[ {\rm n++}; \right.\right.\right.\\
&& \quad\ \ \left. {\rm If} [{\rm ff[i,\ j,\ k]}==0,\ {\rm n=n-1},\ {\rm Ff[n]}={\rm ff[i,\ j,\ k]}==0,\ {\rm Ff[n]}={\rm ff[i,\ j,\ k]}==0],\
   \{{\rm i},\ 27\}\right],\\
&& \quad\left.\left. {\rm \{ j,\ 27\}}\right],\ {\rm \{k,\ 27\}} \right];\\
&& {\rm s[1]}=\{\};\\
&& \Do[{\rm s[i+1]}=\Append {\rm [s[i],\ Ff[i]]},\ \{{\rm i,\ n}\}];\\
&& {\rm m=0};\\
&& \Do[\Do[\{{\rm m++,\ gg[m]=mm[i,j]} \},\ \{{\rm i,\ 27}\}],\ \{{\rm j,\ 27}\}];\\
&& {\rm v[1]}=\{\};\\
&& \Do[{\rm v[i+1]}=\Append{\rm [v[i],\ gg[i]]}, \{{\rm i,\ 729}\}];\\
&& {\rm sol=Solve[s[n],\ v[730]]};\\
&& \Do[\Do[{\rm mat[i,j]=mm[i,j],\ \{i,\ 27\}}],\ \{{\rm j,\ 27}\}];\\
&& \Do\left[\Do\left[\Do\left[{\rm If}\left[\Part[\Part[\Part{\rm [sol,\ 1,\ i,\ 1]}]]=={\rm mm[j,\ k],\ mat[j,\ k]}=\right.\right.
   \right.\right. \\
&& \qquad \quad \left.\left.\left.\left. {\rm mm[j,\ k]}/.\Part[\Part[{\rm sol,\ 1,\ i}]]\right],\ \{{\rm i,\ 677}\}\right],\
   \{{\rm j,\ 27}\}\right],\ \{{\rm k,\ 27}\}\right];\\
&& {\rm MM}=\Array[ {\rm mat,\ \{27,\ 27\}}];\\
&& {\rm n=0};\\
&& \Do\left[\Do\left[{\rm n++;\ If}\left[{\rm D[MM,\ mm[i,j]]}=={\rm DiagonalMatrix\left[\right.}\right.\right.\right.\\
&& \qquad \quad \left. \{0,\ 0,\ 0,\ 0,\ 0,\ 0,\ 0,\ 0,\ 0,\ 0,\ 0,\ 0,\ 0,\ 0,\ 0,\ 0,\ 0,\ 0,\ 0,\ 0,\ 0,\ 0,\ 0,\ 0,\ 0,\ 0,\ 0\}\right],\\
&& \quad\ \left.\left.\left. {\rm n=n-1,\ Md[n]=D[MM,\ mm[i,j]],\ Md[n]=D[MM,\ mm[i,j]]}\right], \rm{\{i,\ j\}}\right],\
   \rm{\{j,\ 27\}}\right];\\
&& {\rm Mn[1]=Md[1]};\\
&& \Do\left[\Do\left[\right.\right.\\
&& \quad \left.\left. {\rm AA[j,i]=0,\ \{i,\ 52\}}\right],\ {\rm \{j,\ 52\}}\right]\\
&& \Do\left[ \Do[{\rm AA[j,\ i]=-Tr[Md[j].Mn[i]]/Tr[Mn[i].Mn[i]],\ \{i,\ j-1\}}];\right.\\
&& \ \ \left. {\rm Mn[j]=Md[j]+Sum[AA[j,\ i]Mn[i],\ \{i,\ j-1\}]},\ {\rm \{j,\ 2,\ 52\}}\right];\\
&& \Do[c[i]=-{\rm Sqrt[6]}Mn[i] /{\rm Sqrt[Tr[Mn[i].Mn[i]]],\ \{i,\ 52\}}];\\
&& \%\%\%\% \mbox{     the structure constants}\\
&& \Do[\Do[{\rm CC[i,\ j]=c[i].c[j]-c[j].c[i],\ \{i,\ 52\}}], \{j,\ 52\}];\\
&& \Do[\Do[\Do[{\rm coeff[i,\ j,\ k]=-Tr[CC[i,j].c[k]]/6 ,\ \{k,\ 52\}}], {\rm\{i,\ 52\}}], {\rm\{j,\ 52\}}];
\end{eqnarray*}
\begin{eqnarray*}
&& \%\%\% \mbox{     rotation to give the irreducible $26$ representation       }\\
&& XXX=\{\{1,\ 0,\ 0,\ 0,\ 0,\ 0,\ 0,\ 0,\ 0,\ 0,\ 0,\ 0,\ 0,\ 0,\ 0,\ 0,\ 0,\ -1,\ 0,\ 0,\ 0,\ 0,\ 0,\ 0,\ 0,\ 0,\ 0 \}/{\rm Sqrt}[2],\\
&& \qquad\qquad\  \{0,\ 1,\ 0,\ 0,\ 0,\ 0,\ 0,\ 0,\ 0,\ 0,\ 0,\ 0,\ 0,\ 0,\ 0,\ 0,\ 0,\ 0,\ 0,\ 0,\ 0,\ 0,\ 0,\ 0,\ 0,\ 0,\ 0\}, \\
&& \qquad\qquad\  \{0,\ 0,\ 1,\ 0,\ 0,\ 0,\ 0,\ 0,\ 0,\ 0,\ 0,\ 0,\ 0,\ 0,\ 0,\ 0,\ 0,\ 0,\ 0,\ 0,\ 0,\ 0,\ 0,\ 0,\ 0,\ 0,\ 0\}, \\
&& \qquad\qquad\  \{0,\ 0,\ 0,\ 1,\ 0,\ 0,\ 0,\ 0,\ 0,\ 0,\ 0,\ 0,\ 0,\ 0,\ 0,\ 0,\ 0,\ 0,\ 0,\ 0,\ 0,\ 0,\ 0,\ 0,\ 0,\ 0,\ 0\}, \\
&& \qquad\qquad\  \{0,\ 0,\ 0,\ 0,\ 1,\ 0,\ 0,\ 0,\ 0,\ 0,\ 0,\ 0,\ 0,\ 0,\ 0,\ 0,\ 0,\ 0,\ 0,\ 0,\ 0,\ 0,\ 0,\ 0,\ 0,\ 0,\ 0\}, \\
&& \qquad\qquad\  \{0,\ 0,\ 0,\ 0,\ 0,\ 1,\ 0,\ 0,\ 0,\ 0,\ 0,\ 0,\ 0,\ 0,\ 0,\ 0,\ 0,\ 0,\ 0,\ 0,\ 0,\ 0,\ 0,\ 0,\ 0,\ 0,\ 0\}, \\
&& \qquad\qquad\  \{0,\ 0,\ 0,\ 0,\ 0,\ 0,\ 1,\ 0,\ 0,\ 0,\ 0,\ 0,\ 0,\ 0,\ 0,\ 0,\ 0,\ 0,\ 0,\ 0,\ 0,\ 0,\ 0,\ 0,\ 0,\ 0,\ 0\}, \\
&& \qquad\qquad\  \{0,\ 0,\ 0,\ 0,\ 0,\ 0,\ 0,\ 1,\ 0,\ 0,\ 0,\ 0,\ 0,\ 0,\ 0,\ 0,\ 0,\ 0,\ 0,\ 0,\ 0,\ 0,\ 0,\ 0,\ 0,\ 0,\ 0\}, \\
&& \qquad\qquad\  \{0,\ 0,\ 0,\ 0,\ 0,\ 0,\ 0,\ 0,\ 1,\ 0,\ 0,\ 0,\ 0,\ 0,\ 0,\ 0,\ 0,\ 0,\ 0,\ 0,\ 0,\ 0,\ 0,\ 0,\ 0,\ 0,\ 0\}, \\
&& \qquad\qquad\  \{0,\ 0,\ 0,\ 0,\ 0,\ 0,\ 0,\ 0,\ 0,\ 1,\ 0,\ 0,\ 0,\ 0,\ 0,\ 0,\ 0,\ 0,\ 0,\ 0,\ 0,\ 0,\ 0,\ 0,\ 0,\ 0,\ 0\}, \\
&& \qquad\qquad\  \{0,\ 0,\ 0,\ 0,\ 0,\ 0,\ 0,\ 0,\ 0,\ 0,\ 1,\ 0,\ 0,\ 0,\ 0,\ 0,\ 0,\ 0,\ 0,\ 0,\ 0,\ 0,\ 0,\ 0,\ 0,\ 0,\ 0\}, \\
&& \qquad\qquad\  \{0,\ 0,\ 0,\ 0,\ 0,\ 0,\ 0,\ 0,\ 0,\ 0,\ 0,\ 1,\ 0,\ 0,\ 0,\ 0,\ 0,\ 0,\ 0,\ 0,\ 0,\ 0,\ 0,\ 0,\ 0,\ 0,\ 0\}, \\
&& \qquad\qquad\  \{0,\ 0,\ 0,\ 0,\ 0,\ 0,\ 0,\ 0,\ 0,\ 0,\ 0,\ 0,\ 1,\ 0,\ 0,\ 0,\ 0,\ 0,\ 0,\ 0,\ 0,\ 0,\ 0,\ 0,\ 0,\ 0,\ 0\}, \\
&& \qquad\qquad\  \{0,\ 0,\ 0,\ 0,\ 0,\ 0,\ 0,\ 0,\ 0,\ 0,\ 0,\ 0,\ 0,\ 1,\ 0,\ 0,\ 0,\ 0,\ 0,\ 0,\ 0,\ 0,\ 0,\ 0,\ 0,\ 0,\ 0\}, \\
&& \qquad\qquad\  \{0,\ 0,\ 0,\ 0,\ 0,\ 0,\ 0,\ 0,\ 0,\ 0,\ 0,\ 0,\ 0,\ 0,\ 1,\ 0,\ 0,\ 0,\ 0,\ 0,\ 0,\ 0,\ 0,\ 0,\ 0,\ 0,\ 0\}, \\
&& \qquad\qquad\  \{0,\ 0,\ 0,\ 0,\ 0,\ 0,\ 0,\ 0,\ 0,\ 0,\ 0,\ 0,\ 0,\ 0,\ 0,\ 1,\ 0,\ 0,\ 0,\ 0,\ 0,\ 0,\ 0,\ 0,\ 0,\ 0,\ 0\}, \\
&& \qquad\qquad\  \{0,\ 0,\ 0,\ 0,\ 0,\ 0,\ 0,\ 0,\ 0,\ 0,\ 0,\ 0,\ 0,\ 0,\ 0,\ 0,\ 1,\ 0,\ 0,\ 0,\ 0,\ 0,\ 0,\ 0,\ 0,\ 0,\ 0\}, \\
&& \qquad\qquad\  \{1,\ 0,\ 0,\ 0,\ 0,\ 0,\ 0,\ 0,\ 0,\ 0,\ 0,\ 0,\ 0,\ 0,\ 0,\ 0,\ 0,\ 1,\ 0,\ 0,\ 0,\ 0,\ 0,\ 0,\ 0,\ 0,\ -2\}/{\rm Sqrt}[6], \\
&& \qquad\qquad\  \{0,\ 0,\ 0,\ 0,\ 0,\ 0,\ 0,\ 0,\ 0,\ 0,\ 0,\ 0,\ 0,\ 0,\ 0,\ 0,\ 0,\ 0,\ 1,\ 0,\ 0,\ 0,\ 0,\ 0,\ 0,\ 0,\ 0\}, \\
&& \qquad\qquad\  \{0,\ 0,\ 0,\ 0,\ 0,\ 0,\ 0,\ 0,\ 0,\ 0,\ 0,\ 0,\ 0,\ 0,\ 0,\ 0,\ 0,\ 0,\ 0,\ 1,\ 0,\ 0,\ 0,\ 0,\ 0,\ 0,\ 0\}, \\
&& \qquad\qquad\  \{0,\ 0,\ 0,\ 0,\ 0,\ 0,\ 0,\ 0,\ 0,\ 0,\ 0,\ 0,\ 0,\ 0,\ 0,\ 0,\ 0,\ 0,\ 0,\ 0,\ 1,\ 0,\ 0,\ 0,\ 0,\ 0,\ 0\}, \\
&& \qquad\qquad\  \{0,\ 0,\ 0,\ 0,\ 0,\ 0,\ 0,\ 0,\ 0,\ 0,\ 0,\ 0,\ 0,\ 0,\ 0,\ 0,\ 0,\ 0,\ 0,\ 0,\ 0,\ 1,\ 0,\ 0,\ 0,\ 0,\ 0\}, \\
&& \qquad\qquad\  \{0,\ 0,\ 0,\ 0,\ 0,\ 0,\ 0,\ 0,\ 0,\ 0,\ 0,\ 0,\ 0,\ 0,\ 0,\ 0,\ 0,\ 0,\ 0,\ 0,\ 0,\ 0,\ 1,\ 0,\ 0,\ 0,\ 0\}, \\
&& \qquad\qquad\  \{0,\ 0,\ 0,\ 0,\ 0,\ 0,\ 0,\ 0,\ 0,\ 0,\ 0,\ 0,\ 0,\ 0,\ 0,\ 0,\ 0,\ 0,\ 0,\ 0,\ 0,\ 0,\ 0,\ 1,\ 0,\ 0,\ 0\}, \\
&& \qquad\qquad\  \{0,\ 0,\ 0,\ 0,\ 0,\ 0,\ 0,\ 0,\ 0,\ 0,\ 0,\ 0,\ 0,\ 0,\ 0,\ 0,\ 0,\ 0,\ 0,\ 0,\ 0,\ 0,\ 0,\ 0,\ 1,\ 0,\ 0\}, \\
&& \qquad\qquad\  \{0,\ 0,\ 0,\ 0,\ 0,\ 0,\ 0,\ 0,\ 0,\ 0,\ 0,\ 0,\ 0,\ 0,\ 0,\ 0,\ 0,\ 0,\ 0,\ 0,\ 0,\ 0,\ 0,\ 0,\ 0,\ 1,\ 0\}, \\
&& \qquad\qquad\  \{1,\ 0,\ 0,\ 0,\ 0,\ 0,\ 0,\ 0,\ 0,\ 0,\ 0,\ 0,\ 0,\ 0,\ 0,\ 0,\ 0,\ 1,\ 0,\ 0,\ 0,\ 0,\ 0,\ 0,\ 0,\ 0,\ 1\}/{\rm Sqrt}[3] \};\\
&& \Do[cc[i]=XXX.c[i].{\rm Transpose}[XXX],\ \{i,\ 52\}];
\end{eqnarray*}

\

\noindent c[i] are the matrices before the rotation, and cc[i] are the $27\times 27$ matrices after the rotations, all having the last row and
the last column vanishing. The irreducible $26$ dimensional representation is obtained dropping the last row and the last column.
The structure constants are provided by coeff[i,j,k].

\newpage
%%%%%%%%%%%%%%%%%%%%%%%%%%%%%%%%%%%%%%%%%%%%%%%%%%%%%%%%%%%%%%%%%%%%%%%%
\section{The \boldmath{$f_4$} matrices.}
\setcounter{equation}{0}
The matrices we found using Mathematica, and orthonormalized with respect to the scalar product
$\langle a , b\rangle :=-\frac 16 Trace (ab)$, are
{\tiny
\eqn
c_{1}=\lp
\begin{array}{ccccccccccccccccccccccccccc}
0,\!\!\!\!\!\!\!\! &0,\!\!\!\!\!\!\!\! &0,\!\!\!\!\!\!\!\! &0,\!\!\!\!\!\!\!\! &0,\!\!\!\!\!\!\!\! &0,\!\!\!\!\!\!\!\! &0,\!\!\!\!\!\!\!\! &0,\!\!\!\!\!\!\!\! &0,\!\!\!\!\!\!\!\! &0,\!\!\!\!\!\!\!\! &0,\!\!\!\!\!\!\!\! &0,\!\!\!\!\!\!\!\! &0,\!\!\!\!\!\!\!\! &0,\!\!\!\!\!\!\!\! &0,\!\!\!\!\!\!\!\! &0,\!\!\!\!\!\!\!\! &0,\!\!\!\!\!\!\!\! &0,\!\!\!\!\!\!\!\! &0,\!\!\!\!\!\!\!\! &0,\!\!\!\!\!\!\!\! &0,\!\!\!\!\!\!\!\! &0,\!\!\!\!\!\!\!\! &0,\!\!\!\!\!\!\!\! &0,\!\!\!\!\!\!\!\! &0,\!\!\!\!\!\!\!\! &0,\!\!\!\!\!\!\!\! &0 \\
0,\!\!\!\!\!\!\!\! &0,\!\!\!\!\!\!\!\! &-\frac 12,\!\!\!\!\!\!\!\! &0,\!\!\!\!\!\!\!\! &0,\!\!\!\!\!\!\!\! &0,\!\!\!\!\!\!\!\! &0,\!\!\!\!\!\!\!\! &0,\!\!\!\!\!\!\!\! &0,\!\!\!\!\!\!\!\! &0,\!\!\!\!\!\!\!\! &0,\!\!\!\!\!\!\!\! &0,\!\!\!\!\!\!\!\! &0,\!\!\!\!\!\!\!\! &0,\!\!\!\!\!\!\!\! &0,\!\!\!\!\!\!\!\! &0,\!\!\!\!\!\!\!\! &0,\!\!\!\!\!\!\!\! &0,\!\!\!\!\!\!\!\! &0,\!\!\!\!\!\!\!\! &0,\!\!\!\!\!\!\!\! &0,\!\!\!\!\!\!\!\! &0,\!\!\!\!\!\!\!\! &0,\!\!\!\!\!\!\!\! &0,\!\!\!\!\!\!\!\! &0,\!\!\!\!\!\!\!\! &0,\!\!\!\!\!\!\!\! &0 \\
0,\!\!\!\!\!\!\!\! &\frac 12,\!\!\!\!\!\!\!\! &0,\!\!\!\!\!\!\!\! &0,\!\!\!\!\!\!\!\! &0,\!\!\!\!\!\!\!\! &0,\!\!\!\!\!\!\!\! &0,\!\!\!\!\!\!\!\! &0,\!\!\!\!\!\!\!\! &0,\!\!\!\!\!\!\!\! &0,\!\!\!\!\!\!\!\! &0,\!\!\!\!\!\!\!\! &0,\!\!\!\!\!\!\!\! &0,\!\!\!\!\!\!\!\! &0,\!\!\!\!\!\!\!\! &0,\!\!\!\!\!\!\!\! &0,\!\!\!\!\!\!\!\! &0,\!\!\!\!\!\!\!\! &0,\!\!\!\!\!\!\!\! &0,\!\!\!\!\!\!\!\! &0,\!\!\!\!\!\!\!\! &0,\!\!\!\!\!\!\!\! &0,\!\!\!\!\!\!\!\! &0,\!\!\!\!\!\!\!\! &0,\!\!\!\!\!\!\!\! &0,\!\!\!\!\!\!\!\! &0,\!\!\!\!\!\!\!\! &0 \\
\end{array}
\rp
\feqn
\eqn
c_{2}=\lp
\begin{array}{ccccccccccccccccccccccccccc}
0,\!\!\!\!\!\!\!\! &0,\!\!\!\!\!\!\!\! &0,\!\!\!\!\!\!\!\! &0,\!\!\!\!\!\!\!\! &0,\!\!\!\!\!\!\!\! &0,\!\!\!\!\!\!\!\! &0,\!\!\!\!\!\!\!\! &0,\!\!\!\!\!\!\!\! &0,\!\!\!\!\!\!\!\! &0,\!\!\!\!\!\!\!\! &0,\!\!\!\!\!\!\!\! &0,\!\!\!\!\!\!\!\! &0,\!\!\!\!\!\!\!\! &0,\!\!\!\!\!\!\!\! &0,\!\!\!\!\!\!\!\! &0,\!\!\!\!\!\!\!\! &0,\!\!\!\!\!\!\!\! &0,\!\!\!\!\!\!\!\! &0,\!\!\!\!\!\!\!\! &0,\!\!\!\!\!\!\!\! &0,\!\!\!\!\!\!\!\! &0,\!\!\!\!\!\!\!\! &0,\!\!\!\!\!\!\!\! &0,\!\!\!\!\!\!\!\! &0,\!\!\!\!\!\!\!\! &0,\!\!\!\!\!\!\!\! &0 \\
0,\!\!\!\!\!\!\!\! &0,\!\!\!\!\!\!\!\! &0,\!\!\!\!\!\!\!\! &-\frac 12,\!\!\!\!\!\!\!\! &0,\!\!\!\!\!\!\!\! &0,\!\!\!\!\!\!\!\! &0,\!\!\!\!\!\!\!\! &0,\!\!\!\!\!\!\!\! &0,\!\!\!\!\!\!\!\! &0,\!\!\!\!\!\!\!\! &0,\!\!\!\!\!\!\!\! &0,\!\!\!\!\!\!\!\! &0,\!\!\!\!\!\!\!\! &0,\!\!\!\!\!\!\!\! &0,\!\!\!\!\!\!\!\! &0,\!\!\!\!\!\!\!\! &0,\!\!\!\!\!\!\!\! &0,\!\!\!\!\!\!\!\! &0,\!\!\!\!\!\!\!\! &0,\!\!\!\!\!\!\!\! &0,\!\!\!\!\!\!\!\! &0,\!\!\!\!\!\!\!\! &0,\!\!\!\!\!\!\!\! &0,\!\!\!\!\!\!\!\! &0,\!\!\!\!\!\!\!\! &0,\!\!\!\!\!\!\!\! &0 \\
0,\!\!\!\!\!\!\!\! &0,\!\!\!\!\!\!\!\! &0,\!\!\!\!\!\!\!\! &0,\!\!\!\!\!\!\!\! &0,\!\!\!\!\!\!\!\! &-\frac 12,\!\!\!\!\!\!\!\! &0,\!\!\!\!\!\!\!\! &0,\!\!\!\!\!\!\!\! &0,\!\!\!\!\!\!\!\! &0,\!\!\!\!\!\!\!\! &0,\!\!\!\!\!\!\!\! &0,\!\!\!\!\!\!\!\! &0,\!\!\!\!\!\!\!\! &0,\!\!\!\!\!\!\!\! &0,\!\!\!\!\!\!\!\! &0,\!\!\!\!\!\!\!\! &0,\!\!\!\!\!\!\!\! &0,\!\!\!\!\!\!\!\! &0,\!\!\!\!\!\!\!\! &0,\!\!\!\!\!\!\!\! &0,\!\!\!\!\!\!\!\! &0,\!\!\!\!\!\!\!\! &0,\!\!\!\!\!\!\!\! &0,\!\!\!\!\!\!\!\! &0,\!\!\!\!\!\!\!\! &0,\!\!\!\!\!\!\!\! &0 \\
\end{array}
\rp
\feqn
\eqn
c_{3}=\lp
\begin{array}{ccccccccccccccccccccccccccc}
0,\!\!\!\!\!\!\!\! &0,\!\!\!\!\!\!\!\! &0,\!\!\!\!\!\!\!\! &0,\!\!\!\!\!\!\!\! &0,\!\!\!\!\!\!\!\! &0,\!\!\!\!\!\!\!\! &0,\!\!\!\!\!\!\!\! &0,\!\!\!\!\!\!\!\! &0,\!\!\!\!\!\!\!\! &0,\!\!\!\!\!\!\!\! &0,\!\!\!\!\!\!\!\! &0,\!\!\!\!\!\!\!\! &0,\!\!\!\!\!\!\!\! &0,\!\!\!\!\!\!\!\! &0,\!\!\!\!\!\!\!\! &0,\!\!\!\!\!\!\!\! &0,\!\!\!\!\!\!\!\! &0,\!\!\!\!\!\!\!\! &0,\!\!\!\!\!\!\!\! &0,\!\!\!\!\!\!\!\! &0,\!\!\!\!\!\!\!\! &0,\!\!\!\!\!\!\!\! &0,\!\!\!\!\!\!\!\! &0,\!\!\!\!\!\!\!\! &0,\!\!\!\!\!\!\!\! &0,\!\!\!\!\!\!\!\! &0 \\
0,\!\!\!\!\!\!\!\! &0,\!\!\!\!\!\!\!\! &0,\!\!\!\!\!\!\!\! &0,\!\!\!\!\!\!\!\! &0,\!\!\!\!\!\!\!\! &-\frac 12,\!\!\!\!\!\!\!\! &0,\!\!\!\!\!\!\!\! &0,\!\!\!\!\!\!\!\! &0,\!\!\!\!\!\!\!\! &0,\!\!\!\!\!\!\!\! &0,\!\!\!\!\!\!\!\! &0,\!\!\!\!\!\!\!\! &0,\!\!\!\!\!\!\!\! &0,\!\!\!\!\!\!\!\! &0,\!\!\!\!\!\!\!\! &0,\!\!\!\!\!\!\!\! &0,\!\!\!\!\!\!\!\! &0,\!\!\!\!\!\!\!\! &0,\!\!\!\!\!\!\!\! &0,\!\!\!\!\!\!\!\! &0,\!\!\!\!\!\!\!\! &0,\!\!\!\!\!\!\!\! &0,\!\!\!\!\!\!\!\! &0,\!\!\!\!\!\!\!\! &0,\!\!\!\!\!\!\!\! &0,\!\!\!\!\!\!\!\! &0 \\
0,\!\!\!\!\!\!\!\! &0,\!\!\!\!\!\!\!\! &0,\!\!\!\!\!\!\!\! &\frac 12,\!\!\!\!\!\!\!\! &0,\!\!\!\!\!\!\!\! &0,\!\!\!\!\!\!\!\! &0,\!\!\!\!\!\!\!\! &0,\!\!\!\!\!\!\!\! &0,\!\!\!\!\!\!\!\! &0,\!\!\!\!\!\!\!\! &0,\!\!\!\!\!\!\!\! &0,\!\!\!\!\!\!\!\! &0,\!\!\!\!\!\!\!\! &0,\!\!\!\!\!\!\!\! &0,\!\!\!\!\!\!\!\! &0,\!\!\!\!\!\!\!\! &0,\!\!\!\!\!\!\!\! &0,\!\!\!\!\!\!\!\! &0,\!\!\!\!\!\!\!\! &0,\!\!\!\!\!\!\!\! &0,\!\!\!\!\!\!\!\! &0,\!\!\!\!\!\!\!\! &0,\!\!\!\!\!\!\!\! &0,\!\!\!\!\!\!\!\! &0,\!\!\!\!\!\!\!\! &0,\!\!\!\!\!\!\!\! &0 \\
\end{array}
\rp
\feqn
\eqn
c_{4}=\lp
\begin{array}{ccccccccccccccccccccccccccc}
0,\!\!\!\!\!\!\!\! &0,\!\!\!\!\!\!\!\! &0,\!\!\!\!\!\!\!\! &0,\!\!\!\!\!\!\!\! &0,\!\!\!\!\!\!\!\! &0,\!\!\!\!\!\!\!\! &0,\!\!\!\!\!\!\!\! &0,\!\!\!\!\!\!\!\! &0,\!\!\!\!\!\!\!\! &0,\!\!\!\!\!\!\!\! &0,\!\!\!\!\!\!\!\! &0,\!\!\!\!\!\!\!\! &0,\!\!\!\!\!\!\!\! &0,\!\!\!\!\!\!\!\! &0,\!\!\!\!\!\!\!\! &0,\!\!\!\!\!\!\!\! &0,\!\!\!\!\!\!\!\! &0,\!\!\!\!\!\!\!\! &0,\!\!\!\!\!\!\!\! &0,\!\!\!\!\!\!\!\! &0,\!\!\!\!\!\!\!\! &0,\!\!\!\!\!\!\!\! &0,\!\!\!\!\!\!\!\! &0,\!\!\!\!\!\!\!\! &0,\!\!\!\!\!\!\!\! &0,\!\!\!\!\!\!\!\! &0 \\
0,\!\!\!\!\!\!\!\! &0,\!\!\!\!\!\!\!\! &0,\!\!\!\!\!\!\!\! &0,\!\!\!\!\!\!\!\! &-\frac 12,\!\!\!\!\!\!\!\! &0,\!\!\!\!\!\!\!\! &0,\!\!\!\!\!\!\!\! &0,\!\!\!\!\!\!\!\! &0,\!\!\!\!\!\!\!\! &0,\!\!\!\!\!\!\!\! &0,\!\!\!\!\!\!\!\! &0,\!\!\!\!\!\!\!\! &0,\!\!\!\!\!\!\!\! &0,\!\!\!\!\!\!\!\! &0,\!\!\!\!\!\!\!\! &0,\!\!\!\!\!\!\!\! &0,\!\!\!\!\!\!\!\! &0,\!\!\!\!\!\!\!\! &0,\!\!\!\!\!\!\!\! &0,\!\!\!\!\!\!\!\! &0,\!\!\!\!\!\!\!\! &0,\!\!\!\!\!\!\!\! &0,\!\!\!\!\!\!\!\! &0,\!\!\!\!\!\!\!\! &0,\!\!\!\!\!\!\!\! &0,\!\!\!\!\!\!\!\! &0 \\
0,\!\!\!\!\!\!\!\! &0,\!\!\!\!\!\!\!\! &0,\!\!\!\!\!\!\!\! &0,\!\!\!\!\!\!\!\! &0,\!\!\!\!\!\!\!\! &0,\!\!\!\!\!\!\!\! &0,\!\!\!\!\!\!\!\! &0,\!\!\!\!\!\!\!\! &-\frac 12,\!\!\!\!\!\!\!\! &0,\!\!\!\!\!\!\!\! &0,\!\!\!\!\!\!\!\! &0,\!\!\!\!\!\!\!\! &0,\!\!\!\!\!\!\!\! &0,\!\!\!\!\!\!\!\! &0,\!\!\!\!\!\!\!\! &0,\!\!\!\!\!\!\!\! &0,\!\!\!\!\!\!\!\! &0,\!\!\!\!\!\!\!\! &0,\!\!\!\!\!\!\!\! &0,\!\!\!\!\!\!\!\! &0,\!\!\!\!\!\!\!\! &0,\!\!\!\!\!\!\!\! &0,\!\!\!\!\!\!\!\! &0,\!\!\!\!\!\!\!\! &0,\!\!\!\!\!\!\!\! &0,\!\!\!\!\!\!\!\! &0 \\
\end{array}
\rp
\feqn
\eqn
c_{5}=\lp
\begin{array}{ccccccccccccccccccccccccccc}
0,\!\!\!\!\!\!\!\! &0,\!\!\!\!\!\!\!\! &0,\!\!\!\!\!\!\!\! &0,\!\!\!\!\!\!\!\! &0,\!\!\!\!\!\!\!\! &0,\!\!\!\!\!\!\!\! &0,\!\!\!\!\!\!\!\! &0,\!\!\!\!\!\!\!\! &0,\!\!\!\!\!\!\!\! &0,\!\!\!\!\!\!\!\! &0,\!\!\!\!\!\!\!\! &0,\!\!\!\!\!\!\!\! &0,\!\!\!\!\!\!\!\! &0,\!\!\!\!\!\!\!\! &0,\!\!\!\!\!\!\!\! &0,\!\!\!\!\!\!\!\! &0,\!\!\!\!\!\!\!\! &0,\!\!\!\!\!\!\!\! &0,\!\!\!\!\!\!\!\! &0,\!\!\!\!\!\!\!\! &0,\!\!\!\!\!\!\!\! &0,\!\!\!\!\!\!\!\! &0,\!\!\!\!\!\!\!\! &0,\!\!\!\!\!\!\!\! &0,\!\!\!\!\!\!\!\! &0,\!\!\!\!\!\!\!\! &0 \\
0,\!\!\!\!\!\!\!\! &0,\!\!\!\!\!\!\!\! &0,\!\!\!\!\!\!\!\! &0,\!\!\!\!\!\!\!\! &0,\!\!\!\!\!\!\!\! &0,\!\!\!\!\!\!\!\! &0,\!\!\!\!\!\!\!\! &0,\!\!\!\!\!\!\!\! &-\frac 12,\!\!\!\!\!\!\!\! &0,\!\!\!\!\!\!\!\! &0,\!\!\!\!\!\!\!\! &0,\!\!\!\!\!\!\!\! &0,\!\!\!\!\!\!\!\! &0,\!\!\!\!\!\!\!\! &0,\!\!\!\!\!\!\!\! &0,\!\!\!\!\!\!\!\! &0,\!\!\!\!\!\!\!\! &0,\!\!\!\!\!\!\!\! &0,\!\!\!\!\!\!\!\! &0,\!\!\!\!\!\!\!\! &0,\!\!\!\!\!\!\!\! &0,\!\!\!\!\!\!\!\! &0,\!\!\!\!\!\!\!\! &0,\!\!\!\!\!\!\!\! &0,\!\!\!\!\!\!\!\! &0,\!\!\!\!\!\!\!\! &0 \\
0,\!\!\!\!\!\!\!\! &0,\!\!\!\!\!\!\!\! &0,\!\!\!\!\!\!\!\! &0,\!\!\!\!\!\!\!\! &\frac 12,\!\!\!\!\!\!\!\! &0,\!\!\!\!\!\!\!\! &0,\!\!\!\!\!\!\!\! &0,\!\!\!\!\!\!\!\! &0,\!\!\!\!\!\!\!\! &0,\!\!\!\!\!\!\!\! &0,\!\!\!\!\!\!\!\! &0,\!\!\!\!\!\!\!\! &0,\!\!\!\!\!\!\!\! &0,\!\!\!\!\!\!\!\! &0,\!\!\!\!\!\!\!\! &0,\!\!\!\!\!\!\!\! &0,\!\!\!\!\!\!\!\! &0,\!\!\!\!\!\!\!\! &0,\!\!\!\!\!\!\!\! &0,\!\!\!\!\!\!\!\! &0,\!\!\!\!\!\!\!\! &0,\!\!\!\!\!\!\!\! &0,\!\!\!\!\!\!\!\! &0,\!\!\!\!\!\!\!\! &0,\!\!\!\!\!\!\!\! &0,\!\!\!\!\!\!\!\! &0 \\
\end{array}
\rp
\feqn
\eqn
c_{6}=\lp
\begin{array}{ccccccccccccccccccccccccccc}
0,\!\!\!\!\!\!\!\! &0,\!\!\!\!\!\!\!\! &0,\!\!\!\!\!\!\!\! &0,\!\!\!\!\!\!\!\! &0,\!\!\!\!\!\!\!\! &0,\!\!\!\!\!\!\!\! &0,\!\!\!\!\!\!\!\! &0,\!\!\!\!\!\!\!\! &0,\!\!\!\!\!\!\!\! &0,\!\!\!\!\!\!\!\! &0,\!\!\!\!\!\!\!\! &0,\!\!\!\!\!\!\!\! &0,\!\!\!\!\!\!\!\! &0,\!\!\!\!\!\!\!\! &0,\!\!\!\!\!\!\!\! &0,\!\!\!\!\!\!\!\! &0,\!\!\!\!\!\!\!\! &0,\!\!\!\!\!\!\!\! &0,\!\!\!\!\!\!\!\! &0,\!\!\!\!\!\!\!\! &0,\!\!\!\!\!\!\!\! &0,\!\!\!\!\!\!\!\! &0,\!\!\!\!\!\!\!\! &0,\!\!\!\!\!\!\!\! &0,\!\!\!\!\!\!\!\! &0,\!\!\!\!\!\!\!\! &0 \\
0,\!\!\!\!\!\!\!\! &0,\!\!\!\!\!\!\!\! &0,\!\!\!\!\!\!\!\! &0,\!\!\!\!\!\!\!\! &0,\!\!\!\!\!\!\!\! &0,\!\!\!\!\!\!\!\! &-\frac 12,\!\!\!\!\!\!\!\! &0,\!\!\!\!\!\!\!\! &0,\!\!\!\!\!\!\!\! &0,\!\!\!\!\!\!\!\! &0,\!\!\!\!\!\!\!\! &0,\!\!\!\!\!\!\!\! &0,\!\!\!\!\!\!\!\! &0,\!\!\!\!\!\!\!\! &0,\!\!\!\!\!\!\!\! &0,\!\!\!\!\!\!\!\! &0,\!\!\!\!\!\!\!\! &0,\!\!\!\!\!\!\!\! &0,\!\!\!\!\!\!\!\! &0,\!\!\!\!\!\!\!\! &0,\!\!\!\!\!\!\!\! &0,\!\!\!\!\!\!\!\! &0,\!\!\!\!\!\!\!\! &0,\!\!\!\!\!\!\!\! &0,\!\!\!\!\!\!\!\! &0,\!\!\!\!\!\!\!\! &0 \\
0,\!\!\!\!\!\!\!\! &0,\!\!\!\!\!\!\!\! &0,\!\!\!\!\!\!\!\! &0,\!\!\!\!\!\!\!\! &0,\!\!\!\!\!\!\!\! &0,\!\!\!\!\!\!\!\! &0,\!\!\!\!\!\!\!\! &\frac 12,\!\!\!\!\!\!\!\! &0,\!\!\!\!\!\!\!\! &0,\!\!\!\!\!\!\!\! &0,\!\!\!\!\!\!\!\! &0,\!\!\!\!\!\!\!\! &0,\!\!\!\!\!\!\!\! &0,\!\!\!\!\!\!\!\! &0,\!\!\!\!\!\!\!\! &0,\!\!\!\!\!\!\!\! &0,\!\!\!\!\!\!\!\! &0,\!\!\!\!\!\!\!\! &0,\!\!\!\!\!\!\!\! &0,\!\!\!\!\!\!\!\! &0,\!\!\!\!\!\!\!\! &0,\!\!\!\!\!\!\!\! &0,\!\!\!\!\!\!\!\! &0,\!\!\!\!\!\!\!\! &0,\!\!\!\!\!\!\!\! &0,\!\!\!\!\!\!\!\! &0 \\
\end{array}
\rp
\feqn
\eqn
c_{7}=\lp
\begin{array}{ccccccccccccccccccccccccccc}
0,\!\!\!\!\!\!\!\! &0,\!\!\!\!\!\!\!\! &0,\!\!\!\!\!\!\!\! &0,\!\!\!\!\!\!\!\! &0,\!\!\!\!\!\!\!\! &0,\!\!\!\!\!\!\!\! &0,\!\!\!\!\!\!\!\! &0,\!\!\!\!\!\!\!\! &0,\!\!\!\!\!\!\!\! &0,\!\!\!\!\!\!\!\! &0,\!\!\!\!\!\!\!\! &0,\!\!\!\!\!\!\!\! &0,\!\!\!\!\!\!\!\! &0,\!\!\!\!\!\!\!\! &0,\!\!\!\!\!\!\!\! &0,\!\!\!\!\!\!\!\! &0,\!\!\!\!\!\!\!\! &0,\!\!\!\!\!\!\!\! &0,\!\!\!\!\!\!\!\! &0,\!\!\!\!\!\!\!\! &0,\!\!\!\!\!\!\!\! &0,\!\!\!\!\!\!\!\! &0,\!\!\!\!\!\!\!\! &0,\!\!\!\!\!\!\!\! &0,\!\!\!\!\!\!\!\! &0,\!\!\!\!\!\!\!\! &0 \\
0,\!\!\!\!\!\!\!\! &0,\!\!\!\!\!\!\!\! &0,\!\!\!\!\!\!\!\! &0,\!\!\!\!\!\!\!\! &0,\!\!\!\!\!\!\!\! &-\frac 12,\!\!\!\!\!\!\!\! &0,\!\!\!\!\!\!\!\! &0,\!\!\!\!\!\!\!\! &0,\!\!\!\!\!\!\!\! &0,\!\!\!\!\!\!\!\! &0,\!\!\!\!\!\!\!\! &0,\!\!\!\!\!\!\!\! &0,\!\!\!\!\!\!\!\! &0,\!\!\!\!\!\!\!\! &0,\!\!\!\!\!\!\!\! &0,\!\!\!\!\!\!\!\! &0,\!\!\!\!\!\!\!\! &0,\!\!\!\!\!\!\!\! &0,\!\!\!\!\!\!\!\! &0,\!\!\!\!\!\!\!\! &0,\!\!\!\!\!\!\!\! &0,\!\!\!\!\!\!\!\! &0,\!\!\!\!\!\!\!\! &0,\!\!\!\!\!\!\!\! &0,\!\!\!\!\!\!\!\! &0,\!\!\!\!\!\!\!\! &0 \\
0,\!\!\!\!\!\!\!\! &0,\!\!\!\!\!\!\!\! &0,\!\!\!\!\!\!\!\! &\frac 12,\!\!\!\!\!\!\!\! &0,\!\!\!\!\!\!\!\! &0,\!\!\!\!\!\!\!\! &0,\!\!\!\!\!\!\!\! &0,\!\!\!\!\!\!\!\! &0,\!\!\!\!\!\!\!\! &0,\!\!\!\!\!\!\!\! &0,\!\!\!\!\!\!\!\! &0,\!\!\!\!\!\!\!\! &0,\!\!\!\!\!\!\!\! &0,\!\!\!\!\!\!\!\! &0,\!\!\!\!\!\!\!\! &0,\!\!\!\!\!\!\!\! &0,\!\!\!\!\!\!\!\! &0,\!\!\!\!\!\!\!\! &0,\!\!\!\!\!\!\!\! &0,\!\!\!\!\!\!\!\! &0,\!\!\!\!\!\!\!\! &0,\!\!\!\!\!\!\!\! &0,\!\!\!\!\!\!\!\! &0,\!\!\!\!\!\!\!\! &0,\!\!\!\!\!\!\!\! &0,\!\!\!\!\!\!\!\! &0 \\
\end{array}
\rp
\feqn
\eqn
c_{8}=\lp
\begin{array}{ccccccccccccccccccccccccccc}
0,\!\!\!\!\!\!\!\! &0,\!\!\!\!\!\!\!\! &0,\!\!\!\!\!\!\!\! &0,\!\!\!\!\!\!\!\! &0,\!\!\!\!\!\!\!\! &0,\!\!\!\!\!\!\!\! &0,\!\!\!\!\!\!\!\! &0,\!\!\!\!\!\!\!\! &0,\!\!\!\!\!\!\!\! &0,\!\!\!\!\!\!\!\! &0,\!\!\!\!\!\!\!\! &0,\!\!\!\!\!\!\!\! &0,\!\!\!\!\!\!\!\! &0,\!\!\!\!\!\!\!\! &0,\!\!\!\!\!\!\!\! &0,\!\!\!\!\!\!\!\! &0,\!\!\!\!\!\!\!\! &0,\!\!\!\!\!\!\!\! &0,\!\!\!\!\!\!\!\! &0,\!\!\!\!\!\!\!\! &0,\!\!\!\!\!\!\!\! &0,\!\!\!\!\!\!\!\! &0,\!\!\!\!\!\!\!\! &0,\!\!\!\!\!\!\!\! &0,\!\!\!\!\!\!\!\! &0,\!\!\!\!\!\!\!\! &0 \\
0,\!\!\!\!\!\!\!\! &0,\!\!\!\!\!\!\!\! &0,\!\!\!\!\!\!\!\! &\frac 12,\!\!\!\!\!\!\!\! &0,\!\!\!\!\!\!\!\! &0,\!\!\!\!\!\!\!\! &0,\!\!\!\!\!\!\!\! &0,\!\!\!\!\!\!\!\! &0,\!\!\!\!\!\!\!\! &0,\!\!\!\!\!\!\!\! &0,\!\!\!\!\!\!\!\! &0,\!\!\!\!\!\!\!\! &0,\!\!\!\!\!\!\!\! &0,\!\!\!\!\!\!\!\! &0,\!\!\!\!\!\!\!\! &0,\!\!\!\!\!\!\!\! &0,\!\!\!\!\!\!\!\! &0,\!\!\!\!\!\!\!\! &0,\!\!\!\!\!\!\!\! &0,\!\!\!\!\!\!\!\! &0,\!\!\!\!\!\!\!\! &0,\!\!\!\!\!\!\!\! &0,\!\!\!\!\!\!\!\! &0,\!\!\!\!\!\!\!\! &0,\!\!\!\!\!\!\!\! &0,\!\!\!\!\!\!\!\! &0 \\
0,\!\!\!\!\!\!\!\! &0,\!\!\!\!\!\!\!\! &0,\!\!\!\!\!\!\!\! &0,\!\!\!\!\!\!\!\! &0,\!\!\!\!\!\!\!\! &\frac 12,\!\!\!\!\!\!\!\! &0,\!\!\!\!\!\!\!\! &0,\!\!\!\!\!\!\!\! &0,\!\!\!\!\!\!\!\! &0,\!\!\!\!\!\!\!\! &0,\!\!\!\!\!\!\!\! &0,\!\!\!\!\!\!\!\! &0,\!\!\!\!\!\!\!\! &0,\!\!\!\!\!\!\!\! &0,\!\!\!\!\!\!\!\! &0,\!\!\!\!\!\!\!\! &0,\!\!\!\!\!\!\!\! &0,\!\!\!\!\!\!\!\! &0,\!\!\!\!\!\!\!\! &0,\!\!\!\!\!\!\!\! &0,\!\!\!\!\!\!\!\! &0,\!\!\!\!\!\!\!\! &0,\!\!\!\!\!\!\!\! &0,\!\!\!\!\!\!\!\! &0,\!\!\!\!\!\!\!\! &0,\!\!\!\!\!\!\!\! &0 \\
\end{array}
\rp
\feqn
\eqn
c_{9}=\lp
\begin{array}{ccccccccccccccccccccccccccc}
0,\!\!\!\!\!\!\!\! &0,\!\!\!\!\!\!\!\! &0,\!\!\!\!\!\!\!\! &0,\!\!\!\!\!\!\!\! &0,\!\!\!\!\!\!\!\! &0,\!\!\!\!\!\!\!\! &0,\!\!\!\!\!\!\!\! &0,\!\!\!\!\!\!\!\! &0,\!\!\!\!\!\!\!\! &0,\!\!\!\!\!\!\!\! &0,\!\!\!\!\!\!\!\! &0,\!\!\!\!\!\!\!\! &0,\!\!\!\!\!\!\!\! &0,\!\!\!\!\!\!\!\! &0,\!\!\!\!\!\!\!\! &0,\!\!\!\!\!\!\!\! &0,\!\!\!\!\!\!\!\! &0,\!\!\!\!\!\!\!\! &0,\!\!\!\!\!\!\!\! &0,\!\!\!\!\!\!\!\! &0,\!\!\!\!\!\!\!\! &0,\!\!\!\!\!\!\!\! &0,\!\!\!\!\!\!\!\! &0,\!\!\!\!\!\!\!\! &0,\!\!\!\!\!\!\!\! &0,\!\!\!\!\!\!\!\! &0 \\
0,\!\!\!\!\!\!\!\! &0,\!\!\!\!\!\!\!\! &-\dem,\!\!\!\!\!\!\!\! &0,\!\!\!\!\!\!\!\! &0,\!\!\!\!\!\!\!\! &0,\!\!\!\!\!\!\!\! &0,\!\!\!\!\!\!\!\! &0,\!\!\!\!\!\!\!\! &0,\!\!\!\!\!\!\!\! &0,\!\!\!\!\!\!\!\! &0,\!\!\!\!\!\!\!\! &0,\!\!\!\!\!\!\!\! &0,\!\!\!\!\!\!\!\! &0,\!\!\!\!\!\!\!\! &0,\!\!\!\!\!\!\!\! &0,\!\!\!\!\!\!\!\! &0,\!\!\!\!\!\!\!\! &0,\!\!\!\!\!\!\!\! &0,\!\!\!\!\!\!\!\! &0,\!\!\!\!\!\!\!\! &0,\!\!\!\!\!\!\!\! &0,\!\!\!\!\!\!\!\! &0,\!\!\!\!\!\!\!\! &0,\!\!\!\!\!\!\!\! &0,\!\!\!\!\!\!\!\! &0,\!\!\!\!\!\!\!\! &0 \\
0,\!\!\!\!\!\!\!\! &\dem,\!\!\!\!\!\!\!\! &0,\!\!\!\!\!\!\!\! &0,\!\!\!\!\!\!\!\! &0,\!\!\!\!\!\!\!\! &0,\!\!\!\!\!\!\!\! &0,\!\!\!\!\!\!\!\! &0,\!\!\!\!\!\!\!\! &0,\!\!\!\!\!\!\!\! &0,\!\!\!\!\!\!\!\! &0,\!\!\!\!\!\!\!\! &0,\!\!\!\!\!\!\!\! &0,\!\!\!\!\!\!\!\! &0,\!\!\!\!\!\!\!\! &0,\!\!\!\!\!\!\!\! &0,\!\!\!\!\!\!\!\! &0,\!\!\!\!\!\!\!\! &0,\!\!\!\!\!\!\!\! &0,\!\!\!\!\!\!\!\! &0,\!\!\!\!\!\!\!\! &0,\!\!\!\!\!\!\!\! &0,\!\!\!\!\!\!\!\! &0,\!\!\!\!\!\!\!\! &0,\!\!\!\!\!\!\!\! &0,\!\!\!\!\!\!\!\! &0,\!\!\!\!\!\!\!\! &0 \\
\end{array}
\rp
\feqn
\eqn
c_{10}=\lp
\begin{array}{ccccccccccccccccccccccccccc}
0,\!\!\!\!\!\!\!\! &0,\!\!\!\!\!\!\!\! &0,\!\!\!\!\!\!\!\! &0,\!\!\!\!\!\!\!\! &0,\!\!\!\!\!\!\!\! &0,\!\!\!\!\!\!\!\! &0,\!\!\!\!\!\!\!\! &0,\!\!\!\!\!\!\!\! &0,\!\!\!\!\!\!\!\! &0,\!\!\!\!\!\!\!\! &0,\!\!\!\!\!\!\!\! &0,\!\!\!\!\!\!\!\! &0,\!\!\!\!\!\!\!\! &0,\!\!\!\!\!\!\!\! &0,\!\!\!\!\!\!\!\! &0,\!\!\!\!\!\!\!\! &0,\!\!\!\!\!\!\!\! &0,\!\!\!\!\!\!\!\! &0,\!\!\!\!\!\!\!\! &0,\!\!\!\!\!\!\!\! &0,\!\!\!\!\!\!\!\! &0,\!\!\!\!\!\!\!\! &0,\!\!\!\!\!\!\!\! &0,\!\!\!\!\!\!\!\! &0,\!\!\!\!\!\!\!\! &0,\!\!\!\!\!\!\!\! &0 \\
0,\!\!\!\!\!\!\!\! &0,\!\!\!\!\!\!\!\! &0,\!\!\!\!\!\!\!\! &0,\!\!\!\!\!\!\!\! &0,\!\!\!\!\!\!\!\! &0,\!\!\!\!\!\!\!\! &0,\!\!\!\!\!\!\!\! &-\dem,\!\!\!\!\!\!\!\! &0,\!\!\!\!\!\!\!\! &0,\!\!\!\!\!\!\!\! &0,\!\!\!\!\!\!\!\! &0,\!\!\!\!\!\!\!\! &0,\!\!\!\!\!\!\!\! &0,\!\!\!\!\!\!\!\! &0,\!\!\!\!\!\!\!\! &0,\!\!\!\!\!\!\!\! &0,\!\!\!\!\!\!\!\! &0,\!\!\!\!\!\!\!\! &0,\!\!\!\!\!\!\!\! &0,\!\!\!\!\!\!\!\! &0,\!\!\!\!\!\!\!\! &0,\!\!\!\!\!\!\!\! &0,\!\!\!\!\!\!\!\! &0,\!\!\!\!\!\!\!\! &0,\!\!\!\!\!\!\!\! &0,\!\!\!\!\!\!\!\! &0 \\
0,\!\!\!\!\!\!\!\! &0,\!\!\!\!\!\!\!\! &0,\!\!\!\!\!\!\!\! &0,\!\!\!\!\!\!\!\! &0,\!\!\!\!\!\!\!\! &0,\!\!\!\!\!\!\!\! &-\dem,\!\!\!\!\!\!\!\! &0,\!\!\!\!\!\!\!\! &0,\!\!\!\!\!\!\!\! &0,\!\!\!\!\!\!\!\! &0,\!\!\!\!\!\!\!\! &0,\!\!\!\!\!\!\!\! &0,\!\!\!\!\!\!\!\! &0,\!\!\!\!\!\!\!\! &0,\!\!\!\!\!\!\!\! &0,\!\!\!\!\!\!\!\! &0,\!\!\!\!\!\!\!\! &0,\!\!\!\!\!\!\!\! &0,\!\!\!\!\!\!\!\! &0,\!\!\!\!\!\!\!\! &0,\!\!\!\!\!\!\!\! &0,\!\!\!\!\!\!\!\! &0,\!\!\!\!\!\!\!\! &0,\!\!\!\!\!\!\!\! &0,\!\!\!\!\!\!\!\! &0,\!\!\!\!\!\!\!\! &0 \\
\end{array}
\rp
\feqn
\eqn
c_{11}=\lp
\begin{array}{ccccccccccccccccccccccccccc}
0,\!\!\!\!\!\!\!\! &0,\!\!\!\!\!\!\!\! &0,\!\!\!\!\!\!\!\! &0,\!\!\!\!\!\!\!\! &0,\!\!\!\!\!\!\!\! &0,\!\!\!\!\!\!\!\! &0,\!\!\!\!\!\!\!\! &0,\!\!\!\!\!\!\!\! &0,\!\!\!\!\!\!\!\! &0,\!\!\!\!\!\!\!\! &0,\!\!\!\!\!\!\!\! &0,\!\!\!\!\!\!\!\! &0,\!\!\!\!\!\!\!\! &0,\!\!\!\!\!\!\!\! &0,\!\!\!\!\!\!\!\! &0,\!\!\!\!\!\!\!\! &0,\!\!\!\!\!\!\!\! &0,\!\!\!\!\!\!\!\! &0,\!\!\!\!\!\!\!\! &0,\!\!\!\!\!\!\!\! &0,\!\!\!\!\!\!\!\! &0,\!\!\!\!\!\!\!\! &0,\!\!\!\!\!\!\!\! &0,\!\!\!\!\!\!\!\! &0,\!\!\!\!\!\!\!\! &0,\!\!\!\!\!\!\!\! &0 \\
0,\!\!\!\!\!\!\!\! &0,\!\!\!\!\!\!\!\! &0,\!\!\!\!\!\!\!\! &0,\!\!\!\!\!\!\!\! &0,\!\!\!\!\!\!\!\! &0,\!\!\!\!\!\!\!\! &-\dem,\!\!\!\!\!\!\!\! &0,\!\!\!\!\!\!\!\! &0,\!\!\!\!\!\!\!\! &0,\!\!\!\!\!\!\!\! &0,\!\!\!\!\!\!\!\! &0,\!\!\!\!\!\!\!\! &0,\!\!\!\!\!\!\!\! &0,\!\!\!\!\!\!\!\! &0,\!\!\!\!\!\!\!\! &0,\!\!\!\!\!\!\!\! &0,\!\!\!\!\!\!\!\! &0,\!\!\!\!\!\!\!\! &0,\!\!\!\!\!\!\!\! &0,\!\!\!\!\!\!\!\! &0,\!\!\!\!\!\!\!\! &0,\!\!\!\!\!\!\!\! &0,\!\!\!\!\!\!\!\! &0,\!\!\!\!\!\!\!\! &0,\!\!\!\!\!\!\!\! &0,\!\!\!\!\!\!\!\! &0 \\
0,\!\!\!\!\!\!\!\! &0,\!\!\!\!\!\!\!\! &0,\!\!\!\!\!\!\!\! &0,\!\!\!\!\!\!\!\! &0,\!\!\!\!\!\!\!\! &0,\!\!\!\!\!\!\!\! &0,\!\!\!\!\!\!\!\! &-\dem,\!\!\!\!\!\!\!\! &0,\!\!\!\!\!\!\!\! &0,\!\!\!\!\!\!\!\! &0,\!\!\!\!\!\!\!\! &0,\!\!\!\!\!\!\!\! &0,\!\!\!\!\!\!\!\! &0,\!\!\!\!\!\!\!\! &0,\!\!\!\!\!\!\!\! &0,\!\!\!\!\!\!\!\! &0,\!\!\!\!\!\!\!\! &0,\!\!\!\!\!\!\!\! &0,\!\!\!\!\!\!\!\! &0,\!\!\!\!\!\!\!\! &0,\!\!\!\!\!\!\!\! &0,\!\!\!\!\!\!\!\! &0,\!\!\!\!\!\!\!\! &0,\!\!\!\!\!\!\!\! &0,\!\!\!\!\!\!\!\! &0,\!\!\!\!\!\!\!\! &0 \\
\end{array}
\rp
\feqn
\eqn
c_{12}=\lp
\begin{array}{ccccccccccccccccccccccccccc}
0,\!\!\!\!\!\!\!\! &0,\!\!\!\!\!\!\!\! &0,\!\!\!\!\!\!\!\! &0,\!\!\!\!\!\!\!\! &0,\!\!\!\!\!\!\!\! &0,\!\!\!\!\!\!\!\! &0,\!\!\!\!\!\!\!\! &0,\!\!\!\!\!\!\!\! &0,\!\!\!\!\!\!\!\! &0,\!\!\!\!\!\!\!\! &0,\!\!\!\!\!\!\!\! &0,\!\!\!\!\!\!\!\! &0,\!\!\!\!\!\!\!\! &0,\!\!\!\!\!\!\!\! &0,\!\!\!\!\!\!\!\! &0,\!\!\!\!\!\!\!\! &0,\!\!\!\!\!\!\!\! &0,\!\!\!\!\!\!\!\! &0,\!\!\!\!\!\!\!\! &0,\!\!\!\!\!\!\!\! &0,\!\!\!\!\!\!\!\! &0,\!\!\!\!\!\!\!\! &0,\!\!\!\!\!\!\!\! &0,\!\!\!\!\!\!\!\! &0,\!\!\!\!\!\!\!\! &0,\!\!\!\!\!\!\!\! &0 \\
0,\!\!\!\!\!\!\!\! &0,\!\!\!\!\!\!\!\! &0,\!\!\!\!\!\!\!\! &0,\!\!\!\!\!\!\!\! &0,\!\!\!\!\!\!\!\! &0,\!\!\!\!\!\!\!\! &0,\!\!\!\!\!\!\!\! &-\dem,\!\!\!\!\!\!\!\! &0,\!\!\!\!\!\!\!\! &0,\!\!\!\!\!\!\!\! &0,\!\!\!\!\!\!\!\! &0,\!\!\!\!\!\!\!\! &0,\!\!\!\!\!\!\!\! &0,\!\!\!\!\!\!\!\! &0,\!\!\!\!\!\!\!\! &0,\!\!\!\!\!\!\!\! &0,\!\!\!\!\!\!\!\! &0,\!\!\!\!\!\!\!\! &0,\!\!\!\!\!\!\!\! &0,\!\!\!\!\!\!\!\! &0,\!\!\!\!\!\!\!\! &0,\!\!\!\!\!\!\!\! &0,\!\!\!\!\!\!\!\! &0,\!\!\!\!\!\!\!\! &0,\!\!\!\!\!\!\!\! &0,\!\!\!\!\!\!\!\! &0 \\
0,\!\!\!\!\!\!\!\! &0,\!\!\!\!\!\!\!\! &0,\!\!\!\!\!\!\!\! &0,\!\!\!\!\!\!\!\! &0,\!\!\!\!\!\!\!\! &0,\!\!\!\!\!\!\!\! &\dem,\!\!\!\!\!\!\!\! &0,\!\!\!\!\!\!\!\! &0,\!\!\!\!\!\!\!\! &0,\!\!\!\!\!\!\!\! &0,\!\!\!\!\!\!\!\! &0,\!\!\!\!\!\!\!\! &0,\!\!\!\!\!\!\!\! &0,\!\!\!\!\!\!\!\! &0,\!\!\!\!\!\!\!\! &0,\!\!\!\!\!\!\!\! &0,\!\!\!\!\!\!\!\! &0,\!\!\!\!\!\!\!\! &0,\!\!\!\!\!\!\!\! &0,\!\!\!\!\!\!\!\! &0,\!\!\!\!\!\!\!\! &0,\!\!\!\!\!\!\!\! &0,\!\!\!\!\!\!\!\! &0,\!\!\!\!\!\!\!\! &0,\!\!\!\!\!\!\!\! &0,\!\!\!\!\!\!\!\! &0 \\
\end{array}
\rp
\feqn
\eqn
c_{13}=\lp
\begin{array}{ccccccccccccccccccccccccccc}
0,\!\!\!\!\!\!\!\! &0,\!\!\!\!\!\!\!\! &0,\!\!\!\!\!\!\!\! &0,\!\!\!\!\!\!\!\! &0,\!\!\!\!\!\!\!\! &0,\!\!\!\!\!\!\!\! &0,\!\!\!\!\!\!\!\! &0,\!\!\!\!\!\!\!\! &0,\!\!\!\!\!\!\!\! &0,\!\!\!\!\!\!\!\! &0,\!\!\!\!\!\!\!\! &0,\!\!\!\!\!\!\!\! &0,\!\!\!\!\!\!\!\! &0,\!\!\!\!\!\!\!\! &0,\!\!\!\!\!\!\!\! &0,\!\!\!\!\!\!\!\! &0,\!\!\!\!\!\!\!\! &0,\!\!\!\!\!\!\!\! &0,\!\!\!\!\!\!\!\! &0,\!\!\!\!\!\!\!\! &0,\!\!\!\!\!\!\!\! &0,\!\!\!\!\!\!\!\! &0,\!\!\!\!\!\!\!\! &0,\!\!\!\!\!\!\!\! &0,\!\!\!\!\!\!\!\! &0,\!\!\!\!\!\!\!\! &0 \\
0,\!\!\!\!\!\!\!\! &0,\!\!\!\!\!\!\!\! &0,\!\!\!\!\!\!\!\! &0,\!\!\!\!\!\!\!\! &\dem,\!\!\!\!\!\!\!\! &0,\!\!\!\!\!\!\!\! &0,\!\!\!\!\!\!\!\! &0,\!\!\!\!\!\!\!\! &0,\!\!\!\!\!\!\!\! &0,\!\!\!\!\!\!\!\! &0,\!\!\!\!\!\!\!\! &0,\!\!\!\!\!\!\!\! &0,\!\!\!\!\!\!\!\! &0,\!\!\!\!\!\!\!\! &0,\!\!\!\!\!\!\!\! &0,\!\!\!\!\!\!\!\! &0,\!\!\!\!\!\!\!\! &0,\!\!\!\!\!\!\!\! &0,\!\!\!\!\!\!\!\! &0,\!\!\!\!\!\!\!\! &0,\!\!\!\!\!\!\!\! &0,\!\!\!\!\!\!\!\! &0,\!\!\!\!\!\!\!\! &0,\!\!\!\!\!\!\!\! &0,\!\!\!\!\!\!\!\! &0,\!\!\!\!\!\!\!\! &0 \\
0,\!\!\!\!\!\!\!\! &0,\!\!\!\!\!\!\!\! &0,\!\!\!\!\!\!\!\! &0,\!\!\!\!\!\!\!\! &0,\!\!\!\!\!\!\!\! &0,\!\!\!\!\!\!\!\! &0,\!\!\!\!\!\!\!\! &0,\!\!\!\!\!\!\!\! &-\dem,\!\!\!\!\!\!\!\! &0,\!\!\!\!\!\!\!\! &0,\!\!\!\!\!\!\!\! &0,\!\!\!\!\!\!\!\! &0,\!\!\!\!\!\!\!\! &0,\!\!\!\!\!\!\!\! &0,\!\!\!\!\!\!\!\! &0,\!\!\!\!\!\!\!\! &0,\!\!\!\!\!\!\!\! &0,\!\!\!\!\!\!\!\! &0,\!\!\!\!\!\!\!\! &0,\!\!\!\!\!\!\!\! &0,\!\!\!\!\!\!\!\! &0,\!\!\!\!\!\!\!\! &0,\!\!\!\!\!\!\!\! &0,\!\!\!\!\!\!\!\! &0,\!\!\!\!\!\!\!\! &0,\!\!\!\!\!\!\!\! &0 \\
\end{array}
\rp
\feqn
\eqn
c_{14}=\lp
\begin{array}{ccccccccccccccccccccccccccc}
0,\!\!\!\!\!\!\!\! &0,\!\!\!\!\!\!\!\! &0,\!\!\!\!\!\!\!\! &0,\!\!\!\!\!\!\!\! &0,\!\!\!\!\!\!\!\! &0,\!\!\!\!\!\!\!\! &0,\!\!\!\!\!\!\!\! &0,\!\!\!\!\!\!\!\! &0,\!\!\!\!\!\!\!\! &0,\!\!\!\!\!\!\!\! &0,\!\!\!\!\!\!\!\! &0,\!\!\!\!\!\!\!\! &0,\!\!\!\!\!\!\!\! &0,\!\!\!\!\!\!\!\! &0,\!\!\!\!\!\!\!\! &0,\!\!\!\!\!\!\!\! &0,\!\!\!\!\!\!\!\! &0,\!\!\!\!\!\!\!\! &0,\!\!\!\!\!\!\!\! &0,\!\!\!\!\!\!\!\! &0,\!\!\!\!\!\!\!\! &0,\!\!\!\!\!\!\!\! &0,\!\!\!\!\!\!\!\! &0,\!\!\!\!\!\!\!\! &0,\!\!\!\!\!\!\!\! &0,\!\!\!\!\!\!\!\! &0 \\
0,\!\!\!\!\!\!\!\! &0,\!\!\!\!\!\!\!\! &0,\!\!\!\!\!\!\!\! &-\dem,\!\!\!\!\!\!\!\! &0,\!\!\!\!\!\!\!\! &0,\!\!\!\!\!\!\!\! &0,\!\!\!\!\!\!\!\! &0,\!\!\!\!\!\!\!\! &0,\!\!\!\!\!\!\!\! &0,\!\!\!\!\!\!\!\! &0,\!\!\!\!\!\!\!\! &0,\!\!\!\!\!\!\!\! &0,\!\!\!\!\!\!\!\! &0,\!\!\!\!\!\!\!\! &0,\!\!\!\!\!\!\!\! &0,\!\!\!\!\!\!\!\! &0,\!\!\!\!\!\!\!\! &0,\!\!\!\!\!\!\!\! &0,\!\!\!\!\!\!\!\! &0,\!\!\!\!\!\!\!\! &0,\!\!\!\!\!\!\!\! &0,\!\!\!\!\!\!\!\! &0,\!\!\!\!\!\!\!\! &0,\!\!\!\!\!\!\!\! &0,\!\!\!\!\!\!\!\! &0,\!\!\!\!\!\!\!\! &0 \\
0,\!\!\!\!\!\!\!\! &0,\!\!\!\!\!\!\!\! &0,\!\!\!\!\!\!\!\! &0,\!\!\!\!\!\!\!\! &0,\!\!\!\!\!\!\!\! &\dem,\!\!\!\!\!\!\!\! &0,\!\!\!\!\!\!\!\! &0,\!\!\!\!\!\!\!\! &0,\!\!\!\!\!\!\!\! &0,\!\!\!\!\!\!\!\! &0,\!\!\!\!\!\!\!\! &0,\!\!\!\!\!\!\!\! &0,\!\!\!\!\!\!\!\! &0,\!\!\!\!\!\!\!\! &0,\!\!\!\!\!\!\!\! &0,\!\!\!\!\!\!\!\! &0,\!\!\!\!\!\!\!\! &0,\!\!\!\!\!\!\!\! &0,\!\!\!\!\!\!\!\! &0,\!\!\!\!\!\!\!\! &0,\!\!\!\!\!\!\!\! &0,\!\!\!\!\!\!\!\! &0,\!\!\!\!\!\!\!\! &0,\!\!\!\!\!\!\!\! &0,\!\!\!\!\!\!\!\! &0,\!\!\!\!\!\!\!\! &0 \\
\end{array}
\rp
\feqn
\eqn
c_{15}=\lp
\begin{array}{ccccccccccccccccccccccccccc}
0,\!\!\!\!\!\!\!\! &0,\!\!\!\!\!\!\!\! &0,\!\!\!\!\!\!\!\! &0,\!\!\!\!\!\!\!\! &0,\!\!\!\!\!\!\!\! &0,\!\!\!\!\!\!\!\! &0,\!\!\!\!\!\!\!\! &0,\!\!\!\!\!\!\!\! &0,\!\!\!\!\!\!\!\! &0,\!\!\!\!\!\!\!\! &0,\!\!\!\!\!\!\!\! &0,\!\!\!\!\!\!\!\! &0,\!\!\!\!\!\!\!\! &0,\!\!\!\!\!\!\!\! &0,\!\!\!\!\!\!\!\! &0,\!\!\!\!\!\!\!\! &0,\!\!\!\!\!\!\!\! &0,\!\!\!\!\!\!\!\! &0,\!\!\!\!\!\!\!\! &0,\!\!\!\!\!\!\!\! &0,\!\!\!\!\!\!\!\! &0,\!\!\!\!\!\!\!\! &0,\!\!\!\!\!\!\!\! &0,\!\!\!\!\!\!\!\! &0,\!\!\!\!\!\!\!\! &0,\!\!\!\!\!\!\!\! &0 \\
0,\!\!\!\!\!\!\!\! &0,\!\!\!\!\!\!\!\! &0,\!\!\!\!\!\!\!\! &0,\!\!\!\!\!\!\!\! &0,\!\!\!\!\!\!\!\! &0,\!\!\!\!\!\!\!\! &0,\!\!\!\!\!\!\!\! &0,\!\!\!\!\!\!\!\! &-\dem,\!\!\!\!\!\!\!\! &0,\!\!\!\!\!\!\!\! &0,\!\!\!\!\!\!\!\! &0,\!\!\!\!\!\!\!\! &0,\!\!\!\!\!\!\!\! &0,\!\!\!\!\!\!\!\! &0,\!\!\!\!\!\!\!\! &0,\!\!\!\!\!\!\!\! &0,\!\!\!\!\!\!\!\! &0,\!\!\!\!\!\!\!\! &0,\!\!\!\!\!\!\!\! &0,\!\!\!\!\!\!\!\! &0,\!\!\!\!\!\!\!\! &0,\!\!\!\!\!\!\!\! &0,\!\!\!\!\!\!\!\! &0,\!\!\!\!\!\!\!\! &0,\!\!\!\!\!\!\!\! &0,\!\!\!\!\!\!\!\! &0 \\
0,\!\!\!\!\!\!\!\! &0,\!\!\!\!\!\!\!\! &0,\!\!\!\!\!\!\!\! &0,\!\!\!\!\!\!\!\! &-\dem,\!\!\!\!\!\!\!\! &0,\!\!\!\!\!\!\!\! &0,\!\!\!\!\!\!\!\! &0,\!\!\!\!\!\!\!\! &0,\!\!\!\!\!\!\!\! &0,\!\!\!\!\!\!\!\! &0,\!\!\!\!\!\!\!\! &0,\!\!\!\!\!\!\!\! &0,\!\!\!\!\!\!\!\! &0,\!\!\!\!\!\!\!\! &0,\!\!\!\!\!\!\!\! &0,\!\!\!\!\!\!\!\! &0,\!\!\!\!\!\!\!\! &0,\!\!\!\!\!\!\!\! &0,\!\!\!\!\!\!\!\! &0,\!\!\!\!\!\!\!\! &0,\!\!\!\!\!\!\!\! &0,\!\!\!\!\!\!\!\! &0,\!\!\!\!\!\!\!\! &0,\!\!\!\!\!\!\!\! &0,\!\!\!\!\!\!\!\! &0,\!\!\!\!\!\!\!\! &0 \\
\end{array}
\rp
\feqn
\eqn
c_{16}=\lp
\begin{array}{ccccccccccccccccccccccccccc}
0,\!\!\!\!\!\!\!\! &0,\!\!\!\!\!\!\!\! &0,\!\!\!\!\!\!\!\! &0,\!\!\!\!\!\!\!\! &0,\!\!\!\!\!\!\!\! &0,\!\!\!\!\!\!\!\! &0,\!\!\!\!\!\!\!\! &0,\!\!\!\!\!\!\!\! &0,\!\!\!\!\!\!\!\! &0,\!\!\!\!\!\!\!\! &0,\!\!\!\!\!\!\!\! &0,\!\!\!\!\!\!\!\! &0,\!\!\!\!\!\!\!\! &0,\!\!\!\!\!\!\!\! &0,\!\!\!\!\!\!\!\! &0,\!\!\!\!\!\!\!\! &0,\!\!\!\!\!\!\!\! &0,\!\!\!\!\!\!\!\! &0,\!\!\!\!\!\!\!\! &0,\!\!\!\!\!\!\!\! &0,\!\!\!\!\!\!\!\! &0,\!\!\!\!\!\!\!\! &0,\!\!\!\!\!\!\!\! &0,\!\!\!\!\!\!\!\! &0,\!\!\!\!\!\!\!\! &0,\!\!\!\!\!\!\!\! &0 \\
0,\!\!\!\!\!\!\!\! &0,\!\!\!\!\!\!\!\! &0,\!\!\!\!\!\!\!\! &0,\!\!\!\!\!\!\!\! &0,\!\!\!\!\!\!\!\! &0,\!\!\!\!\!\!\!\! &0,\!\!\!\!\!\!\!\! &-\dem,\!\!\!\!\!\!\!\! &0,\!\!\!\!\!\!\!\! &0,\!\!\!\!\!\!\!\! &0,\!\!\!\!\!\!\!\! &0,\!\!\!\!\!\!\!\! &0,\!\!\!\!\!\!\!\! &0,\!\!\!\!\!\!\!\! &0,\!\!\!\!\!\!\!\! &0,\!\!\!\!\!\!\!\! &0,\!\!\!\!\!\!\!\! &0,\!\!\!\!\!\!\!\! &0,\!\!\!\!\!\!\!\! &0,\!\!\!\!\!\!\!\! &0,\!\!\!\!\!\!\!\! &0,\!\!\!\!\!\!\!\! &0,\!\!\!\!\!\!\!\! &0,\!\!\!\!\!\!\!\! &0,\!\!\!\!\!\!\!\! &0,\!\!\!\!\!\!\!\! &0 \\
0,\!\!\!\!\!\!\!\! &0,\!\!\!\!\!\!\!\! &0,\!\!\!\!\!\!\!\! &0,\!\!\!\!\!\!\!\! &0,\!\!\!\!\!\!\!\! &0,\!\!\!\!\!\!\!\! &\dem,\!\!\!\!\!\!\!\! &0,\!\!\!\!\!\!\!\! &0,\!\!\!\!\!\!\!\! &0,\!\!\!\!\!\!\!\! &0,\!\!\!\!\!\!\!\! &0,\!\!\!\!\!\!\!\! &0,\!\!\!\!\!\!\!\! &0,\!\!\!\!\!\!\!\! &0,\!\!\!\!\!\!\!\! &0,\!\!\!\!\!\!\!\! &0,\!\!\!\!\!\!\!\! &0,\!\!\!\!\!\!\!\! &0,\!\!\!\!\!\!\!\! &0,\!\!\!\!\!\!\!\! &0,\!\!\!\!\!\!\!\! &0,\!\!\!\!\!\!\!\! &0,\!\!\!\!\!\!\!\! &0,\!\!\!\!\!\!\!\! &0,\!\!\!\!\!\!\!\! &0,\!\!\!\!\!\!\!\! &0 \\
\end{array}
\rp
\feqn
\eqn
c_{17}=\lp
\begin{array}{ccccccccccccccccccccccccccc}
0,\!\!\!\!\!\!\!\! &0,\!\!\!\!\!\!\!\! &0,\!\!\!\!\!\!\!\! &0,\!\!\!\!\!\!\!\! &0,\!\!\!\!\!\!\!\! &0,\!\!\!\!\!\!\!\! &0,\!\!\!\!\!\!\!\! &0,\!\!\!\!\!\!\!\! &0,\!\!\!\!\!\!\!\! &0,\!\!\!\!\!\!\!\! &0,\!\!\!\!\!\!\!\! &0,\!\!\!\!\!\!\!\! &0,\!\!\!\!\!\!\!\! &0,\!\!\!\!\!\!\!\! &0,\!\!\!\!\!\!\!\! &0,\!\!\!\!\!\!\!\! &0,\!\!\!\!\!\!\!\! &0,\!\!\!\!\!\!\!\! &0,\!\!\!\!\!\!\!\! &0,\!\!\!\!\!\!\!\! &0,\!\!\!\!\!\!\!\! &0,\!\!\!\!\!\!\!\! &0,\!\!\!\!\!\!\!\! &0,\!\!\!\!\!\!\!\! &0,\!\!\!\!\!\!\!\! &0,\!\!\!\!\!\!\!\! &0 \\
0,\!\!\!\!\!\!\!\! &0,\!\!\!\!\!\!\!\! &0,\!\!\!\!\!\!\!\! &0,\!\!\!\!\!\!\!\! &0,\!\!\!\!\!\!\!\! &0,\!\!\!\!\!\!\!\! &\dem,\!\!\!\!\!\!\!\! &0,\!\!\!\!\!\!\!\! &0,\!\!\!\!\!\!\!\! &0,\!\!\!\!\!\!\!\! &0,\!\!\!\!\!\!\!\! &0,\!\!\!\!\!\!\!\! &0,\!\!\!\!\!\!\!\! &0,\!\!\!\!\!\!\!\! &0,\!\!\!\!\!\!\!\! &0,\!\!\!\!\!\!\!\! &0,\!\!\!\!\!\!\!\! &0,\!\!\!\!\!\!\!\! &0,\!\!\!\!\!\!\!\! &0,\!\!\!\!\!\!\!\! &0,\!\!\!\!\!\!\!\! &0,\!\!\!\!\!\!\!\! &0,\!\!\!\!\!\!\!\! &0,\!\!\!\!\!\!\!\! &0,\!\!\!\!\!\!\!\! &0,\!\!\!\!\!\!\!\! &0 \\
0,\!\!\!\!\!\!\!\! &0,\!\!\!\!\!\!\!\! &0,\!\!\!\!\!\!\!\! &0,\!\!\!\!\!\!\!\! &0,\!\!\!\!\!\!\!\! &0,\!\!\!\!\!\!\!\! &0,\!\!\!\!\!\!\!\! &\dem,\!\!\!\!\!\!\!\! &0,\!\!\!\!\!\!\!\! &0,\!\!\!\!\!\!\!\! &0,\!\!\!\!\!\!\!\! &0,\!\!\!\!\!\!\!\! &0,\!\!\!\!\!\!\!\! &0,\!\!\!\!\!\!\!\! &0,\!\!\!\!\!\!\!\! &0,\!\!\!\!\!\!\!\! &0,\!\!\!\!\!\!\!\! &0,\!\!\!\!\!\!\!\! &0,\!\!\!\!\!\!\!\! &0,\!\!\!\!\!\!\!\! &0,\!\!\!\!\!\!\!\! &0,\!\!\!\!\!\!\!\! &0,\!\!\!\!\!\!\!\! &0,\!\!\!\!\!\!\!\! &0,\!\!\!\!\!\!\!\! &0,\!\!\!\!\!\!\!\! &0 \\
\end{array}
\rp
\feqn
\eqn
c_{18}=\lp
\begin{array}{ccccccccccccccccccccccccccc}
0,\!\!\!\!\!\!\!\! &0,\!\!\!\!\!\!\!\! &0,\!\!\!\!\!\!\!\! &0,\!\!\!\!\!\!\!\! &0,\!\!\!\!\!\!\!\! &0,\!\!\!\!\!\!\!\! &0,\!\!\!\!\!\!\!\! &0,\!\!\!\!\!\!\!\! &0,\!\!\!\!\!\!\!\! &0,\!\!\!\!\!\!\!\! &0,\!\!\!\!\!\!\!\! &0,\!\!\!\!\!\!\!\! &0,\!\!\!\!\!\!\!\! &0,\!\!\!\!\!\!\!\! &0,\!\!\!\!\!\!\!\! &0,\!\!\!\!\!\!\!\! &0,\!\!\!\!\!\!\!\! &0,\!\!\!\!\!\!\!\! &0,\!\!\!\!\!\!\!\! &0,\!\!\!\!\!\!\!\! &0,\!\!\!\!\!\!\!\! &0,\!\!\!\!\!\!\!\! &0,\!\!\!\!\!\!\!\! &0,\!\!\!\!\!\!\!\! &0,\!\!\!\!\!\!\!\! &0,\!\!\!\!\!\!\!\! &0 \\
0,\!\!\!\!\!\!\!\! &0,\!\!\!\!\!\!\!\! &0,\!\!\!\!\!\!\!\! &0,\!\!\!\!\!\!\!\! &0,\!\!\!\!\!\!\!\! &0,\!\!\!\!\!\!\!\! &0,\!\!\!\!\!\!\!\! &0,\!\!\!\!\!\!\!\! &-\dem,\!\!\!\!\!\!\!\! &0,\!\!\!\!\!\!\!\! &0,\!\!\!\!\!\!\!\! &0,\!\!\!\!\!\!\!\! &0,\!\!\!\!\!\!\!\! &0,\!\!\!\!\!\!\!\! &0,\!\!\!\!\!\!\!\! &0,\!\!\!\!\!\!\!\! &0,\!\!\!\!\!\!\!\! &0,\!\!\!\!\!\!\!\! &0,\!\!\!\!\!\!\!\! &0,\!\!\!\!\!\!\!\! &0,\!\!\!\!\!\!\!\! &0,\!\!\!\!\!\!\!\! &0,\!\!\!\!\!\!\!\! &0,\!\!\!\!\!\!\!\! &0,\!\!\!\!\!\!\!\! &0,\!\!\!\!\!\!\!\! &0 \\
0,\!\!\!\!\!\!\!\! &0,\!\!\!\!\!\!\!\! &0,\!\!\!\!\!\!\!\! &0,\!\!\!\!\!\!\!\! &-\dem,\!\!\!\!\!\!\!\! &0,\!\!\!\!\!\!\!\! &0,\!\!\!\!\!\!\!\! &0,\!\!\!\!\!\!\!\! &0,\!\!\!\!\!\!\!\! &0,\!\!\!\!\!\!\!\! &0,\!\!\!\!\!\!\!\! &0,\!\!\!\!\!\!\!\! &0,\!\!\!\!\!\!\!\! &0,\!\!\!\!\!\!\!\! &0,\!\!\!\!\!\!\!\! &0,\!\!\!\!\!\!\!\! &0,\!\!\!\!\!\!\!\! &0,\!\!\!\!\!\!\!\! &0,\!\!\!\!\!\!\!\! &0,\!\!\!\!\!\!\!\! &0,\!\!\!\!\!\!\!\! &0,\!\!\!\!\!\!\!\! &0,\!\!\!\!\!\!\!\! &0,\!\!\!\!\!\!\!\! &0,\!\!\!\!\!\!\!\! &0,\!\!\!\!\!\!\!\! &0 \\
\end{array}
\rp
\feqn
\eqn
c_{19}=\lp
\begin{array}{ccccccccccccccccccccccccccc}
0,\!\!\!\!\!\!\!\! &0,\!\!\!\!\!\!\!\! &0,\!\!\!\!\!\!\!\! &0,\!\!\!\!\!\!\!\! &0,\!\!\!\!\!\!\!\! &0,\!\!\!\!\!\!\!\! &0,\!\!\!\!\!\!\!\! &0,\!\!\!\!\!\!\!\! &0,\!\!\!\!\!\!\!\! &0,\!\!\!\!\!\!\!\! &0,\!\!\!\!\!\!\!\! &0,\!\!\!\!\!\!\!\! &0,\!\!\!\!\!\!\!\! &0,\!\!\!\!\!\!\!\! &0,\!\!\!\!\!\!\!\! &0,\!\!\!\!\!\!\!\! &0,\!\!\!\!\!\!\!\! &0,\!\!\!\!\!\!\!\! &0,\!\!\!\!\!\!\!\! &0,\!\!\!\!\!\!\!\! &0,\!\!\!\!\!\!\!\! &0,\!\!\!\!\!\!\!\! &0,\!\!\!\!\!\!\!\! &0,\!\!\!\!\!\!\!\! &0,\!\!\!\!\!\!\!\! &0,\!\!\!\!\!\!\!\! &0 \\
0,\!\!\!\!\!\!\!\! &0,\!\!\!\!\!\!\!\! &0,\!\!\!\!\!\!\!\! &0,\!\!\!\!\!\!\!\! &0,\!\!\!\!\!\!\!\! &\dem,\!\!\!\!\!\!\!\! &0,\!\!\!\!\!\!\!\! &0,\!\!\!\!\!\!\!\! &0,\!\!\!\!\!\!\!\! &0,\!\!\!\!\!\!\!\! &0,\!\!\!\!\!\!\!\! &0,\!\!\!\!\!\!\!\! &0,\!\!\!\!\!\!\!\! &0,\!\!\!\!\!\!\!\! &0,\!\!\!\!\!\!\!\! &0,\!\!\!\!\!\!\!\! &0,\!\!\!\!\!\!\!\! &0,\!\!\!\!\!\!\!\! &0,\!\!\!\!\!\!\!\! &0,\!\!\!\!\!\!\!\! &0,\!\!\!\!\!\!\!\! &0,\!\!\!\!\!\!\!\! &0,\!\!\!\!\!\!\!\! &0,\!\!\!\!\!\!\!\! &0,\!\!\!\!\!\!\!\! &0,\!\!\!\!\!\!\!\! &0 \\
0,\!\!\!\!\!\!\!\! &0,\!\!\!\!\!\!\!\! &0,\!\!\!\!\!\!\!\! &\dem,\!\!\!\!\!\!\!\! &0,\!\!\!\!\!\!\!\! &0,\!\!\!\!\!\!\!\! &0,\!\!\!\!\!\!\!\! &0,\!\!\!\!\!\!\!\! &0,\!\!\!\!\!\!\!\! &0,\!\!\!\!\!\!\!\! &0,\!\!\!\!\!\!\!\! &0,\!\!\!\!\!\!\!\! &0,\!\!\!\!\!\!\!\! &0,\!\!\!\!\!\!\!\! &0,\!\!\!\!\!\!\!\! &0,\!\!\!\!\!\!\!\! &0,\!\!\!\!\!\!\!\! &0,\!\!\!\!\!\!\!\! &0,\!\!\!\!\!\!\!\! &0,\!\!\!\!\!\!\!\! &0,\!\!\!\!\!\!\!\! &0,\!\!\!\!\!\!\!\! &0,\!\!\!\!\!\!\!\! &0,\!\!\!\!\!\!\!\! &0,\!\!\!\!\!\!\!\! &0,\!\!\!\!\!\!\!\! &0 \\
\end{array}
\rp
\feqn
\eqn
c_{20}=\lp
\begin{array}{ccccccccccccccccccccccccccc}
0,\!\!\!\!\!\!\!\! &0,\!\!\!\!\!\!\!\! &0,\!\!\!\!\!\!\!\! &0,\!\!\!\!\!\!\!\! &0,\!\!\!\!\!\!\!\! &0,\!\!\!\!\!\!\!\! &0,\!\!\!\!\!\!\!\! &0,\!\!\!\!\!\!\!\! &0,\!\!\!\!\!\!\!\! &0,\!\!\!\!\!\!\!\! &0,\!\!\!\!\!\!\!\! &0,\!\!\!\!\!\!\!\! &0,\!\!\!\!\!\!\!\! &0,\!\!\!\!\!\!\!\! &0,\!\!\!\!\!\!\!\! &0,\!\!\!\!\!\!\!\! &0,\!\!\!\!\!\!\!\! &0,\!\!\!\!\!\!\!\! &0,\!\!\!\!\!\!\!\! &0,\!\!\!\!\!\!\!\! &0,\!\!\!\!\!\!\!\! &0,\!\!\!\!\!\!\!\! &0,\!\!\!\!\!\!\!\! &0,\!\!\!\!\!\!\!\! &0,\!\!\!\!\!\!\!\! &0,\!\!\!\!\!\!\!\! &0 \\
0,\!\!\!\!\!\!\!\! &0,\!\!\!\!\!\!\!\! &0,\!\!\!\!\!\!\!\! &0,\!\!\!\!\!\!\!\! &-\dem,\!\!\!\!\!\!\!\! &0,\!\!\!\!\!\!\!\! &0,\!\!\!\!\!\!\!\! &0,\!\!\!\!\!\!\!\! &0,\!\!\!\!\!\!\!\! &0,\!\!\!\!\!\!\!\! &0,\!\!\!\!\!\!\!\! &0,\!\!\!\!\!\!\!\! &0,\!\!\!\!\!\!\!\! &0,\!\!\!\!\!\!\!\! &0,\!\!\!\!\!\!\!\! &0,\!\!\!\!\!\!\!\! &0,\!\!\!\!\!\!\!\! &0,\!\!\!\!\!\!\!\! &0,\!\!\!\!\!\!\!\! &0,\!\!\!\!\!\!\!\! &0,\!\!\!\!\!\!\!\! &0,\!\!\!\!\!\!\!\! &0,\!\!\!\!\!\!\!\! &0,\!\!\!\!\!\!\!\! &0,\!\!\!\!\!\!\!\! &0,\!\!\!\!\!\!\!\! &0 \\
0,\!\!\!\!\!\!\!\! &0,\!\!\!\!\!\!\!\! &0,\!\!\!\!\!\!\!\! &0,\!\!\!\!\!\!\!\! &0,\!\!\!\!\!\!\!\! &0,\!\!\!\!\!\!\!\! &0,\!\!\!\!\!\!\!\! &0,\!\!\!\!\!\!\!\! &\dem,\!\!\!\!\!\!\!\! &0,\!\!\!\!\!\!\!\! &0,\!\!\!\!\!\!\!\! &0,\!\!\!\!\!\!\!\! &0,\!\!\!\!\!\!\!\! &0,\!\!\!\!\!\!\!\! &0,\!\!\!\!\!\!\!\! &0,\!\!\!\!\!\!\!\! &0,\!\!\!\!\!\!\!\! &0,\!\!\!\!\!\!\!\! &0,\!\!\!\!\!\!\!\! &0,\!\!\!\!\!\!\!\! &0,\!\!\!\!\!\!\!\! &0,\!\!\!\!\!\!\!\! &0,\!\!\!\!\!\!\!\! &0,\!\!\!\!\!\!\!\! &0,\!\!\!\!\!\!\!\! &0,\!\!\!\!\!\!\!\! &0 \\
\end{array}
\rp
\feqn
\eqn
c_{21}=\lp
\begin{array}{ccccccccccccccccccccccccccc}
0,\!\!\!\!\!\!\!\! &0,\!\!\!\!\!\!\!\! &0,\!\!\!\!\!\!\!\! &0,\!\!\!\!\!\!\!\! &0,\!\!\!\!\!\!\!\! &0,\!\!\!\!\!\!\!\! &0,\!\!\!\!\!\!\!\! &0,\!\!\!\!\!\!\!\! &0,\!\!\!\!\!\!\!\! &0,\!\!\!\!\!\!\!\! &0,\!\!\!\!\!\!\!\! &0,\!\!\!\!\!\!\!\! &0,\!\!\!\!\!\!\!\! &0,\!\!\!\!\!\!\!\! &0,\!\!\!\!\!\!\!\! &0,\!\!\!\!\!\!\!\! &0,\!\!\!\!\!\!\!\! &0,\!\!\!\!\!\!\!\! &0,\!\!\!\!\!\!\!\! &0,\!\!\!\!\!\!\!\! &0,\!\!\!\!\!\!\!\! &0,\!\!\!\!\!\!\!\! &0,\!\!\!\!\!\!\!\! &0,\!\!\!\!\!\!\!\! &0,\!\!\!\!\!\!\!\! &0,\!\!\!\!\!\!\!\! &0 \\
0,\!\!\!\!\!\!\!\! &0,\!\!\!\!\!\!\!\! &-\dem,\!\!\!\!\!\!\!\! &0,\!\!\!\!\!\!\!\! &0,\!\!\!\!\!\!\!\! &0,\!\!\!\!\!\!\!\! &0,\!\!\!\!\!\!\!\! &0,\!\!\!\!\!\!\!\! &0,\!\!\!\!\!\!\!\! &0,\!\!\!\!\!\!\!\! &0,\!\!\!\!\!\!\!\! &0,\!\!\!\!\!\!\!\! &0,\!\!\!\!\!\!\!\! &0,\!\!\!\!\!\!\!\! &0,\!\!\!\!\!\!\!\! &0,\!\!\!\!\!\!\!\! &0,\!\!\!\!\!\!\!\! &0,\!\!\!\!\!\!\!\! &0,\!\!\!\!\!\!\!\! &0,\!\!\!\!\!\!\!\! &0,\!\!\!\!\!\!\!\! &0,\!\!\!\!\!\!\!\! &0,\!\!\!\!\!\!\!\! &0,\!\!\!\!\!\!\!\! &0,\!\!\!\!\!\!\!\! &0,\!\!\!\!\!\!\!\! &0 \\
0,\!\!\!\!\!\!\!\! &\dem,\!\!\!\!\!\!\!\! &0,\!\!\!\!\!\!\!\! &0,\!\!\!\!\!\!\!\! &0,\!\!\!\!\!\!\!\! &0,\!\!\!\!\!\!\!\! &0,\!\!\!\!\!\!\!\! &0,\!\!\!\!\!\!\!\! &0,\!\!\!\!\!\!\!\! &0,\!\!\!\!\!\!\!\! &0,\!\!\!\!\!\!\!\! &0,\!\!\!\!\!\!\!\! &0,\!\!\!\!\!\!\!\! &0,\!\!\!\!\!\!\!\! &0,\!\!\!\!\!\!\!\! &0,\!\!\!\!\!\!\!\! &0,\!\!\!\!\!\!\!\! &0,\!\!\!\!\!\!\!\! &0,\!\!\!\!\!\!\!\! &0,\!\!\!\!\!\!\!\! &0,\!\!\!\!\!\!\!\! &0,\!\!\!\!\!\!\!\! &0,\!\!\!\!\!\!\!\! &0,\!\!\!\!\!\!\!\! &0,\!\!\!\!\!\!\!\! &0,\!\!\!\!\!\!\!\! &0 \\
\end{array}
\rp
\feqn
\eqn
c_{22}=\lp
\begin{array}{ccccccccccccccccccccccccccc}
0,\!\!\!\!\!\!\!\! &0,\!\!\!\!\!\!\!\! &0,\!\!\!\!\!\!\!\! &0,\!\!\!\!\!\!\!\! &0,\!\!\!\!\!\!\!\! &0,\!\!\!\!\!\!\!\! &0,\!\!\!\!\!\!\!\! &0,\!\!\!\!\!\!\!\! &-\sqrt 2,\!\!\!\!\!\!\!\! &0,\!\!\!\!\!\!\!\! &0,\!\!\!\!\!\!\!\! &0,\!\!\!\!\!\!\!\! &0,\!\!\!\!\!\!\!\! &0,\!\!\!\!\!\!\!\! &0,\!\!\!\!\!\!\!\! &0,\!\!\!\!\!\!\!\! &0,\!\!\!\!\!\!\!\! &0,\!\!\!\!\!\!\!\! &0,\!\!\!\!\!\!\!\! &0,\!\!\!\!\!\!\!\! &0,\!\!\!\!\!\!\!\! &0,\!\!\!\!\!\!\!\! &0,\!\!\!\!\!\!\!\! &0,\!\!\!\!\!\!\!\! &0,\!\!\!\!\!\!\!\! &0,\!\!\!\!\!\!\!\! &0 \\
0,\!\!\!\!\!\!\!\! &0,\!\!\!\!\!\!\!\! &0,\!\!\!\!\!\!\!\! &0,\!\!\!\!\!\!\!\! &0,\!\!\!\!\!\!\!\! &0,\!\!\!\!\!\!\!\! &0,\!\!\!\!\!\!\!\! &0,\!\!\!\!\!\!\!\! &0,\!\!\!\!\!\!\!\! &0,\!\!\!\!\!\!\!\! &0,\!\!\!\!\!\!\!\! &0,\!\!\!\!\!\!\!\! &0,\!\!\!\!\!\!\!\! &0,\!\!\!\!\!\!\!\! &0,\!\!\!\!\!\!\!\! &0,\!\!\!\!\!\!\!\! &0,\!\!\!\!\!\!\!\! &0,\!\!\!\!\!\!\!\! &0,\!\!\!\!\!\!\!\! &0,\!\!\!\!\!\!\!\! &0,\!\!\!\!\!\!\!\! &0,\!\!\!\!\!\!\!\! &0,\!\!\!\!\!\!\!\! &0,\!\!\!\!\!\!\!\! &0,\!\!\!\!\!\!\!\! &0,\!\!\!\!\!\!\!\! &0 \\
0,\!\!\!\!\!\!\!\! &0,\!\!\!\!\!\!\!\! &0,\!\!\!\!\!\!\!\! &0,\!\!\!\!\!\!\!\! &0,\!\!\!\!\!\!\!\! &0,\!\!\!\!\!\!\!\! &0,\!\!\!\!\!\!\!\! &0,\!\!\!\!\!\!\!\! &0,\!\!\!\!\!\!\!\! &0,\!\!\!\!\!\!\!\! &0,\!\!\!\!\!\!\!\! &0,\!\!\!\!\!\!\!\! &0,\!\!\!\!\!\!\!\! &0,\!\!\!\!\!\!\!\! &0,\!\!\!\!\!\!\!\! &0,\!\!\!\!\!\!\!\! &0,\!\!\!\!\!\!\!\! &0,\!\!\!\!\!\!\!\! &0,\!\!\!\!\!\!\!\! &0,\!\!\!\!\!\!\!\! &0,\!\!\!\!\!\!\!\! &0,\!\!\!\!\!\!\!\! &0,\!\!\!\!\!\!\!\! &0,\!\!\!\!\!\!\!\! &0,\!\!\!\!\!\!\!\! &0,\!\!\!\!\!\!\!\! &0 \\
\end{array}
\rp
\feqn
\eqn
c_{23}=\lp
\begin{array}{ccccccccccccccccccccccccccc}
0,\!\!\!\!\!\!\!\! &0,\!\!\!\!\!\!\!\! &0,\!\!\!\!\!\!\!\! &0,\!\!\!\!\!\!\!\! &\sqrt 2,\!\!\!\!\!\!\!\! &0,\!\!\!\!\!\!\!\! &0,\!\!\!\!\!\!\!\! &0,\!\!\!\!\!\!\!\! &0,\!\!\!\!\!\!\!\! &0,\!\!\!\!\!\!\!\! &0,\!\!\!\!\!\!\!\! &0,\!\!\!\!\!\!\!\! &0,\!\!\!\!\!\!\!\! &0,\!\!\!\!\!\!\!\! &0,\!\!\!\!\!\!\!\! &0,\!\!\!\!\!\!\!\! &0,\!\!\!\!\!\!\!\! &0,\!\!\!\!\!\!\!\! &0,\!\!\!\!\!\!\!\! &0,\!\!\!\!\!\!\!\! &0,\!\!\!\!\!\!\!\! &0,\!\!\!\!\!\!\!\! &0,\!\!\!\!\!\!\!\! &0,\!\!\!\!\!\!\!\! &0,\!\!\!\!\!\!\!\! &0,\!\!\!\!\!\!\!\! &0 \\
0,\!\!\!\!\!\!\!\! &0,\!\!\!\!\!\!\!\! &0,\!\!\!\!\!\!\!\! &0,\!\!\!\!\!\!\!\! &0,\!\!\!\!\!\!\!\! &0,\!\!\!\!\!\!\!\! &0,\!\!\!\!\!\!\!\! &0,\!\!\!\!\!\!\!\! &0,\!\!\!\!\!\!\!\! &0,\!\!\!\!\!\!\!\! &0,\!\!\!\!\!\!\!\! &0,\!\!\!\!\!\!\!\! &0,\!\!\!\!\!\!\!\! &0,\!\!\!\!\!\!\!\! &0,\!\!\!\!\!\!\!\! &0,\!\!\!\!\!\!\!\! &0,\!\!\!\!\!\!\!\! &0,\!\!\!\!\!\!\!\! &0,\!\!\!\!\!\!\!\! &0,\!\!\!\!\!\!\!\! &0,\!\!\!\!\!\!\!\! &0,\!\!\!\!\!\!\!\! &0,\!\!\!\!\!\!\!\! &0,\!\!\!\!\!\!\!\! &0,\!\!\!\!\!\!\!\! &0,\!\!\!\!\!\!\!\! &0 \\
0,\!\!\!\!\!\!\!\! &0,\!\!\!\!\!\!\!\! &0,\!\!\!\!\!\!\!\! &0,\!\!\!\!\!\!\!\! &0,\!\!\!\!\!\!\!\! &0,\!\!\!\!\!\!\!\! &0,\!\!\!\!\!\!\!\! &0,\!\!\!\!\!\!\!\! &0,\!\!\!\!\!\!\!\! &0,\!\!\!\!\!\!\!\! &0,\!\!\!\!\!\!\!\! &0,\!\!\!\!\!\!\!\! &0,\!\!\!\!\!\!\!\! &0,\!\!\!\!\!\!\!\! &0,\!\!\!\!\!\!\!\! &0,\!\!\!\!\!\!\!\! &0,\!\!\!\!\!\!\!\! &0,\!\!\!\!\!\!\!\! &0,\!\!\!\!\!\!\!\! &0,\!\!\!\!\!\!\!\! &0,\!\!\!\!\!\!\!\! &0,\!\!\!\!\!\!\!\! &0,\!\!\!\!\!\!\!\! &0,\!\!\!\!\!\!\!\! &0,\!\!\!\!\!\!\!\! &0,\!\!\!\!\!\!\!\! &0 \\
\end{array}
\rp
\feqn
\eqn
c_{24}=\lp
\begin{array}{ccccccccccccccccccccccccccc}
0,\!\!\!\!\!\!\!\! &0,\!\!\!\!\!\!\!\! &0,\!\!\!\!\!\!\!\! &0,\!\!\!\!\!\!\!\! &0,\!\!\!\!\!\!\!\! &0,\!\!\!\!\!\!\!\! &0,\!\!\!\!\!\!\!\! &\sqrt2,\!\!\!\!\!\!\!\! &0,\!\!\!\!\!\!\!\! &0,\!\!\!\!\!\!\!\! &0,\!\!\!\!\!\!\!\! &0,\!\!\!\!\!\!\!\! &0,\!\!\!\!\!\!\!\! &0,\!\!\!\!\!\!\!\! &0,\!\!\!\!\!\!\!\! &0,\!\!\!\!\!\!\!\! &0,\!\!\!\!\!\!\!\! &0,\!\!\!\!\!\!\!\! &0,\!\!\!\!\!\!\!\! &0,\!\!\!\!\!\!\!\! &0,\!\!\!\!\!\!\!\! &0,\!\!\!\!\!\!\!\! &0,\!\!\!\!\!\!\!\! &0,\!\!\!\!\!\!\!\! &0,\!\!\!\!\!\!\!\! &0,\!\!\!\!\!\!\!\! &0 \\
0,\!\!\!\!\!\!\!\! &0,\!\!\!\!\!\!\!\! &0,\!\!\!\!\!\!\!\! &0,\!\!\!\!\!\!\!\! &0,\!\!\!\!\!\!\!\! &0,\!\!\!\!\!\!\!\! &0,\!\!\!\!\!\!\!\! &0,\!\!\!\!\!\!\!\! &0,\!\!\!\!\!\!\!\! &0,\!\!\!\!\!\!\!\! &0,\!\!\!\!\!\!\!\! &0,\!\!\!\!\!\!\!\! &0,\!\!\!\!\!\!\!\! &0,\!\!\!\!\!\!\!\! &0,\!\!\!\!\!\!\!\! &0,\!\!\!\!\!\!\!\! &0,\!\!\!\!\!\!\!\! &0,\!\!\!\!\!\!\!\! &0,\!\!\!\!\!\!\!\! &0,\!\!\!\!\!\!\!\! &0,\!\!\!\!\!\!\!\! &0,\!\!\!\!\!\!\!\! &0,\!\!\!\!\!\!\!\! &0,\!\!\!\!\!\!\!\! &0,\!\!\!\!\!\!\!\! &0,\!\!\!\!\!\!\!\! &0 \\
0,\!\!\!\!\!\!\!\! &0,\!\!\!\!\!\!\!\! &0,\!\!\!\!\!\!\!\! &0,\!\!\!\!\!\!\!\! &0,\!\!\!\!\!\!\!\! &0,\!\!\!\!\!\!\!\! &0,\!\!\!\!\!\!\!\! &0,\!\!\!\!\!\!\!\! &0,\!\!\!\!\!\!\!\! &0,\!\!\!\!\!\!\!\! &0,\!\!\!\!\!\!\!\! &0,\!\!\!\!\!\!\!\! &0,\!\!\!\!\!\!\!\! &0,\!\!\!\!\!\!\!\! &0,\!\!\!\!\!\!\!\! &0,\!\!\!\!\!\!\!\! &0,\!\!\!\!\!\!\!\! &0,\!\!\!\!\!\!\!\! &0,\!\!\!\!\!\!\!\! &0,\!\!\!\!\!\!\!\! &0,\!\!\!\!\!\!\!\! &0,\!\!\!\!\!\!\!\! &0,\!\!\!\!\!\!\!\! &0,\!\!\!\!\!\!\!\! &0,\!\!\!\!\!\!\!\! &0,\!\!\!\!\!\!\!\! &0 \\
\end{array}
\rp
\feqn
\eqn
c_{25}=\lp
\begin{array}{ccccccccccccccccccccccccccc}
0,\!\!\!\!\!\!\!\! &0,\!\!\!\!\!\!\!\! &-\sqrt2,\!\!\!\!\!\!\!\! &0,\!\!\!\!\!\!\!\! &0,\!\!\!\!\!\!\!\! &0,\!\!\!\!\!\!\!\! &0,\!\!\!\!\!\!\!\! &0,\!\!\!\!\!\!\!\! &0,\!\!\!\!\!\!\!\! &0,\!\!\!\!\!\!\!\! &0,\!\!\!\!\!\!\!\! &0,\!\!\!\!\!\!\!\! &0,\!\!\!\!\!\!\!\! &0,\!\!\!\!\!\!\!\! &0,\!\!\!\!\!\!\!\! &0,\!\!\!\!\!\!\!\! &0,\!\!\!\!\!\!\!\! &0,\!\!\!\!\!\!\!\! &0,\!\!\!\!\!\!\!\! &0,\!\!\!\!\!\!\!\! &0,\!\!\!\!\!\!\!\! &0,\!\!\!\!\!\!\!\! &0,\!\!\!\!\!\!\!\! &0,\!\!\!\!\!\!\!\! &0,\!\!\!\!\!\!\!\! &0,\!\!\!\!\!\!\!\! &0 \\
0,\!\!\!\!\!\!\!\! &0,\!\!\!\!\!\!\!\! &0,\!\!\!\!\!\!\!\! &0,\!\!\!\!\!\!\!\! &0,\!\!\!\!\!\!\!\! &0,\!\!\!\!\!\!\!\! &0,\!\!\!\!\!\!\!\! &0,\!\!\!\!\!\!\!\! &0,\!\!\!\!\!\!\!\! &0,\!\!\!\!\!\!\!\! &0,\!\!\!\!\!\!\!\! &0,\!\!\!\!\!\!\!\! &0,\!\!\!\!\!\!\!\! &0,\!\!\!\!\!\!\!\! &0,\!\!\!\!\!\!\!\! &0,\!\!\!\!\!\!\!\! &0,\!\!\!\!\!\!\!\! &0,\!\!\!\!\!\!\!\! &0,\!\!\!\!\!\!\!\! &0,\!\!\!\!\!\!\!\! &0,\!\!\!\!\!\!\!\! &0,\!\!\!\!\!\!\!\! &0,\!\!\!\!\!\!\!\! &0,\!\!\!\!\!\!\!\! &0,\!\!\!\!\!\!\!\! &0,\!\!\!\!\!\!\!\! &0 \\
\frac 1{\sqrt 2},\!\!\!\!\!\!\!\! &0,\!\!\!\!\!\!\!\! &0,\!\!\!\!\!\!\!\! &0,\!\!\!\!\!\!\!\! &0,\!\!\!\!\!\!\!\! &0,\!\!\!\!\!\!\!\! &0,\!\!\!\!\!\!\!\! &0,\!\!\!\!\!\!\!\! &0,\!\!\!\!\!\!\!\! &0,\!\!\!\!\!\!\!\! &0,\!\!\!\!\!\!\!\! &0,\!\!\!\!\!\!\!\! &0,\!\!\!\!\!\!\!\! &0,\!\!\!\!\!\!\!\! &0,\!\!\!\!\!\!\!\! &0,\!\!\!\!\!\!\!\! &0,\!\!\!\!\!\!\!\! &0,\!\!\!\!\!\!\!\! &0,\!\!\!\!\!\!\!\! &0,\!\!\!\!\!\!\!\! &0,\!\!\!\!\!\!\!\! &0,\!\!\!\!\!\!\!\! &0,\!\!\!\!\!\!\!\! &0,\!\!\!\!\!\!\!\! &0,\!\!\!\!\!\!\!\! &0,\!\!\!\!\!\!\!\! &0 \\
\end{array}
\rp
\feqn
\eqn
c_{26}=\lp
\begin{array}{ccccccccccccccccccccccccccc}
0,\!\!\!\!\!\!\!\! &0,\!\!\!\!\!\!\!\! &0,\!\!\!\!\!\!\!\! &0,\!\!\!\!\!\!\!\! &0,\!\!\!\!\!\!\!\! &0,\!\!\!\!\!\!\!\! &\sqrt2,\!\!\!\!\!\!\!\! &0,\!\!\!\!\!\!\!\! &0,\!\!\!\!\!\!\!\! &0,\!\!\!\!\!\!\!\! &0,\!\!\!\!\!\!\!\! &0,\!\!\!\!\!\!\!\! &0,\!\!\!\!\!\!\!\! &0,\!\!\!\!\!\!\!\! &0,\!\!\!\!\!\!\!\! &0,\!\!\!\!\!\!\!\! &0,\!\!\!\!\!\!\!\! &0,\!\!\!\!\!\!\!\! &0,\!\!\!\!\!\!\!\! &0,\!\!\!\!\!\!\!\! &0,\!\!\!\!\!\!\!\! &0,\!\!\!\!\!\!\!\! &0,\!\!\!\!\!\!\!\! &0,\!\!\!\!\!\!\!\! &0,\!\!\!\!\!\!\!\! &0,\!\!\!\!\!\!\!\! &0 \\
0,\!\!\!\!\!\!\!\! &0,\!\!\!\!\!\!\!\! &0,\!\!\!\!\!\!\!\! &0,\!\!\!\!\!\!\!\! &0,\!\!\!\!\!\!\!\! &0,\!\!\!\!\!\!\!\! &0,\!\!\!\!\!\!\!\! &0,\!\!\!\!\!\!\!\! &0,\!\!\!\!\!\!\!\! &0,\!\!\!\!\!\!\!\! &0,\!\!\!\!\!\!\!\! &0,\!\!\!\!\!\!\!\! &0,\!\!\!\!\!\!\!\! &0,\!\!\!\!\!\!\!\! &0,\!\!\!\!\!\!\!\! &0,\!\!\!\!\!\!\!\! &0,\!\!\!\!\!\!\!\! &0,\!\!\!\!\!\!\!\! &0,\!\!\!\!\!\!\!\! &0,\!\!\!\!\!\!\!\! &0,\!\!\!\!\!\!\!\! &0,\!\!\!\!\!\!\!\! &0,\!\!\!\!\!\!\!\! &0,\!\!\!\!\!\!\!\! &0,\!\!\!\!\!\!\!\! &0,\!\!\!\!\!\!\!\! &0 \\
0,\!\!\!\!\!\!\!\! &0,\!\!\!\!\!\!\!\! &0,\!\!\!\!\!\!\!\! &0,\!\!\!\!\!\!\!\! &0,\!\!\!\!\!\!\!\! &0,\!\!\!\!\!\!\!\! &0,\!\!\!\!\!\!\!\! &0,\!\!\!\!\!\!\!\! &0,\!\!\!\!\!\!\!\! &0,\!\!\!\!\!\!\!\! &0,\!\!\!\!\!\!\!\! &0,\!\!\!\!\!\!\!\! &0,\!\!\!\!\!\!\!\! &0,\!\!\!\!\!\!\!\! &0,\!\!\!\!\!\!\!\! &0,\!\!\!\!\!\!\!\! &0,\!\!\!\!\!\!\!\! &0,\!\!\!\!\!\!\!\! &0,\!\!\!\!\!\!\!\! &0,\!\!\!\!\!\!\!\! &0,\!\!\!\!\!\!\!\! &0,\!\!\!\!\!\!\!\! &0,\!\!\!\!\!\!\!\! &0,\!\!\!\!\!\!\!\! &0,\!\!\!\!\!\!\!\! &0,\!\!\!\!\!\!\!\! &0 \\
\end{array}
\rp
\feqn
\eqn
c_{27}=\lp
\begin{array}{ccccccccccccccccccccccccccc}
0,\!\!\!\!\!\!\!\! &0,\!\!\!\!\!\!\!\! &0,\!\!\!\!\!\!\!\! &0,\!\!\!\!\!\!\!\! &0,\!\!\!\!\!\!\!\! &-\sqrt2,\!\!\!\!\!\!\!\! &0,\!\!\!\!\!\!\!\! &0,\!\!\!\!\!\!\!\! &0,\!\!\!\!\!\!\!\! &0,\!\!\!\!\!\!\!\! &0,\!\!\!\!\!\!\!\! &0,\!\!\!\!\!\!\!\! &0,\!\!\!\!\!\!\!\! &0,\!\!\!\!\!\!\!\! &0,\!\!\!\!\!\!\!\! &0,\!\!\!\!\!\!\!\! &0,\!\!\!\!\!\!\!\! &0,\!\!\!\!\!\!\!\! &0,\!\!\!\!\!\!\!\! &0,\!\!\!\!\!\!\!\! &0,\!\!\!\!\!\!\!\! &0,\!\!\!\!\!\!\!\! &0,\!\!\!\!\!\!\!\! &0,\!\!\!\!\!\!\!\! &0,\!\!\!\!\!\!\!\! &0,\!\!\!\!\!\!\!\! &0 \\
0,\!\!\!\!\!\!\!\! &0,\!\!\!\!\!\!\!\! &0,\!\!\!\!\!\!\!\! &0,\!\!\!\!\!\!\!\! &0,\!\!\!\!\!\!\!\! &0,\!\!\!\!\!\!\!\! &0,\!\!\!\!\!\!\!\! &0,\!\!\!\!\!\!\!\! &0,\!\!\!\!\!\!\!\! &0,\!\!\!\!\!\!\!\! &0,\!\!\!\!\!\!\!\! &0,\!\!\!\!\!\!\!\! &0,\!\!\!\!\!\!\!\! &0,\!\!\!\!\!\!\!\! &0,\!\!\!\!\!\!\!\! &0,\!\!\!\!\!\!\!\! &0,\!\!\!\!\!\!\!\! &0,\!\!\!\!\!\!\!\! &0,\!\!\!\!\!\!\!\! &0,\!\!\!\!\!\!\!\! &0,\!\!\!\!\!\!\!\! &0,\!\!\!\!\!\!\!\! &0,\!\!\!\!\!\!\!\! &0,\!\!\!\!\!\!\!\! &0,\!\!\!\!\!\!\!\! &0,\!\!\!\!\!\!\!\! &0 \\
0,\!\!\!\!\!\!\!\! &0,\!\!\!\!\!\!\!\! &0,\!\!\!\!\!\!\!\! &0,\!\!\!\!\!\!\!\! &0,\!\!\!\!\!\!\!\! &0,\!\!\!\!\!\!\!\! &0,\!\!\!\!\!\!\!\! &0,\!\!\!\!\!\!\!\! &0,\!\!\!\!\!\!\!\! &0,\!\!\!\!\!\!\!\! &0,\!\!\!\!\!\!\!\! &0,\!\!\!\!\!\!\!\! &0,\!\!\!\!\!\!\!\! &0,\!\!\!\!\!\!\!\! &0,\!\!\!\!\!\!\!\! &0,\!\!\!\!\!\!\!\! &0,\!\!\!\!\!\!\!\! &0,\!\!\!\!\!\!\!\! &0,\!\!\!\!\!\!\!\! &0,\!\!\!\!\!\!\!\! &0,\!\!\!\!\!\!\!\! &0,\!\!\!\!\!\!\!\! &0,\!\!\!\!\!\!\!\! &0,\!\!\!\!\!\!\!\! &0,\!\!\!\!\!\!\!\! &0,\!\!\!\!\!\!\!\! &0 \\
\end{array}
\rp
\feqn
\eqn
c_{28}=\lp
\begin{array}{ccccccccccccccccccccccccccc}
0,\!\!\!\!\!\!\!\! &0,\!\!\!\!\!\!\!\! &0,\!\!\!\!\!\!\!\! &-\sqrt2,\!\!\!\!\!\!\!\! &0,\!\!\!\!\!\!\!\! &0,\!\!\!\!\!\!\!\! &0,\!\!\!\!\!\!\!\! &0,\!\!\!\!\!\!\!\! &0,\!\!\!\!\!\!\!\! &0,\!\!\!\!\!\!\!\! &0,\!\!\!\!\!\!\!\! &0,\!\!\!\!\!\!\!\! &0,\!\!\!\!\!\!\!\! &0,\!\!\!\!\!\!\!\! &0,\!\!\!\!\!\!\!\! &0,\!\!\!\!\!\!\!\! &0,\!\!\!\!\!\!\!\! &0,\!\!\!\!\!\!\!\! &0,\!\!\!\!\!\!\!\! &0,\!\!\!\!\!\!\!\! &0,\!\!\!\!\!\!\!\! &0,\!\!\!\!\!\!\!\! &0,\!\!\!\!\!\!\!\! &0,\!\!\!\!\!\!\!\! &0,\!\!\!\!\!\!\!\! &0,\!\!\!\!\!\!\!\! &0 \\
0,\!\!\!\!\!\!\!\! &0,\!\!\!\!\!\!\!\! &0,\!\!\!\!\!\!\!\! &0,\!\!\!\!\!\!\!\! &0,\!\!\!\!\!\!\!\! &0,\!\!\!\!\!\!\!\! &0,\!\!\!\!\!\!\!\! &0,\!\!\!\!\!\!\!\! &0,\!\!\!\!\!\!\!\! &0,\!\!\!\!\!\!\!\! &0,\!\!\!\!\!\!\!\! &0,\!\!\!\!\!\!\!\! &0,\!\!\!\!\!\!\!\! &0,\!\!\!\!\!\!\!\! &0,\!\!\!\!\!\!\!\! &0,\!\!\!\!\!\!\!\! &0,\!\!\!\!\!\!\!\! &0,\!\!\!\!\!\!\!\! &0,\!\!\!\!\!\!\!\! &0,\!\!\!\!\!\!\!\! &0,\!\!\!\!\!\!\!\! &0,\!\!\!\!\!\!\!\! &0,\!\!\!\!\!\!\!\! &0,\!\!\!\!\!\!\!\! &0,\!\!\!\!\!\!\!\! &0,\!\!\!\!\!\!\!\! &0 \\
0,\!\!\!\!\!\!\!\! &0,\!\!\!\!\!\!\!\! &0,\!\!\!\!\!\!\!\! &0,\!\!\!\!\!\!\!\! &0,\!\!\!\!\!\!\!\! &0,\!\!\!\!\!\!\!\! &0,\!\!\!\!\!\!\!\! &0,\!\!\!\!\!\!\!\! &0,\!\!\!\!\!\!\!\! &0,\!\!\!\!\!\!\!\! &0,\!\!\!\!\!\!\!\! &0,\!\!\!\!\!\!\!\! &0,\!\!\!\!\!\!\!\! &0,\!\!\!\!\!\!\!\! &0,\!\!\!\!\!\!\!\! &0,\!\!\!\!\!\!\!\! &0,\!\!\!\!\!\!\!\! &0,\!\!\!\!\!\!\!\! &0,\!\!\!\!\!\!\!\! &0,\!\!\!\!\!\!\!\! &0,\!\!\!\!\!\!\!\! &0,\!\!\!\!\!\!\!\! &0,\!\!\!\!\!\!\!\! &0,\!\!\!\!\!\!\!\! &0,\!\!\!\!\!\!\!\! &0,\!\!\!\!\!\!\!\! &0 \\
\end{array}
\rp
\feqn
\eqn
c_{29}=\lp
\begin{array}{ccccccccccccccccccccccccccc}
0,\!\!\!\!\!\!\!\! &\sqrt2,\!\!\!\!\!\!\!\! &0,\!\!\!\!\!\!\!\! &0,\!\!\!\!\!\!\!\! &0,\!\!\!\!\!\!\!\! &0,\!\!\!\!\!\!\!\! &0,\!\!\!\!\!\!\!\! &0,\!\!\!\!\!\!\!\! &0,\!\!\!\!\!\!\!\! &0,\!\!\!\!\!\!\!\! &0,\!\!\!\!\!\!\!\! &0,\!\!\!\!\!\!\!\! &0,\!\!\!\!\!\!\!\! &0,\!\!\!\!\!\!\!\! &0,\!\!\!\!\!\!\!\! &0,\!\!\!\!\!\!\!\! &0,\!\!\!\!\!\!\!\! &0,\!\!\!\!\!\!\!\! &0,\!\!\!\!\!\!\!\! &0,\!\!\!\!\!\!\!\! &0,\!\!\!\!\!\!\!\! &0,\!\!\!\!\!\!\!\! &0,\!\!\!\!\!\!\!\! &0,\!\!\!\!\!\!\!\! &0,\!\!\!\!\!\!\!\! &0,\!\!\!\!\!\!\!\! &0 \\
-\frac 1{\sqrt2},\!\!\!\!\!\!\!\! &0,\!\!\!\!\!\!\!\! &0,\!\!\!\!\!\!\!\! &0,\!\!\!\!\!\!\!\! &0,\!\!\!\!\!\!\!\! &0,\!\!\!\!\!\!\!\! &0,\!\!\!\!\!\!\!\! &0,\!\!\!\!\!\!\!\! &0,\!\!\!\!\!\!\!\! &0,\!\!\!\!\!\!\!\! &0,\!\!\!\!\!\!\!\! &0,\!\!\!\!\!\!\!\! &0,\!\!\!\!\!\!\!\! &0,\!\!\!\!\!\!\!\! &0,\!\!\!\!\!\!\!\! &0,\!\!\!\!\!\!\!\! &0,\!\!\!\!\!\!\!\! &0,\!\!\!\!\!\!\!\! &0,\!\!\!\!\!\!\!\! &0,\!\!\!\!\!\!\!\! &0,\!\!\!\!\!\!\!\! &0,\!\!\!\!\!\!\!\! &0,\!\!\!\!\!\!\!\! &0,\!\!\!\!\!\!\!\! &0,\!\!\!\!\!\!\!\! &0,\!\!\!\!\!\!\!\! &0 \\
0,\!\!\!\!\!\!\!\! &0,\!\!\!\!\!\!\!\! &0,\!\!\!\!\!\!\!\! &0,\!\!\!\!\!\!\!\! &0,\!\!\!\!\!\!\!\! &0,\!\!\!\!\!\!\!\! &0,\!\!\!\!\!\!\!\! &0,\!\!\!\!\!\!\!\! &0,\!\!\!\!\!\!\!\! &0,\!\!\!\!\!\!\!\! &0,\!\!\!\!\!\!\!\! &0,\!\!\!\!\!\!\!\! &0,\!\!\!\!\!\!\!\! &0,\!\!\!\!\!\!\!\! &0,\!\!\!\!\!\!\!\! &0,\!\!\!\!\!\!\!\! &0,\!\!\!\!\!\!\!\! &0,\!\!\!\!\!\!\!\! &0,\!\!\!\!\!\!\!\! &0,\!\!\!\!\!\!\!\! &0,\!\!\!\!\!\!\!\! &0,\!\!\!\!\!\!\!\! &0,\!\!\!\!\!\!\!\! &0,\!\!\!\!\!\!\!\! &0,\!\!\!\!\!\!\!\! &0,\!\!\!\!\!\!\!\! &0 \\
\end{array}
\rp
\feqn
\eqn
c_{30}=\lp
\begin{array}{ccccccccccccccccccccccccccc}
0,\!\!\!\!\!\!\!\! &0,\!\!\!\!\!\!\!\! &0,\!\!\!\!\!\!\!\! &0,\!\!\!\!\!\!\!\! &0,\!\!\!\!\!\!\!\! &0,\!\!\!\!\!\!\!\! &0,\!\!\!\!\!\!\!\! &0,\!\!\!\!\!\!\!\! &0,\!\!\!\!\!\!\!\! &0,\!\!\!\!\!\!\!\! &0,\!\!\!\!\!\!\!\! &0,\!\!\!\!\!\!\!\! &0,\!\!\!\!\!\!\!\! &0,\!\!\!\!\!\!\!\! &0,\!\!\!\!\!\!\!\! &0,\!\!\!\!\!\!\!\! &0,\!\!\!\!\!\!\!\! &0,\!\!\!\!\!\!\!\! &0,\!\!\!\!\!\!\!\! &0,\!\!\!\!\!\!\!\! &0,\!\!\!\!\!\!\!\! &0,\!\!\!\!\!\!\!\! &0,\!\!\!\!\!\!\!\! &0,\!\!\!\!\!\!\!\! &0,\!\!\!\!\!\!\!\! &0,\!\!\!\!\!\!\!\! &0 \\
0,\!\!\!\!\!\!\!\! &0,\!\!\!\!\!\!\!\! &0,\!\!\!\!\!\!\!\! &0,\!\!\!\!\!\!\!\! &0,\!\!\!\!\!\!\!\! &0,\!\!\!\!\!\!\!\! &0,\!\!\!\!\!\!\!\! &0,\!\!\!\!\!\!\!\! &-\frac 12,\!\!\!\!\!\!\!\! &0,\!\!\!\!\!\!\!\! &0,\!\!\!\!\!\!\!\! &0,\!\!\!\!\!\!\!\! &0,\!\!\!\!\!\!\!\! &0,\!\!\!\!\!\!\!\! &0,\!\!\!\!\!\!\!\! &0,\!\!\!\!\!\!\!\! &0,\!\!\!\!\!\!\!\! &0,\!\!\!\!\!\!\!\! &0,\!\!\!\!\!\!\!\! &0,\!\!\!\!\!\!\!\! &0,\!\!\!\!\!\!\!\! &0,\!\!\!\!\!\!\!\! &0,\!\!\!\!\!\!\!\! &0,\!\!\!\!\!\!\!\! &0,\!\!\!\!\!\!\!\! &0,\!\!\!\!\!\!\!\! &0 \\
0,\!\!\!\!\!\!\!\! &0,\!\!\!\!\!\!\!\! &0,\!\!\!\!\!\!\!\! &0,\!\!\!\!\!\!\!\! &\frac 12,\!\!\!\!\!\!\!\! &0,\!\!\!\!\!\!\!\! &0,\!\!\!\!\!\!\!\! &0,\!\!\!\!\!\!\!\! &0,\!\!\!\!\!\!\!\! &0,\!\!\!\!\!\!\!\! &0,\!\!\!\!\!\!\!\! &0,\!\!\!\!\!\!\!\! &0,\!\!\!\!\!\!\!\! &0,\!\!\!\!\!\!\!\! &0,\!\!\!\!\!\!\!\! &0,\!\!\!\!\!\!\!\! &0,\!\!\!\!\!\!\!\! &0,\!\!\!\!\!\!\!\! &0,\!\!\!\!\!\!\!\! &0,\!\!\!\!\!\!\!\! &0,\!\!\!\!\!\!\!\! &0,\!\!\!\!\!\!\!\! &0,\!\!\!\!\!\!\!\! &0,\!\!\!\!\!\!\!\! &0,\!\!\!\!\!\!\!\! &0,\!\!\!\!\!\!\!\! &0 \\
\end{array}%
\rp
\feqn
}

\newpage
%%%%%%%%%%%%%%%%%%%%%%%%%%%%%%%%%%%%%%%%%%%%%%%%%%%%%%%%%%%%%%%%%%%%%%%%%%
\section{Structure constants.} \label{app:struttura}
\setcounter{equation}{0}
The structure constants ${s_{IJ}}^K$ are defined by $[c_I,c_J]=\sum_{K=1}^{52}{s_{IJ}}^K c_K$. We found that the coefficients
$s_{IJK}:={s_{IJ}}^K$ are completely antisymmetric in the indices and the non-vanishing terms, up to symmetries, are
\arr
\bea{cccc}
s_{1,2,3}=-1 \ , & s_{1,4,5}=-1 \ , & s_{1,7,8}=-1 \ , & s_{1,11,12}=-1 \ , \\
s_{1,16,17}=-1 \ , & s_{1,22,23}=-\dem \ , & s_{1,24,26}=\dem \ , & s_{1,25,29}=\dem \ , \\
s_{1,27,28}=\dem \ , & s_{1,30,31}=-1 \ , & s_{1,37,38}=-\dem \ , & s_{1,39,41}=\dem \ , \\
s_{1,40,44}=\dem \ , & s_{1,42,43}=\dem \ , & s_{1,45,46}=-1 \ , & s_{2,4,6}=-1 \ , \\
s_{2,7,9}=-1 \ , & s_{2,11,13}=-1 \ , & s_{2,16,18}=-1 \ , & s_{2,22,24}=-\dem \ , \\
s_{2,23,26}=-\dem \ , & s_{2,25,27}=\dem \ , & s_{2,28,29}=\dem \ , & s_{2,30,32}=-1 \ , \\
s_{2,37,39}=-\dem \ , & s_{2,38,41}=-\dem \ , & s_{2,40,42}=\dem \ , & s_{2,43,44}=\dem \ , \\
s_{2,45,47}=-1 \ , & s_{3,5,6}=-1 \ , & s_{3,8,9}=-1 \ , & s_{3,12,13}=-1 \ , \\
s_{3,17,18}=-1 \ , & s_{3,22,26}=\dem \ , & s_{3,23,24}=-\dem \ , & s_{3,25,28}=-\dem \ , \\
s_{2,27,29}=\dem \ , & s_{3,31,32}=-1 \ , & s_{3,37,41}=\dem \ , & s_{3,38,39}=-\dem \ , \\
s_{3,40,43}=-\dem \ , & s_{3,42,44}=\dem \ , & s_{3,46,47}=-1 \ , & s_{4,7,10}=-1 \ , \\
s_{4,11,14}=-1 \ , & s_{4,16,19}=-1 \ , & s_{4,22,25}=-\dem \ , & s_{4,23,29}=-\dem \ , \\
s_{4,24,27}=-\dem \ , & s_{4,26,28}=\dem \ , & s_{4,30,33}=-1 \ , & s_{4,37,40}=-\dem \ , \\
s_{4,38,44}=-\dem \ , & s_{4,39,42}=-\dem \ , & s_{4,41,43}=\dem \ , & s_{4,45,48}=-1 \ , \\
s_{5,8,10}=-1 \ , & s_{5,12,14}=-1 \ , & s_{5,17,19}=-1 \ , & s_{5,22,29}=\dem \ , \\
s_{5,23,25}=-\dem \ , & s_{5,24,28}=\dem \ , & s_{5,26,27}=\dem \ , & s_{5,31,33}=-1 \ , \\
s_{5,37,44}=\dem \ , & s_{5,38,40}=-\dem \ , & s_{5,39,43}=\dem \ , & s_{5,41,42}=\dem \ , \\
s_{5,46,48}=-1 \ , & s_{6,9,10}=-1 \ , & s_{6,13,14}=-1 \ , & s_{6,18,19}=-1 \ , \\
s_{6,22,27}=\dem \ , & s_{6,23,28}=-\dem \ , & s_{6,24,25}=-\dem \ , & s_{6,26,29}=-\dem \ , \\
s_{6,32,33}=-1 \ , & s_{6,37,42}=\dem \ , & s_{6,38,43}=-\dem \ , & s_{6,39,40}=-\dem \ , \\
s_{6,41,44}=-\dem \ , & s_{6,47,48}=-1 \ , & s_{7,11,15}=-1 \ , & s_{7,16,20}=-1 \ ,\\
s_{7,22,26}=-\dem \ , & s_{7,23,24}=-\dem \ , & s_{7,25,28}=-\dem \ , & s_{7,27,29}=\dem \ , \\
s_{7,30,34}=-1 \ , & s_{7,37,41}=-\dem \ , & s_{7,38,39}=\dem \ , & s_{7,40,43}=-\dem \ , \\
s_{7,42,44}=\dem \ , & s_{7,45,49}=-1 \ , & s_{8,12,15}=-1 \ , & s_{8,17,20}=-1 \ , \\
s_{8,22,24}=-\dem \ , & s_{8,23,26}=-\dem \ , & s_{8,25,27}=-\dem \ , & s_{8,28,29}=-\dem \ , \\
s_{8,31,34}=-1 \ , & s_{8,37,39}=-\dem \ , & s_{8,38,41}=-\dem \ , & s_{8,40,42}=-\dem \ , \\
s_{8,43,44}=-\dem \ , & s_{8,46,49}=-1 \ , & s_{9,13,15}=-1 \ , & s_{9,18,20}=-1 \ , \\
s_{9,22,23}=\dem \ , & s_{9,24,26}=-\dem \ , & s_{9,25,29}=\dem \ , & s_{9,27,28}=\dem \ , \\
s_{9,32,34}=-1 \ , & s_{9,37,38}=\dem \ , & s_{9,39,41}=-\dem \ , & s_{9,40,44}=\dem \ , \\
s_{9,42,43}=\dem \ , & s_{9,47,49}=-1 \ , & s_{10,14,15}=-1 \ , & s_{10,19,20}=-1 \ , \\
s_{10,22,28}=\dem \ , & s_{10,23,27}=\dem \ , & s_{10,24,29}=-\dem \ , & s_{10,25,26}=-\dem \ , \\
s_{10,33,34}=-1 \ , & s_{10,37,43}=\dem \ , & s_{10,38,42}=\dem \ , & s_{10,39,44}=-\dem \ , \\
s_{10,40,41}=-\dem \ , & s_{10,48,49}=-1 \ , & s_{11,16,21}=-1 \ , & s_{11,22,27}=-\dem \ , \\
s_{11,23,28}=-\dem \ , & s_{11,24,25}=\dem \ , & s_{11,26,29}=-\dem \ , & s_{11,30,35}=-1 \ , \\
s_{11,37,42}=-\dem \ , & s_{11,38,43}=-\dem \ , & s_{11,39,40}=\dem \ , & s_{11,41,44}=-\dem \ , \\
s_{11,45,50}=-1 \ , & s_{12,17,21}=-1 \ , & s_{12,22,28}=\dem \ , & s_{12,23,27}=-\dem \ , \\
s_{12,24,29}=-\dem \ , & s_{12,25,26}=\dem \ , & s_{12,31,35}=-1 \ , & s_{12,37,43}=\dem \ , \\
s_{12,38,42}=-\dem \ , & s_{12,39,44}=-\dem \ , & s_{12,40,41}=\dem \ , & s_{12,46,50}=-1 \ , \\
s_{13,18,21}=-1 \ , & s_{13,22,25}=-\dem \ , & s_{13,23,29}=\dem \ , & s_{13,24,27}=-\dem \ , \\
\ea
\farr
\arr
\bea{cccc}
s_{13,26,28}=-\dem \ , & s_{13,32,35}=-1 \ , & s_{13,37,40}=-\dem \ , & s_{13,38,44}=\dem \ , \\
s_{13,39,42}=-\dem \ , & s_{13,41,43}=-\dem \ , & s_{13,47,50}=-1 \ , & s_{14,19,21}=-1 \ , \\
s_{14,22,24}=\dem \ , & s_{14,23,26}=-\dem \ , & s_{14,25,27}=-\dem \ , & s_{14,28,29}=\dem \ , \\
s_{14,33,35}=-1 \ , & s_{14,37,39}=\dem \ , & s_{14,38,41}=-\dem \ , & s_{14,40,42}=-\dem \ , \\
s_{14,43,44}=\dem \ , & s_{14,48,50}=-1 \ , & s_{15,20,21}=-1 \ , & s_{15,22,29}=\dem \ , \\
s_{15,23,25}=\dem \ , & s_{15,24,28}=\dem \ , & s_{15,26,27}=-\dem \ , & s_{15,34,35}=-1 \ , \\
s_{15,37,44}=\dem \ , & s_{15,38,40}=\dem \ , & s_{15,39,43}=\dem \ , & s_{15,41,42}=-\dem \ , \\
s_{15,49,50}=-1 \ , & s_{16,22,28}=-\dem \ , & s_{16,23,27}=\dem \ , & s_{16,24,29}=-\dem \ , \\
s_{16,25,26}=\dem \ , & s_{16,30,36}=-1 \ , & s_{16,37,43}=-\dem \ , & s_{16,38,42}=\dem \ , \\
s_{16,39,44}=-\dem \ , & s_{16,40,41}=\dem \ , & s_{16,45,51}=-1 \ , & s_{17,22,27}=-\dem \ , \\
s_{17,23,28}=-\dem \ , & s_{17,24,25}=-\dem \ , & s_{17,26,29}=\dem \ , & s_{17,31,36}=-1 \ , \\
s_{17,37,42}=-\dem \ , & s_{17,38,43}=-\dem \ , & s_{17,39,40}=-\dem \ , & s_{17,41,44}=\dem \ , \\
s_{17,46,51}=-1 \ , & s_{18,22,29}=\dem \ , & s_{18,23,25}=\dem \ , & s_{18,24,28}=-\dem \ , \\
s_{18,26,27}=\dem \ , & s_{18,32,36}=-1 \ , & s_{18,37,44}=\dem \ , & s_{18,38,40}=\dem \ , \\
s_{18,39,43}=-\dem \ , & s_{18,41,42}=\dem \ , & s_{18,47,51}=-1 \ , & s_{19,22,26}=-\dem \ , \\
s_{19,23,24}=-\dem \ , & s_{19,25,28}=-\dem \ , & s_{19,27,29}=-\dem \ , & s_{19,33,36}=-1 \ , \\
s_{19,37,41}=-\dem \ , & s_{19,38,39}=-\dem \ , & s_{19,40,43}=-\dem \ , & s_{19,42,44}=-\dem \ , \\
s_{19,48,51}=-1 \ , & s_{20,22,25}=\dem \ , & s_{20,23,29}=-\dem \ , & s_{20,24,27}=-\dem \ , \\
s_{20,26,28}=-\dem \ , & s_{20,34,36}=-1 \ , & s_{20,37,40}=\dem \ , & s_{20,38,44}=-\dem \ , \\
s_{20,39,42}=-\dem \ , & s_{20,41,43}=-\dem \ , & s_{20,49,51}=-1 \ , & s_{21,22,23}=\dem \ , \\
s_{21,24,26}=\dem \ , & s_{21,25,29}=\dem \ , & s_{21,27,28}=-\dem \ , &
s_{21,35,36}=-1 \ , \\
s_{21,37,38}=\dem \ , & s_{21,39,41}=\dem \ , & s_{21,40,44}=\dem \ , & s_{21,42,43}=-\dem \ , \\
s_{21,50,51}=-1 \ , & s_{22,23,33}=-\dem \ , & s_{22,24,36}=-\dem \ , & s_{22,25,31}=\dem \ , \\
s_{22,26,35}=-\dem \ , & s_{22,27,34}=-\dem \ , & s_{22,28,32}=\dem \ , & s_{22,29,30}=\dem \ , \\
s_{22,37,52}=-\dem \ , & s_{22,38,48}=-\dem \ , & s_{22,39,51}=-\dem \ , & s_{22,40,46}=\dem \ , \\
s_{22,41,50}=-\dem \ , & s_{22,42,49}=\dem \ , & s_{22,43,47}=\dem \ , & s_{22,44,45}=-\dem \ , \\
s_{23,24,35}=-\dem \ , & s_{23,25,30}=-\dem \ , & s_{23,26,36}=\dem \ , & s_{23,27,32}=\dem \ ,\\
s_{23,28,34}=-\dem \ , & s_{23,29,35}=\dem \ , & s_{23,37,48}=\dem \ , & s_{23,38,52}=-\dem \ , \\
s_{23,39,50}=-\dem \ , & s_{23,40,45}=-\dem \ , & s_{23,41,51}=\dem \ , & s_{23,42,47}=\dem \ , \\
s_{23,43,49}=-\dem \ , & s_{23,44,46}=\dem \ , & s_{24,25,34}=\dem \ , &
s_{24,26,33}=-\dem \ , \\
s_{24,27,31}=-\dem \ , & s_{24,28,30}=-\dem \ , & s_{24,29,32}=\dem \ , & s_{24,37,51}=\dem \ , \\
s_{24,38,50}=\dem \ , & s_{24,39,52}=-\dem \ , & s_{24,40,49}=\dem \ , & s_{24,41,48}=-\dem \ , \\
s_{24,42,46}=-\dem \ , & s_{24,43,45}=-\dem \ , & s_{24,44,47}=\dem \ , & s_{25,26,32}=\dem \ , \\
s_{25,27,36}=-\dem \ , & s_{25,28,35}=\dem \ , & s_{25,29,33}=\dem \ , & s_{25,37,46}=-\dem \ , \\
s_{25,38,45}=\dem \ , & s_{25,39,49}=-\dem \ , & s_{25,40,52}=-\dem \ , & s_{25,41,47}=\dem \ , \\
s_{25,42,51}=-\dem \ , & s_{25,43,50}=\dem \ , & s_{25,44,48}=\dem \ , & s_{26,27,30}=-\dem \ , \\
s_{26,28,31}=\dem \ , & s_{26,29,34}=\dem \ , & s_{26,37,50}=\dem \ , & s_{26,38,51}=-\dem \ , \\
s_{26,39,48}=\dem \ , & s_{26,40,47}=-\dem \ , & s_{26,41,52}=-\dem \ , & s_{26,42,45}=-\dem \ , \\
s_{26,43,46}=\dem \ , & s_{26,44,49}=\dem \ , & s_{27,28,33}=-\dem \ , & s_{27,29,35}=\dem \ , \\
s_{27,37,49}=-\dem \ , & s_{27,38,47}=-\dem \ , & s_{27,39,46}=\dem \ , & s_{27,40,51}=\dem \ , \\
s_{27,41,45}=\dem \ , & s_{27,42,52}=-\dem \ , & s_{27,43,48}=-\dem \ , & s_{27,44,50}=\dem \ , \\
s_{28,29,36}=\dem \ , & s_{28,37,47}=-\dem \ , & s_{28,38,49}=\dem \ , & s_{28,39,45}=\dem \ , \\
s_{28,40,50}=-\dem \ , & s_{28,41,46}=-\dem \ , & s_{28,42,48}=\dem \ , & s_{28,43,52}=-\dem \ , \\
s_{28,44,51}=\dem \ , & s_{29,37,45}=-\dem \ , & s_{29,38,46}=-\dem \ , & s_{29,39,47}=-\dem \ , \\
s_{29,40,48}=-\dem \ , & s_{29,41,49}=-\dem \ , & s_{29,42,50}=-\dem \ , & s_{29,43,51}=-\dem \ , \\
s_{29,44,52}=-\dem \ , & s_{30,37,44}=-\dem \ , & s_{30,38,40}=\dem \ , & s_{30,39,43}=\dem \ , \\
\ea
\farr
\arr
\bea{cccc}
s_{30,41,42}=\dem \ , & s_{30,45,52}=-1 \ , & s_{31,37,40}=-\dem \ , & s_{31,38,44}=-\dem \ , \\
s_{31,39,42}=\dem \ , & s_{31,41,43}=-\dem \ , & s_{31,46,52}=-1 \ , & s_{32,37,43}=-\dem \ , \\
s_{32,38,42}=-\dem \ , & s_{32,39,44}=-\dem \ , & s_{32,40,41}=-\dem \ , & s_{32,47,52}=-1 \ , \\
s_{33,37,38}=\dem \ , & s_{33,39,41}=\dem \ , & s_{33,40,44}=-\dem \ , & s_{33,42,43}=\dem \ , \\
s_{33,48,52}=-1 \ , & s_{34,37,42}=-\dem \ , & s_{34,38,43}=\dem \ , & s_{34,39,40}=-\dem \ , \\
s_{34,41,44}=-\dem \ , & s_{34,49,52}=-1 \ , & s_{35,37,41}=\dem \ , & s_{35,38,39}=\dem \ , \\
s_{35,40,43}=-\dem \ , & s_{35,42,44}=-\dem \ , & s_{35,50,52}=-1 \ , & s_{36,37,39}=\dem \ , \\
s_{36,38,41}=-\dem \ , & s_{36,40,42}=\dem \ , & s_{36,43,44}=-\dem \ , & s_{36,51,52}=-1 \ .
\ea
\farr

%%%%%%%%%%%%%%%%%%%%%%%%%%%%%%%%%%%%%%%%%%%%%%
\section{Surjectivity of the product parametrization.}\label{proof}
\setcounter{equation}{0}
We will follow \cite{chavel}.
First note that $M\equiv G/H$ is a compact homogeneous manifold. Let $\pi:G\longrightarrow G/H$ and
$\pi_\cP :\cG \longrightarrow \cP$ be the natural projections. If $G$ is provided with a bi-invariant Riemannian metric
(as it happens for simple compact Lie groups) then $M$ can also be provided with such an invariant metric. In particular
for compact semisimple Lie groups we can use the Killing metric. The Levi-Civita connection is then exactly the
connection induced by the horizontal distribution defined by taking $(L_g)_*\cP$ as horizontal space at any
$g\in G$, where $L_g$ is the left multiplication on $G$ and the lower $*$ indicate the push-forward.\footnote{Here we
are using the fact that $G\stackrel{\pi}{\longrightarrow} G/H$ is a principal bundle over
$M$, see \cite{KN}. In particular if $T_eG\simeq \cG$ is the tangent space to the identity $e$ of $G$, then
$\cP$ is its horizontal component.}
The invariant metric on $M$ is then obtained by requiring for $\pi_*:T_g G \longrightarrow T_{\pi(g)}M$ to be
an isometry between the horizontal component of $T_g G$ and $T_{\pi(g)}M$ for any $g\in G$. Thus $M$ is geodesically
complete and $\pi$ becomes
a Riemannian submersion. From this, if $o:=\pi (H)$, $Exp_o :T_oM \longrightarrow M$ is the exponential map generated
by the geodesic flow from $o$ and $\exp: \cG \longrightarrow G$ is the exponential map of the Lie group, then it can
be shown that $Exp_o (a)=\pi \exp (a)$ for any $a\in \cP$ (\cite{chavel}, p.47 exercise 1.21). But $Exp_o$ is surjective,
as follows from the Hopf-Rinow theorem, (\cite{chavel}, theorem 1.9). This completes the proof.

%%%%%%%%%%%%%%%%%%%%%%%%%%%%%%%%%%%%%%%%%%%%%%
%\section{The Macdonald formula.} \label{app:macdonald}

%%%%%%%%%%%%%%%%%%%%%%%%%%%%%%%%%%%%%%%%%%%%%%
\section{The orthogonal subgroups.} \label{app:macdonald}
\setcounter{equation}{0}
In this section we collect the volumes of the orthogonal subgroups obtained by means of the Macdonald formula.
The rational homology groups of all simple Lie groups are known to be the homology of the product of odd dimensional spheres:
$H_* (G)=H_*(S^{d_1}\times \ldots \times S^{d_r})$, $r$ being the rank of the group \cite{hopf}. For the subgroups of $F_4$
we find
\begin{eqnarray*}
&& SO(3) :  \qquad d_1=3 \ ;\\
&& SO(4) :  \qquad d_1=3, \ d_2=3 \ ; \\
&& SO(5) :  \qquad d_1=3, \ d_2=7 \ ; \\
&& SO(6) :  \qquad d_1=3, \ d_2=5, \ d_3=7 \ ; \\
&& SO(7) :  \qquad d_1=3, \ d_2=7, \ d_3=11 \ ; \\
&& SO(8) :  \qquad d_1=3, \ d_2=7, \ d_3=7, \ d_4=11 \ ; \\
&& SO(9) :  \qquad d_1=3, \ d_2=7, \ d_3=11, \ d_4=15 \ .
\end{eqnarray*}
The roots of the subgroups are the ones given in \cite{Fulton-harris}, with $L_i=e_i$ identified with the usual orthonormal bases
of an Euclidean space. The volumes can then easily computed by mean of the Macdonald formula, giving
\begin{eqnarray*}
&& SO(3) :  \qquad V=2^4 \cdot \pi^2 \ ;\\
&& SO(4) :  \qquad V=2^5 \cdot \pi^4 \ ; \\
&& SO(5) :  \qquad V=\frac {2^8 \cdot \pi^{6}}3 \ ; \\
&& SO(6) :  \qquad V=\frac {2^8 \cdot \pi^{9}}{3} \ ; \\
&& SO(7) :  \qquad V=\frac {2^{12} \cdot \pi^{12}}{3^2 \cdot 5} \ ; \\
&& SO(8) :  \qquad V=\frac {2^{12} \cdot \pi^{16}}{3^3 \cdot 5} \ ; \\
&& SO(9) :  \qquad V=\frac {2^{17} \cdot \pi^{20}}{3^4 \cdot 5^2 \cdot 7} \ .
\end{eqnarray*}

%%%%%%%%%%%%%%%%%%%%%%%%%%%%%%%%%%%%%%%%%%%%%%
\section{More on the subalgebras.}\label{app:subalgebra}
\setcounter{equation}{0}
In section \ref{algebra} we observed that our $27$ dimensional representation of $F_4$ has a decomposition
${\bf 26}\oplus {\bf 1}$ in irreducible representations. Similar decompositions can be obtained for the subgroups
simply by a direct computation of the weights. For example we computed the decomposition of $so(i)$ for $i=9,8,7,6$
finding:
\begin{itemize}
\item for $so(9)$
$$
so(9)={\bf 16}\oplus {\bf 9} \oplus {\bf 1}^2 \,
$$
where ${\bf 16}$ is the spin representation and {\bf 9} the vector representation;
\item for $so(8)$
$$
so(8)= {\bf 8}_v \oplus {\bf 8}_+ \oplus {\bf 8}_- \oplus {\bf 1}^3 \ .
$$
Here ${\bf 8}_v$ is the vector representation and ${\bf 8}_\pm$ are the spin representations with positive and negative
chirality;
\item for $so(7)$
$$
so(7)={\bf 8}^2\oplus {\bf 7} \oplus {\bf 1}^4\ ,
$$
where ${\bf 8}$ is the spin representation and ${\bf 7}$ is the vector one;
\item for $so(6)$
$$
so(6)={\bf 6}\oplus {\bf 4}^2_+ \oplus {\bf 4}^2_- \oplus {\bf 1}^5 \ ,
$$
where ${\bf 4}_\pm$ are the chiral spin representations and ${\bf 6}$ the vector one.
\end{itemize}

\end{appendix}
%%%%%%%%%%%%%%%%%%%%%%%%%%%%%%%%%%%%%%%%%%%%%%%%%

\end{document}